\documentclass{aa}  

\usepackage{graphicx}
\usepackage{txfonts}
\usepackage{multirow}
\usepackage{amssymb}
\usepackage{amsmath}
\usepackage{wasysym}
\usepackage{commath}
\usepackage{bigdelim}
\usepackage{subfigure}
\usepackage{lscape}
\usepackage{url}
\begin{document}

\title{The UM, a fully--compressible, non--hydrostatic, deep
  atmosphere GCM, applied to hot Jupiters.}
 \titlerunning{The UM applied to hot Jupiters} 
 \subtitle{ENDGame for a HD 209458b test case.}

   \author{N. J. Mayne\inst{1}\fnmsep\thanks{E-mail:
       nathan@astro.ex.ac.uk},
          I. Baraffe\inst{1}, 
          David M. Acreman\inst{1}, 
          Chris Smith\inst{2}, 
          Matthew K. Browning\inst{1},
          David Sk\aa lid Amundsen\inst{1},
          Nigel Wood\inst{2},
          John Thuburn\inst{3}
          \and
          David R. Jackson\inst{2}
          }

          \institute{Physics and Astronomy, College of Engineering,
            Mathematics and Physical Sciences, University of Exeter,
            EX4 4QL.  \and 
            Met Office, FitzRoy Road, Exeter, Devon EX1 3PB, UK.  \and
            Applied Mathematics Group, University of Exeter, Exeter,
            EX4 4QL, United Kingdom.  }

          \authorrunning{Mayne et al.}

   \date{Received September 15, 1996; accepted March 16, 1997}

% \abstract{}{}{}{}{} 
% 5 {} token are mandatory
 
   \abstract{We are adapting the Global Circulation Model (GCM) of the
     UK Met Office, the so--called Unified Model (UM), for the study
     of hot Jupiters. In this work we demonstrate the successful
     adaptation of the most sophisticated dynamical core, the
     component of the GCM which solves the equations of motion for the
     atmosphere, available within the UM, ENDGame (Even Newer Dynamics
     for General atmospheric modelling of the environment). Within the
     same numerical scheme ENDGame supports solution to the dynamical
     equations under varying degrees of simplification. We present
     results from a simple, shallow (in atmospheric domain) hot
     Jupiter model (SHJ), and a more realistic (with a deeper
     atmosphere) HD 209458b test case. For both test cases we find
     that the large--scale, time--averaged (over the 1200 days
     prescribed test period), dynamical state of the atmosphere is
     relatively insensitive to the level of simplification of the
     dynamical equations. However, problems exist when attempting to
     reproduce the results for these test cases derived from other
     models. For the SHJ case the lower (and upper) boundary
     intersects the dominant dynamical features of the atmosphere
     meaning the results are heavily dependent on the boundary
     conditions. For the HD 209458b test case, when using the more
     complete dynamical models, the atmosphere is still clearly
     evolving after 1200 days, and in a transient state. Solving the
     complete (deep atmosphere and non--hydrostatic) dynamical
     equations allows exchange between the vertical and horizontal
     momentum of the atmosphere, via Coriolis and metric
     terms. Subsequently, interaction between the upper atmosphere and
     the deeper more slowly evolving (radiatively inactive) atmosphere
     significantly alters the results, and acts over timescales longer
     than 1200 days.}
 
   \keywords{Hydrodynamics -- Planets and satellites: atmospheres -- Methods: numerical}

   \maketitle
%
%________________________________________________________________
\section{Introduction}
\label{introduction}
Observations made over the last 20 years have enabled the detection of
several hundred exoplanets \citep[the first around a Solar mass star
by][]{mayor_1995} and several thousand candidate systems
\citep[identified, for instance, by the Kepler mission including the
discovery of a system of six planets and a sub--Mercury sized planet,
see][respectively]{lissauer_2011,barclay_2013}. Surveys of variability
(detecting planetary transits) and radial velocity have also provided
estimates of the mass and orbital radii of these exoplanets. Such
surveys are most sensitive to giant ($\sim$Jovian mass) planets which
orbit close to their parent stars, experience intense radiation
($10^3-10^5$ times that received by Jupiter, exacerbating problems
involved with simplified radiative transfer schemes), and are termed
`hot Jupiters'. The strong tidal forces experienced by these planets
is thought to lead to rapid synchronisation of their rotation period
with their orbital period (with the adoption of a reasonable
dissipation parameter). This `tidal--locking' provides a strong
constraint on the planetary rotation rate and means the planet has a
permanent day and night side, experiencing net heating and cooling,
respectively \citep[see][ and references therein]{baraffe_2010}.

Furthermore, precise observations of the luminosity as a function of
time and wavelength (transit spectroscopy) of a transiting
star--planet system can be used to probe the planet's atmospheric
conditions \citep[see][for review]{seager_2010}. Observations of the
primary eclipse (when the planet transits in front of the star) have
provided the detection of specific species in the atmospheres of hot
Jupiters \citep[see for example][who detected potassium and sodium,
respectively, in the atmosphere of Xo--2b]{sing_2011,sing_2012} as
well as the detection of possible dust or hazes \citep[for example as
found in HD 189733b,][]{pont_2012}. Additionally observations of both
the primary and secondary eclipses (when the planet moves behind the
star) have allowed the derivation of day and night side atmospheric
temperatures \citep[for example $\sim$1250 K and $\sim$1000 K for HD
189733b as found by][]{knutson_2007,knutson_2009}. Moreover, using the
full orbital luminosity phase--curve, the atmospheric temperature as a
function of planetary longitude can be inferred \citep[see for
example][]{knutson_2007,knutson_2009}. Analysis of these temperature
`maps' has revealed offsets of the hottest part of the atmosphere or
`hot spot' (at a given depth) from the sub-stellar points. This offset
was predicted by \citet{showman_2002} as a consequence of the expected
fast circulations induced by the large--scale heating. The presence of
such fast winds has been suggested for HD 209458b using Doppler
shifting of molecular CO bands \citep{snellen_2010}, where
$\sim$km$\,$s$^{-1}$ wind speeds were derived.

Although this is not a complete review of the observational results,
the observations have produced several challenges for our theoretical
models of planetary evolution. Many of these challenges require an
understanding of the full three dimensional circulation, including the
vertical transport. Firstly, comparison of the derived radii with the
predictions of planetary interior models (as a function of age) has
shown that some hot Jupiters appear inflated. \citet{guillot_2002}
suggested that the vertical transport of $\sim1$ percent of the
incident stellar flux, from the top of the atmosphere deep into the
planet interior, could halt the planet's gravitational contraction
sufficiently to explain the observations. \citet{showman_2002} then
suggested that the required levels of kinetic energy could be
generated by the large--scale forcing expected in hot Jupiter
atmospheres. Secondly, as is evident for Solar system planets,
significant abundances of scattering particles can dominate the global
heat balance of a planet's atmosphere \citep[see for
  discussion][]{sanchez_lavega_2004}, which is difficult to capture
using simplistic radiative transfer schemes.

The possible presence of scattering particles \citep[suggested to be
MgSiO$_3$ by][]{etangs_2008} in the observable atmospheres of hot
Jupiters requires they be supported against gravitational settling
and/or be replenished via circulations. Therefore, vertical transport
modeled over a large range of pressures is vital to interpreting these
observations, in addition to a non-grey radiative transfer
scheme. Finally, comparison of day and night side temperatures has
revealed a possible dichotomy separating hot Jupiters with efficient
heat redistribution (from day to night side), which are generally more
intensely irradiated, from those exhibiting less efficient heat
redistribution. The efficiency of redistribution has been linked to
the existence of a region of the hot Jupiter upper atmosphere where
temperature increases with height, a thermal inversion, as inferred
from model fitting for HD 209458b \citep{knutson_2008}, and thereby
the presence of absorbing substances such as VO and TiO
\citep{hubeny_2003,fortney_2008}. A correct description of all these
processes requires a non--grey radiative transfer scheme coupled to a
dynamical model of the atmospheric redistribution.

The glimpses into the atmospheres of hot Jupiters provided by the
observations, and the associated puzzles, have motivated the
application of Global Circulation Models (GCMs), usually developed for
the study of Earth's weather and climate, to hot Jupiters. GCMs are
generally comprised of many components or modules which handle
different aspects of the atmosphere. Many of these components are
highly optimised for conditions on Earth, for example treatments of
the surface boundary layer. To apply GCMs to planets other than Earth
adaptation of the most fundamental components i.e. treating radiative
transfer and dynamical motions, is required. Further adaptations to
more detailed atmospheric process can then occur when merited by
observations.

GCMs have been successfully applied to model other Solar system
planets \citep[see for example models of Jupiter, Saturn, Mars and
Venus:][respectively]{yamazaki_2004,muller_wodarg_2006,hollingsworth_2011,lebonnois_2011},
but hot Jupiters present a very different regime. The latter receive
significantly more radiation and rotate much more slowly than the
giant planets in our Solar system. Therefore, the characteristic
scales of atmospheric features such as the expected vortex size, the
Rossby deformation radius and the elongation in the east--west
direction of wind structures, the Rhines scale are both approximately
the size of the planet \citep[proportionally much larger than for
Solar system planets, see][for a review and comparison with Solar
system planets]{showman_2011}. This effectively means that one might
expect any `weather systems', or jet structures (i.e. prevailing
circumplanetary flows) to be comparable in size with the horizontal
extent of the atmosphere. Despite the exotic nature of the flow
regime, the adaptation of GCMs to hot Jupiters has met with success
as, for example, several models have been able to demonstrate that
offsets in the `hot spot' are consistent with redistribution from
zonal (longitudinal direction) winds
\citep{showman_2009,dobbs_dixon_2008,dobbs_dixon_2009}. The progress
of the modelling has been reviewed in
\citet{showman_2008,showman_2011} and a useful summary of the
different approaches taken can be found in \citet{dobbs_dixon_2012}.

To interpret the observations of hot Jupiters the regime dictates the
model should include solution to the full three--dimensional equations
of motion for a rotating atmosphere coupled to a non--grey radiative
transfer scheme. This will allow exploration of the consequences of
realistic vertical transport and its interaction with the horizontal
advection, and include the effect on the thermal balance of the
atmosphere caused by frequency dependent opacities. Most GCMs applied
to hot Jupiters solve the primitive equations of meteorology,
involving the approximation of vertical hydrostatic equilibrium, and a
`shallow--atmosphere' \citep[combining the constant gravity,
`shallow--fluid' and `traditional' approximations,
see][]{vallis_2006,white_2005}.

The most sophisticated radiative transfer scheme within a GCM, applied
to hot Jupiters, to date is that of \citet{showman_2009} which solves
the primitive equations coupled to a simplified radiative transfer
scheme based on the two-stream, correlated--$k$ method. However, the
approximations involved in the primitive equations neglect the
vertical acceleration of fluid parcels, and the effect of the vertical
velocity on the horizontal momentum. More complete dynamical models,
solving the full Navier--Stokes equations, have been applied to hot
Jupiters by
\citet{dobbs_dixon_2008,dobbs_dixon_2009,dobbs_dixon_2010,dobbs_dixon_2012},
but these models include a radiative transfer scheme more simplified
than the method of \citet{showman_2009}. \citet{dobbs_dixon_2010}
includes frequency dependent radiative transfer via the introduction
of only three opacity bins (and generally runs for short elapsed model
times). \citet{dobbs_dixon_2012} includes a treatment using a similar
number of frequency bins to \citet{showman_2009}, but simply average
the opacity in each bin as opposed to generating opacities via the
correlated--k method. 

Therefore, calculations based on non--grey radiative transfer coupled
to full three--dimensional equations of motion for a rotating
atmosphere do not yet exist. Additionally, current models applied to
hot Jupiters are still missing many other physical processes.
Although not discussed in this paper, treatments of the magnetic
fields, photochemistry and clouds or hazes, may well be required to
create a model capable of meaningful predictions. We are beginning
work on the incorporation of a photochemical network and the simple
modelling of clouds into our model, but this will take some time to
complete.

The ENDGame (Even Newer Dynamics for General atmospheric modelling of
the environment) dynamical core (the part of the GCM solving the
discretised fluid dynamics equations of motion) of the UK Met Office
GCM, the Unified Model (UM) is based on the non--hydrostatic
deep--atmosphere equations
\citep{staniforth_2003,staniforth_2008,wood_2013}, and does not make
the approximations incorporated in the primitive equations
\citep{white_2005}. The UM also includes a two--stream radiative
transfer scheme with correlated--k method. This code has previously
been adapted to studies of Jupiter \citep{yamazaki_2004}, but requires
significant adaptation of the dynamical and radiative transfer schemes
to be applied to hot Jupiters. We have previously presented the
satisfactory completion of several Earth--like test cases of the
dynamical core in \citet{mayne_2013}. Now, we have completed the
adaptation of the dynamical core, and in this work present the first
hot Jupiter test cases. We have completed the Shallow--Hot Jupiter
\citep[SHJ, as prescribed in][]{menou_2009} and the HD 209458b test
case of \citet{heng_2011}. 

Adaptation of the radiative transfer scheme is nearing completion and
coupled models will be presented in a future work. With this paper, we
begin a series in which we will present the details of the model
developments and testing of the UM, as it is adapted for the study of
exoplanets, as well as scientific applications and results.

The structure of the paper is as follows. In Section \ref{model} we
detail the model used including the equations solved and highlight the
important details of our boundary conditions and numerical scheme. We
also discuss the effect of canonical simplifications made to the
dynamical equations. Section \ref{model} also includes an explanation
of the parameterisations used and references to a more detailed
description of the model and previous testing. Section
\ref{test_cases} then describes the setups for two test cases we have
run including the parameter values and temperature and pressure
profiles. We also, in Section \ref{test_cases} demonstrate
satisfactory completion of these tests. Section \ref{discussion}
highlights some problems with the test cases and discusses future
work. Finally, in Section \ref{conclusions} we include a summary of
our conclusions.

%__________________________________________________________________

\section{Model}
\label{model}

The UM dynamical core called ENDGame is explained in detail in
\citet{wood_2013}. The code is based on the non--hydrostatic,
deep--atmosphere (NHD) equations of motion for a planetary atmosphere
\citep{staniforth_2003,staniforth_2008}, including a varying (with
height) gravity and a geometric height vertical grid. Uniquely, the
code allows solution to the non--hydrostatic shallow--atmosphere
\citep[NHS,][]{staniforth_2003,staniforth_2008} equations, or just the
simple assumption of a constant gravity (to create a quasi--NHD
system), within the same numerical scheme.

\subsection{Overview of the numerical scheme}
\label{numerics}

The UM is a finite--difference code where the atmosphere is
discretised onto a latitude--longitude grid (resolutions explained in
Section \ref{test_cases}), using a staggered Arakawa--C grid
\citep{arakawa_1977} and a vertically staggered Charney--Phillips grid
\citep{charney_1953}. The code uses a terrain following height--based
vertical coordinate\footnote{Although for this work we include no
  orography, and have no `surface'.}.

The code is semi--Lagrangian and semi--implicit, where the latter is
based on a Crank--Nicolson scheme. The code employs semi--Lagrangian
advection where the values for advected quantities are derived at
interpolated departure points, and are then used to calculate
quantities within the Eulerian grid. For the semi--implicit scheme the
temporal weighting between the $i$th and the $i+1$th state is set by
the coefficient $\alpha$ which can vary between zero and one, and is
set to 0.55 in this work. For each atmospheric timestep a nested
iteration structure is used. The \textit{outer} iteration performs the
semi--Lagrangian advection (including calculation of the departure
points), and values of the pressure increments from the \textit{inner}
iteration are back substituted to obtain updated values for each
prognostic variable. The \textit{inner} iteration solves the Helmholtz
(elliptical) problem to obtain the pressure increments, and the
Coriolis and nonlinear terms are updated.

The velocity components are staggered such that the meridional
velocity is defined at the pole \citep[see][for a more detailed
explanation]{mayne_2013}, but no other variable is stored at this
location, thereby avoiding the need to solve for pressure at the poles
of the latitude--longitude grid
\citep{wood_2013}. \citet{thuburn_2004} show that mass, angular
momentum and energy are much more readily conserved with a grid
staggered such that $v$ and not $u$ is held at the pole. The stability
afforded by the spatial and temporal discretisation removes the need
for an explicit polar filter (although our diffusion operator has some
aspects in common with a polar filter, see discussion in Section
\ref{diffusion}). The code adopts SI units. A full description of the
code can be found in \citet{wood_2013} and important features relating
to the reproduction of idealised tests are reiterated in
\citet{mayne_2013}.

\subsection{Previous Testing}
\label{testing}
The UM undergoes regular verification for the Earth system, and
\citet{wood_2013} completes several tests from the Dynamical Core
Model Intercomparison Project\footnote{DCMIP, see
  \url{http://earthsystemcog.org/projects/dcmip-2012/}.}, and the
deep--atmosphere baroclinic instability test \citep{ullrich_2013},
using the ENDGame dynamical core. We have also, as part of the
adaptation to exoplanets completed several tests for an Earth--like
model including the Held-Suarez test \citep{held_1994}, the Earth-like
test of \citet{menou_2009} and the Tidally Locked Earth of
\citet{merlis_2010}, the results of which are presented in
\citet{mayne_2013}. Additionally, for each setup used we also complete
a static, non--rotating, hydrostatic isothermal atmosphere test,
ensuring that the horizontal and vertical velocities recorded are
negligibly small and do not grow significantly with time (when run for
a few million iterations).

For simulations we have performed, the longest of which is many
millions of iterations, mass and angular momentum, are conserved to
better than $\lesssim 0.05$\% and $\sim 5$\%, respectively. In the UM
mass is conserved via a correction factor applied after each timestep
\citep[see][for details]{wood_2013}.

\subsection{Equations solved by the dynamical core}
\label{equations_solved}
We model only a section, or spherical shell, of the total atmosphere
and define the material below our inner boundary (discussed in more
detail in Section \ref{boundary}) as the `planet' and the subscript
${\rm p}$ denotes quantities assigned to this region. The dynamical
core solves a set of five equations: one for each momentum component,
a continuity equation for mass and a thermodynamical energy equation,
which are closed by the ideal gas equation. These equations are (using
the ``Full'' equation set, see Table \ref{model_names} for
explanation)
\begin{align}
\frac{Du}{Dt}=\frac{uv\tan\phi}{r}-\frac{uw}{r}+fv-f^{\prime}w-\frac{c_p\theta}{r \cos\phi}\frac{\partial \Pi}{\partial \lambda}+\textbf{D}(u),\\
 \frac{Dv}{Dt}=-\frac{u^2\tan\phi}{r}-\frac{vw}{r}-uf-\frac{c_p\theta}{ r}\frac{\partial \Pi}{\partial \phi}+\textbf{D}(v),\\
 \delta\frac{Dw}{Dt}=\frac{u^2+v^2}{r}+uf^{\prime}-g(r)-c_p\theta\frac{\partial \Pi}{\partial r},\\
 \frac{D\rho}{Dt}=-\rho \left[ \frac{1}{r \cos \phi}\frac{\partial u}{\partial \lambda} \right. +\frac{1}{r \cos\phi}\frac{\partial (v\cos \phi)}{\partial \phi}\left. +\frac{1}{r^2}\frac{\partial (r^2w)}{\partial r} \right],\\
 \frac{D\theta}{Dt}=\frac{Q}{\Pi}+\textbf{D}(\theta),\\
\Pi^{\frac{1-\kappa}{\kappa}}=\frac{R\rho\theta}{p_0},
\label{full_set}
\end{align}
respectively. The coordinates used are $\lambda$, $\phi$, $r$ and $t$,
which are the longitude, latitude (from equator to poles), radial
distance from the centre of the planet and time. The spatial
directions, $\lambda$, $\phi$ and $r$, then have associated wind
components $u$ (zonal), $v$ (meridional) and $w$ (vertical). $c_p$ is
the specific heat capacity, $R$ is the gas constant and $\kappa$ the
ratio of the $R/c_{p}$. $\delta$ is a `switch' (0 or 1) to enable a
quasi--hydrostatic version of the equations \citep[not used in this
  work but detailed in][]{white_2005}. $p_0$ is a chosen reference
pressure and $g(r)$ is the acceleration due to gravity at $(r)$ and is
defined as
\begin{equation}
g(r)=g_{\rm p}\left( \frac{R_{\rm p}}{r}\right)^2,
\label{grav_eg}
\end{equation}
where $g_{\rm p}$ and $R_{\rm p}$ are the gravitational acceleration
and radial position at the inner boundary. $f$ and $f^{\prime}$ are
the Coriolis parameters defined as,
\begin{equation}
f=2\Omega\sin\phi,
\end{equation}
and
\begin{equation}
f^{\prime}=2\Omega\cos\phi,
\end{equation}
where $\Omega$ is the planetary rotation rate. $\rho$ and $\theta$ are
the prognostic variables of density and potential temperature,
respectively. $\Pi$ is the Exner pressure (or function). $\theta$ and
$\Pi$ are then defined, in terms of the temperature, $T$ and pressure,
$p$, as
\begin{equation}
\theta = T\left(\frac{p_0}{p}\right) ^{\frac{R}{c_p}},\\
\end{equation}
and
\begin{equation}
\Pi=\left(\frac{p}{p_0}\right)^{R/c_{p}} = \frac{T}{\theta},\\
\end{equation}
respectively. The material derivative, $\frac{D}{Dt}$ is given by
\begin{equation}
\frac{D}{Dt}=\frac{\partial }{\partial t}+\frac{u}{r\cos\phi}\frac{\partial }{\partial \lambda}+\frac{v}{r}\frac{\partial }{\partial \phi}+w\frac{\partial }{\partial r}.\\
\end{equation}
Finally, $Q$ and $\textbf{D}$ are the heating rate and diffusion
operator (note that diffusion is not applied to the vertical
velocity), respectively. The heating, in this work, is applied using a
temperature relaxation or Newtonian cooling scheme discussed in
Section \ref{radiative_transfer}. The diffusion operator is detailed
in Section \ref{diffusion}.

\subsubsection{Dynamical simplification and variants of the equations of motion}
\label{dynamic_switches}
A quartet of self--consistent governing dynamical equations conserving
axial angular momentum, energy and potential vorticity are described
in detail in \citet{white_2005}. These are the hydrostatic primitive
equations or (hydrostatic) primitive equations (HPEs),
quasi--hydrostatic equations (QHEs), the non--hydrostatic
shallow--atmosphere (NHS) equations and the non--hydrostatic
deep--atmosphere (NHD) equations. For this work we, using ENDGame,
solve the NHD (which are detailed in Section \ref{equations_solved}),
the quasi--NHD (with a constant gravity) and the NHS
equations. \citet{white_2005} includes a full discussion of the
assumptions made in each equation set, the most relevant (for this
work) of which are included in Table \ref{model_names} alongside the
validity criteria, and an estimate for the validity on HD 209458b (an
example hot Jupiter).

Table \ref{model_names} also includes the short reference names we
have used to describe each setup. We adopt the nomenclature of
\citet{white_2005}, where the `shallow--atmosphere' approximation
implies constant gravity, in addition to the adoption of the
`shallow--fluid' and `traditional' approximations. In this work we run
simulations using the ``Full'' (NHD), ``Deep'' (quasi--NHD
i.e. constant gravity) and ``Shallow'' (NHS) equations sets, where the
primitive equations (HPEs) are included as illustrative of the
codes we compare against. However, the GCMs we are comparing with use
either pressure or $\sigma$ ($=p / p_{\rm surf}$, where $p_{\rm surf}$
is the pressure at the inner boundary, which is usually called the
``surface'' for terrestrial planets) as their vertical coordinate. The
key point is that when we compare models using for instance the
``Shallow'' equations with the primitive equations, we are simply
relaxing the assumption of hydrostatic equilibrium. Moving to the
``Deep'' equations then involves further relaxing the `shallow--fluid'
and `traditional' approximation (but retaining a constant gravity) and
finally, the ``Full'' equations include a further relaxation of the
constant gravity approximation.

\begin{table*}
  \caption{Key assumptions in each equation set, the local name used
    to describe the set, and the validity, both in general and for HD
    209458b. $\Phi$ and $G$ are the geopotential and gravitational
    constant respectively. $M_{\rm p}$ and $M_{\rm atm}$ are the total
    masses of the planet below the inner boundary, and of the
    atmosphere, respectively. $z$ is the vertical height from the
    inner boundary. $H$ and $L$ are the vertical and horizontal sizes
    of the atmospheric domain. Finally, $N$ is the buoyancy (or
    Brunt-V\"ais\"al\"a) frequency,
    $N=\sqrt{-\frac{g(r)}{\rho_0}\frac{\partial\rho(r)}{\partial
        r}}$). SI units are used unless otherwise stated.}
\label{model_names}
\centering
\begin{tabular}{ccccllcc}
  \hline\hline
  \multicolumn{4}{c}{Name}&\multicolumn{2}{l}{Approximation}&Formal Condition&HD 209458b\\
  \hline
  \hline
  \multirow{7}{*}{Primitive}\rdelim\{{7}{0mm}[]&\multirow{6}{*}{Shallow}\rdelim\{{6}{0mm}[]&\multirow{3}{*}{Deep}\rdelim\{{3}{0mm}[]&\multirow{2}{*}{Full}\rdelim\{{2}{0mm}[]&Spherical geopotentials$^{(1)}$&$\Phi(\lambda,\phi,r)=\Phi (r)$&$\Omega^2r\ll g$$^{(2)}$&$10^{-2}\ll 10^{1}$\\
  &&&&no self-gravity&$g(r)=\frac{GM_{\rm p}}{r^2}$&$M_{\rm atm}\ll M_{\rm p}$&$\sim 10^{23}\ll 10^{27}$\\
  &&&&constant gravity&$g(r)=g_{\rm p}=\frac{GM_{\rm p}}{R_{\rm p}^2}$&$z\ll R_{\rm p}$&\rdelim\{{2}{4mm}[]\multirow{2}{*}{$\sim 10^{7} < 10^{8}$}\\
  &&&&`shallow--fluid'&$r\rightarrow R_{\rm p}$ \& $\frac{\partial}{\partial r}\rightarrow\frac{\partial}{\partial z}$&$z\ll R_{\rm p}$&\\
  &&&&\multirow{2}{*}{`traditional'}&$\frac{uw}{r}$, $\frac{vw}{r}$, $2\Omega w\cos\phi$ $\rightarrow 0$&\multirow{2}{*}{$N^2\gg \Omega^2$$^{(3)}$}&\multirow{2}{*}{$\sim 10^{-5}\gg 10^{-10}$}\\
  &&&&&$\frac{u^2+v^2}{r}$, $2\Omega u\cos\phi$ $\rightarrow 0$&&\\
  &&&&hydrostasy&$\frac{\partial \Pi}{\partial r}=-\frac{g}{c_p\theta}$ or $\frac{\partial p}{\partial r}=-\rho g$&$H \ll L$&$\sim 10^{7} < 10^{8}$\\
  \hline
\end{tabular}
\tablebib{(1) For a full discussion on the impact of
  the spherical geopotentials approximation see \citet{white_2008}. (2)
  This condition neglects tidal deformation, essentially assuming the
  planetary gravitational field is well isolated (see discussion in
  text). (3) Condition from \citet{phillips_1968}, but may not be
  sufficient \citep[see discussion in][]{white_1995}.}
\end{table*}

The assumptions pertaining to gravity require some
explanation. Firstly, the gravitational potential of the planet is
assumed to be well isolated from external gravity fields \citep[this
  is not the case in some hot Jupiters, for instance
  Wasp--12,][]{li_2010}. Secondly, care must be taken over how to
solve for the centrifugal force, and subsequently construct the
gravitational potential.

When modelling the Earth the acceleration due to gravity can be
measured at the surface, $g_{\rm p}$. This is in effect the
acceleration due to the \textit{apparent} gravity as it includes
contributions from both the gravitational and centrifugal components
\footnote{In reality the surface of the Earth has deformed such that
  the local \textit{apparent} gravity acts normal to the
  surface.}. This combined gravitational and centrifugal potential, or
geopotential is then, in most cases, assumed to be spherically
symmetric. This means, however, that the divergence of the resulting
combined potential is not zero, as should be the case \citep[see][ for
a detailed discussion of the spherical geopotentials
approximation]{white_2008}. Additionally, this spurious divergence in
the combined potential is increased if one adopts a constant
gravitational acceleration throughout the atmosphere, as opposed to
allowing it to fall via an inverse square law
\citep{white_2012}. 

Calculating the acceleration due to the gravitational potential only,
and solving explicitly, as part of the dynamical equations, for the
centrifugal component, however, introduces spurious motions. For
example a modeled hydrostatically balanced and statically stable
atmosphere, at rest, would subsequently have to adjust to the
\textit{apparent} gravity caused by the rotation, which generates a
horizontal force, creating winds. 

For hot Jupiters the acceleration due to gravity cannot be measured,
and there is no surface, in the same sense as on Earth or any other
terrestrial planet. Therefore a value for $g_{\rm p}$ must be
estimated using measurements of the total mass of the planet, $M_{\rm
  p}$ (derived from radial velocity measurements), and assuming this
to be contained within a radius, $R_{\rm p}$ (practically the smallest
available radius derived from observations of the primary
eclipse). The precision to which $g_{\rm p}$ is estimated, or quoted,
is much lower than the magnitude of the expected effect of the
centrifugal component. Therefore, although formally, we absorb the
centrifugal term into the gravity field, due to its negligible,
relative magnitude, we prefer to state that, dynamically we neglect
this term. Finally, most GCMs also neglect the gravity of the
atmosphere itself. For hot Jupiters, whether the gravitational
potential of the atmosphere can be neglected depends on the
distribution of mass between the atmosphere itself and the `planet'
below the inner boundary, whose mass defines $g_{\rm p}$. Formally,
\begin{eqnarray}
  g&=&\frac{GM(r)}{r^2}\\
  &=&\frac{G}{r^2}\left[ M_{\rm p}+M_{\rm atm}(r) \right]\\
  &=&g_{\rm p}\left( \frac{R_{\rm p}}{r}\right)^2+\frac{GM_{\rm atm}(r)}{r^2}\\
  &\sim&g_{\rm p}\left( \frac{R_{\rm p}}{r}\right)^2 \mbox{,}\, M_{\rm atm}(r) \ll M_{\rm p},
\label{grav}
\end{eqnarray}
where $M_{\rm atm}(r)=M(r)-M(R_{\rm p})$. 

The momentum and continuity equations differ depending on the
assumptions made in each of the cases shown in Table
\ref{model_names}. \citet{white_2005} explores, in detail, the form of
the metric and differential operators. In Table \ref{eqn_sets} we
express (in expanded form but omitting the diffusion terms) the
relevant parts of the equations sets which are illustrative of the
main differences, for the three equation sets we use (i.e. ``Full'',
``Deep'' and ``Shallow''), and also the primitive equations for
comparison.

\subsubsection{Consequences of approximations}
\label{approx}
Comparing the terms in the equations in Table \ref{eqn_sets} it is
apparent that each progressive relaxation of an approximation acts to
introduce extra `exchange' terms (and alter existing ones), or terms
in each momentum equation involving the other components of
momenta. Focusing on the $u$ and $v$ components of Table
\ref{eqn_sets} the ``shallow--atmosphere'' approximation neglects the
terms $uw / r$ and $2\Omega w\cos\phi$, and alters the term
$uv\tan\phi / r$. Clearly, regardless of whether $w$ is small compared
to $u$, this assumption, by definition, eliminates the exchange of
vertical and zonal momentum present in a real atmosphere (similarly
for the $v$ component).

The omission of the metric and Coriolis terms is termed the
`traditional' approximation, as explained in Table
\ref{model_names}. Critically, the physical justification for the
adoption of this approximation is weak, and it is largely taken with
the ``shallow--fluid'' approximation to enable conservation of angular
momentum and energy, not for physically motivated reasons. We present,
in Table \ref{model_names} an expression for the validity of this
expression, however, this is debatable and assumes a lack of planetary
scale flows \citep[see discussion in][]{white_1995}. Given that
planetary scale flows are expected for hot Jupiters, this
approximation may well prove crucial to the reliability of the results
of hot Jupiter models. \citet{white_1995} show that the term $2\Omega
w\cos\phi$ in the zonal momentum equation may be neglected if $2\Omega
H\cos\phi/U\ll 1$, which for HD 209458b gives $\sim0.1\ll 1.0$,
suggesting it is marginally valid only for the regions of peak zonal
velocity.

The `traditional' approximation also removes terms from the vertical
momentum equation involving $u$ and $v$, further inhibiting momentum
exchange. Previous attempts have been made to isolate the effect of
this approximation \citep[see for
example][]{cho_2011}. \citet{tokano_2013} show that GCMs adopting the
primitive equations do not correctly represent the dynamics of Titan's
atmosphere (as well as indicating it may be problematic for Venus's
atmosphere). Although \citet{tokano_2013} focus on the assumption of
hydrostatic equilibrium, the term they indicate is dominant,
$(u^2+v^2) / r$, is neglected as part of the `traditional'
approximation. The lack of coupling between the vertical and
horizontal momentum in the HPEs is exacerbated by the adoption of
vertical hydrostatic equilibrium, which neglects the vertical
acceleration of fluid parcels. Vertical velocities are still retained
in the HPEs, derived from the continuity equation, but the lack of
coupling between the vertical and horizontal components of momentum in
the HPEs means these are unlikely to be realistic.

As discussed in Section \ref{introduction} several key physical
problems require a well modeled interaction of the vertical and
horizontal circulations, and between the deep and shallow
atmosphere. Modelling the atmosphere using the NHD equations therefore
allows us to present a much more self--consistent and complete model
of the atmospheric flow. However, vertical velocities are generally
much smaller than the zonal or meridional flows \citep[up to two
orders of magnitude smaller,][]{showman_2002}, and relaxation
times (both radiative and dynamical) in the deeper atmospheres are
generally orders of magnitude longer than those in the shallow
atmosphere. Therefore, the effects of replacing the `exchange' terms
in the dynamical equations may only be appropriate for simulations run
much longer than usual.

Additionally, the introduction of a non--constant (with height)
gravity also subtly affects the stratification, and therefore the
vertical transport. In an atmosphere in hydrostatic equilibrium the
stratification is proportional to the gravitational acceleration, as
the weight of the atmosphere above must be supported from
below. Therefore, allowing gravity to vary (as described by Equation
\ref{grav_eg}) from the value assumed at the `surface' or inner
boundary effectively weakens it throughout the atmosphere, and
therefore weakens the stratification reducing its inhibiting effect on
vertical motions. In our HD 209458b test case over the vertical domain
$\Delta g/g \sim 0.2$.

In summary, the assumption of a `shallow--atmosphere' which includes
the `shallow--fluid', `traditional' and constant gravity
approximations, effectively neglects exchange between the vertical and
horizontal momentum, and is likely to inhibit vertical motions, or
produce inconsistent vertical velocities. Yet, vertical transport and
its interaction with the horizontal advection is believed to be
critical to understanding the major scientific questions regarding hot
Jupiter atmospheres. In this work we find that the results of the test
case, when run using the less simplified dynamical model, diverge from
the literature results. This divergence is caused by the improved
representation of vertical motions and exchange between the horizontal
and vertical components of momentum (discussed in more detail in
Section \ref{discussion}).

\begin{sidewaystable*}
  \caption{List of momenta and continuity equations (where we have
    expanded the material derivative, $\frac{D}{Dt}$, but omitted
    diffusion or damping terms) for each of the named equation sets
    with approximations detailed in Table \ref{model_names}. The
    values of $g$ and $g_{\rm p}$ are discussed in Section
    \ref{equations_solved}.}
\label{eqn_sets}
\centering
%\large
\fontsize{14}{16.8}
\begin{tabular}{cccccccccccccl}
\hline\hline
\multicolumn{2}{l}{Variable}&\multicolumn{4}{c}{Material Derivative $\frac{D}{Dt}$}&\multicolumn{5}{c}{Terms}&Name\\
\hline
\multirow{3}{*}{$u$}&\rdelim\{{2}{0mm}[]&$\left[ \frac{\partial u}{\partial t} \right. $&+$\frac{u}{r\cos\phi}\frac{\partial u}{\partial \lambda}$&$+\frac{v}{r}\frac{\partial u}{\partial \phi}$&$\left. +w\frac{\partial u}{\partial r}\right]=$&$+\frac{uv\tan\phi}{r}$&$-\frac{uw}{r}$&$+2\Omega v\sin\phi$&$-2\Omega w\cos\phi$&$-\frac{c_p\theta}{r \cos\phi}\frac{\partial \Pi}{\partial \lambda}$&\textit{Full \& Deep}\\
&&$\left[ \frac{\partial u}{\partial t}\right. $&+$\frac{u}{R_{\rm p}\cos\phi}\frac{\partial u}{\partial \lambda}$&$+\frac{v}{R_{\rm p}}\frac{\partial u}{\partial \phi}$&$\left. +w\frac{\partial u}{\partial z}\right]=$&$+\frac{uv\tan\phi}{R_{\rm p}}$&&$+2\Omega v\sin\phi$&&$-\frac{c_p\theta}{ R_{\rm p}\cos\phi}\frac{\partial \Pi}{\partial \lambda}$&\textit{Shallow \& Primitive}\\
\hline
\multirow{3}{*}{$v$}&\rdelim\{{2}{0mm}[]&$\left[ \frac{\partial v}{\partial t}\right. $&+$\frac{u}{r\cos\phi}\frac{\partial v}{\partial \lambda}$&$+\frac{v}{r}\frac{\partial v}{\partial \phi}$&$\left. +w\frac{\partial v}{\partial r}\right]=$&$-\frac{u^2\tan\phi}{r}$&$-\frac{vw}{r}$&$-2\Omega u\sin\phi$&&$-\frac{c_p\theta}{ r}\frac{\partial \Pi}{\partial \phi}$&\textit{Full \& Deep}\\
&&$\left[ \frac{\partial v}{\partial t}\right. $&+$\frac{u}{R_{\rm p}\cos\phi}\frac{\partial v}{\partial \lambda}$&$+\frac{v}{R_{\rm p}}\frac{\partial v}{\partial \phi}$&$\left. +w\frac{\partial v}{\partial z}\right]=$&$-\frac{u^2\tan\phi}{R_{\rm p}}$&&$-2\Omega u\sin\phi$&&$-\frac{c_p\theta}{R_{\rm p}}\frac{\partial \Pi}{\partial \phi}$&\textit{Shallow \& Primitive}\\
\hline
\multirow{4}{*}{$w$}&\rdelim\{{4}{0mm}[]&$\delta\left[ \frac{\partial w}{\partial t}\right. $&+$\frac{u}{r\cos\phi}\frac{\partial w}{\partial \lambda}$&$+\frac{v}{r}\frac{\partial w}{\partial \phi}$&$\left.+w\frac{\partial w}{\partial r}\right]=$&\multicolumn{2}{c}{$+\frac{u^2+v^2}{r}$}&$+2\Omega u \cos\phi$&$-g(r)$&$-c_p\theta\frac{\partial \Pi}{\partial r}$&\textit{Full}\\
&&$\delta\left[ \frac{\partial w}{\partial t}\right. $&+$\frac{u}{r\cos\phi}\frac{\partial w}{\partial \lambda}$&$+\frac{v}{r}\frac{\partial w}{\partial \phi}$&$\left. +w\frac{\partial w}{\partial r}\right]=$&\multicolumn{2}{c}{$+\frac{u^2+v^2}{r}$}&$+2\Omega u \cos\phi$&$-g_{\rm p}$&$-c_p\theta\frac{\partial \Pi}{\partial r}$&\textit{Deep}\\
&&$\delta\left[ \frac{\partial w}{\partial t}\right. $&+$\frac{u}{R_{\rm p}\cos\phi}\frac{\partial w}{\partial \lambda}$&$+\frac{v}{R_{\rm p}}\frac{\partial w}{\partial \phi}$&$\left.+w\frac{\partial w}{\partial z}\right]=$&&&&$-g_{\rm p}$&$-c_p\theta\frac{\partial \Pi}{\partial z}$&\textit{Shallow}\\
&&\multicolumn{4}{r}{$0=$}&&&&$-g_{\rm p}$&$-c_p\theta\frac{\partial \Pi}{\partial z}$&\textit{Primitive}\\
\hline
\multirow{2}{*}{$\rho$}&\rdelim\{{2}{0mm}[]&$\left[ \frac{\partial \rho}{\partial t}\right. $&+$\frac{u}{r\cos\phi}\frac{\partial \rho}{\partial \lambda}$&$+\frac{v}{r}\frac{\partial \rho}{\partial \phi}$&$\left. +w\frac{\partial \rho}{\partial r}\right]=$&\multicolumn{2}{c}{$-\rho \left[ \frac{1}{r \cos \phi}\frac{\partial u}{\partial \lambda} \right. $}&\multicolumn{2}{c}{$+\frac{1}{r \cos\phi}\frac{\partial (v\cos \phi)}{\partial \phi}$}&$\left. +\frac{1}{r^2}\frac{\partial (r^2w)}{\partial r} \right]$&\textit{Full \& Deep}\\
&&$\left[\frac{\partial \rho}{\partial t}\right. $&+$\frac{u}{R_{\rm p}\cos\phi}\frac{\partial \rho}{\partial \lambda}$&$+\frac{v}{R_{\rm p}}\frac{\partial \rho}{\partial \phi}$&$\left. +w\frac{\partial \rho}{\partial z}\right]=$&\multicolumn{2}{c}{$-\rho \left[ \frac{1}{R_{\rm p} \cos \phi}\frac{\partial u}{\partial \lambda} \right. $}&\multicolumn{2}{c}{$+\frac{1}{R_{\rm p} \cos\phi}\frac{\partial (v\cos \phi)}{\partial \phi}$}&$\left. +\frac{\partial w}{\partial z} \right]$&\textit{Shallow \& Primitive}\\
\hline
\hline
\end{tabular}
\end{sidewaystable*}

\subsection{Boundary conditions}
\label{boundary}
A full discussion of the boundary conditions used is presented in
\citet{wood_2013}, here we emphasis a few key characteristics. Given
that hot Jupiters do not have a solid surface, we impose an inner
boundary, which is frictionless (placed at $R_{\rm p}$). The inner and
outer boundaries are rigid and impermeable \citep[to ensure energy and
axial momentum conservations,][]{staniforth_2003}. As the boundaries
are rigid they nonphysically act to reflect vertically propagating
waves, such as acoustic or gravity waves, back into the domain. This
is usually only significant during an initial `spin--up' period as
initial transients are produced, in particular waves generated by the
adjustment of the mass distribution in the atmosphere. To solve this
problem the UM incorporates, into the upper boundary, a damping region
(termed a `sponge' layer) high up at the top of the atmosphere to
mitigate the spurious reflection of vertical motions at the upper
boundary. Vertical damping of vertical velocities is
  incorporated using the formulation of \citet{melvin_2010}
  \citep[which follows][]{klemp_2008},
\begin{equation}
w^{t +\Delta t}=w^{t \*}-R_{w}\Delta t w^{t +\Delta t},\\
\end{equation}
where $w^t$ and $w^{t+\Delta t}$ are the vertical velocities at the
current and next timestep, and $\Delta t$ the length of the
timestep. The spatial extent and value of the damping coefficient
($R_{w}$) is then determined by the equation
\begin{multline}
R_{w} =
\begin{cases}
  {\rm C} \sin^2\left( 0.5\pi(\eta-\eta_{\rm s}) \left( \frac{1.0}{1.0-\eta_{\rm s}}\right) \right) \mbox{,}\, &\eta \geq \eta_{\rm s}\\
0 \mbox{,}\, & \eta < \eta_{\rm s},\\
\end{cases}
\label{sponge}
\end{multline}
where, given the absence of orography, $\eta=z / H$
(i.e. non--dimensional height), $\eta_{\rm s}$ is the start height for
the top level damping (set to $\eta_{\rm s}=0.75$) and $C$ is a
coefficient. The value of $C$ is minimized for a given run. Usually,
in Earth based studies one would place the sponge layer high above (or
below) the region where the atmospheric flow is most active (i.e. the
region of interest). However, for these test cases the top boundary
intersects fast flowing features, and the sponge layer will
potentially alter our solution there. While it may alter the solution
this is more desirable than reflecting vertically propagating waves,
artificially, back into the domain. The values assigned to the sponge
layer are stated in Section \ref{hd209458b}. It is important to note
that the damping coefficient $C$ represents the maximum damping
present at the top boundary. Equation \ref{sponge} reduces the damping
$\propto \sin^2$ as we move down from the upper boundary, meaning the
practical damping felt by the vertical velocities reduces
significantly from $C$.

\subsection{Vertical coordinate and model comparison}
\label{vert}
In contrast to most other GCMs applied to hot Jupiters, which use
$\sigma$ or pressure as the vertical coordinate, the UM uses geometric
height coordinates. Ostensibly the choice of vertical coordinates
should not alter the solution to a given equation set. However, due to
large horizontal gradients in pressure, expected in the lower pressure
regions of hot Jupiter atmospheres, surfaces of constant height do not
align with surfaces of constant pressure (isobars). Therefore, to
compare our model to a pressure--based model we must overcome three
problems, namely, matching the boundary conditions and model domain,
matching the vertical resolutions and comparing the results
consistently.

Generally, for both height--based and pressure--based models the inner
boundary is at a set geopotential, and therefore (given the canonical
assumption of spherical geopotentials) a fixed radial position,
$r$. In general, as the inner boundary is deeper in the atmosphere
pressure will not change significantly with time or horizontal
position. Therefore, practically, if we set the pressure on our inner
boundary to the value used in the pressure--based model our inner
boundary conditions will be similar. However, for the upper boundary
we use a constant height surface and pressure--based models use a
constant pressure surface. 

The strong contrast in temperatures expected between the day and night
side of hot Jupiters leads, in the upper atmosphere where the
radiative timescale is short, to a significant gradient in pressure at
a given height. At a given height the atmosphere will be hotter with
higher pressure on the day side and cooler with lower pressures on the
night side. If we are to completely capture the domain of a
pressure--based code, we must set the position of our upper boundary
so as to capture the minimum required pressure on the day side, and
this height surface will sample lower pressures as it moves to the
night side. This effectively means that we include an extra region of
the atmosphere, over the domain modeled by a pressure--based code,
being the region of the night side atmosphere at pressures lower than
the minimum sampled by the pressure--based code (and by the height of
our boundary on the higher pressure day side). The pressures,
temperatures and densities of this material should, however, be small
and therefore its dynamical effect be negligible (i.e. its angular
momentum and kinetic energy contribution), as is shown by the
agreement of our results with those from a pressure--based code (see
Sections \ref{shj_results} and \ref{hd209458b_results}). However, we
do have to alter the formulation of the radiative--equilibrium
temperature--pressure ($T-p$) profiles in this region for stability
(see Section \ref{hd209458b}), from that presented in
\citet{heng_2011}.

Additionally, as the pressure at a given height in an atmosphere will
fluctuate in time it is impossible to exactly match the distribution
of levels in a pressure--based models with one, such as the UM, based
on geometric height. To provide a mapping between height and
approximate pressure (or more specifically $\sigma$), for the SHJ test
case we have completed a simulation using a uniform distribution of
levels with an upper boundary high enough to capture the lowest
required pressure. We then zonally and temporally--average the
pressure structure. This allows us to distribute levels in height so
as to sample $\sigma$ evenly, however, as the pressure will fluctuate
we have increased the number of vertical levels (compared to the
literature cases) to compensate. For HD 209458b, we have used uniform
(in height) levels but, again, have increased the number relative to
\citet{heng_2011} to compensate. We have altered our vertical
resolution and level distributions and show in Section
\ref{hd209458b_results} that it has a negligible effect on the results
of the HD 209458b test case.

Finally, to aid comparison of our results with literature pressure (or
$\sigma$)--based models, we have interpolated the prognostic variables
onto a pressure grid at each output. Horizontal averaging has then
been performed along isobaric surfaces and the plots are presented
with $\sigma$ or pressure as the vertical coordinate.

\subsection{Diffusion, dissipation and artificial viscosity}
\label{diffusion}
In physical flows eddies and turbulence can cause cascades of kinetic
energy from large--scale flows to smaller scales. At the smallest
scales the kinetic energy is converted to thermal energy, heating the
gas, due to the molecular viscosity of the gas. The resolutions
possible with current models of planetary atmospheres (and other
astrophysical models) do not reach the scales associated with
molecular viscosity, and so a numerical scheme is required to mimic
this dissipative process, as previously explained by many authors
\citep{cooper_2005,cho_2008,menou_2009,heng_2011, bending_2013}. Some
effective dissipation is provided by ``numerical viscosity'' inherent
to the computational scheme itself. However, explicit schemes are
included in different codes (both astrophysical and meteorological) to
varying levels of accuracy or sophistication, and using differing
nomenclature.

Many astrophysical codes include an ``artificial viscosity'', where
the controlling parameter can be altered to set the level of eddy or
turbulent dissipation. Correctly formulated, an artificial viscosity
includes the conversion of kinetic energy to heat via terms appearing
in the momentum and thermal energy equation. For GCMs, and in
meteorology, the term ``dissipation'' represents a similar scheme
where losses of kinetic energy are accounted for in the thermal energy
equation. Another scheme also regularly used in GCMs, is termed
``diffusion'', in this case a similar approach is used to remove
kinetic energy, but this is not accounted for in the thermal energy
equation. Such diffusion can be viewed as a numerical tool to remove
grid scale noise. Although the operational\footnote{The version used
  by the UK Met Office for weather and climate prediction will use
  ENDGame from early 2014.} version of the ENDGame dynamical core
includes no explicit diffusion, in our case, as with many other GCMs,
we have incorporated a diffusion scheme. Whichever scheme is used the
loss of kinetic energy can affect the characteristic flow and the
maximum velocities achieved \citep{heng_2011,li_2010b}.

It is possible to use known flows, such as in the boundary layer on
Earth, to tune the form of this dissipation but this is not possible
for hot Jupiters \citep[see discussion in][]{li_2010b}. Therefore, we
do not ``tune'' our diffusion scheme to achieve a required wind speed,
but for each of our test cases keep the diffusion constant for all
simulations. Essentially, diffusion is used to provide numerical
stability, although it will affect the results. Therefore, as with all
other studies, the magnitude of our wind velocities are not robust
predictions of the flow on a given hot Jupiter, rather the relative
flows and patterns are the features to be interpreted. The scalar form
of the diffusion operator $\textbf{D}(X)$ (which operates along
$\eta$, or height as we have no orography, layers), is given by:
\begin{multline}
  \textbf{D}(X)=\left( \frac{1}{r^2\cos\phi}\frac{\partial \eta}{\partial r}\right) \\
  \left\{ \frac{\partial}{\partial \lambda}\left[ \frac{K_{\lambda}}{\cos\phi}\frac{\partial r}{\partial \eta}\frac{\partial }{\partial \lambda}\left( X \right) \right]+\frac{\partial}{\partial \phi} \left[ K_{\phi}\cos\phi \frac{\partial r}{\partial \eta}\frac{\partial}{\partial \phi} \left(X \right) \right] \right\},\\
\label{diffusion_eqn}
\end{multline}
where $X$ is the quantity to be diffused and $K_\lambda$ is given by
\begin{equation}
  \frac{K_\lambda}{\cos\phi}=Kr^2\Delta\lambda^2\frac{\sin^2\left( \frac{\pi}{2}\cos\phi_{p} \right)}{\sin^2\left( \frac{\pi}{2}\cos\phi \right)}
\end{equation}
where $\phi_p=(\pi / 2)- (\Delta\Phi / 2)$ is the latitude of the row
closest to the pole and
$K_\phi=K_\lambda\left(\phi=0\right)(\Delta\phi^2 /
\Delta\lambda^2)$. The value of $K$ is stated for each simulation in
Table \ref{par_par}. In practice, as a further approximation, the
diffusion operator is applied separately to each component of the
vector field, as shown in Equations \ref{full_set} in Section
\ref{equations_solved}. The construction of the diffusion operator
allows the damping of the same physical scales as one approaches the
equator (in practice this means that there is very little diffusion
applied away from the polar regions and that small scale waves that
could accumulate in the polar regions are removed) and also allows for
variable resolution. 

Usually, polar filtering is achieved by applying multiple passes of an
operator similar to that in Equation \ref{diffusion_eqn} from $\sim\pm
85^{\circ}$ to the poles, damping only in the zonal direction (as this
is the scale which decreases towards the poles). In contrast, our
diffusion operator is applied once across the entire globe and in both
the zonal and meridional direction. We do not require an explicit
polar filter, as used in other GCMs or previous versions of the
UM. This is due to the changes in numerical scheme and the fact that
our diffusion scheme will apply some damping, although significantly
reduced, as would result from application of a polar filter.  The
diffusion is applied directly to the $u$, $v$ and $\theta$ fields for
the SHJ test case \citep[][apply diffusion to relative vorticity and
temperature using a $\sigma$ vertical coordinate]{menou_2009}. Whereas
for the HD 209458b test case it is only applied to the $u$ and $v$
fields \citep[][apply diffusion to the $u$, $v$ and $T$ fields, again
using a $\sigma$ vertical coordinate]{heng_2011}. One would ideally
prefer to apply diffusion to the potential temperature for the HD
209458b test case, to match more closely the diffusion scheme of
\citet{heng_2011}. However, firstly our results show some divergence
from the results of \citet{heng_2011} when also applying diffusion to
$\theta$, as shown and discussed in Section
\ref{hd209458b_discuss}. Secondly, it is not actually clear that
diffusing potential temperature along constant height surfaces (our
scheme) is analogous to diffusing temperature along constant pressure
surfaces \citep[scheme of][]{heng_2011}. We postpone a more complete
discussion of this effect for a later work (Mayne et al, in
preparation), and simply note here that the choice of diffusion
scheme, target fields, and its interaction with the choice of vertical
coordinate can potentially alter the results.

\subsection{Radiative transfer}
\label{radiative_transfer}

Radiative transfer, for these tests, has been parameterised using a
Newtonian cooling scheme \citep[as used for many models of hot
  Jupiters, e.g.][]{cooper_2005,menou_2009,heng_2011}. The heating
rate in the thermodynamic equation stated in Section
\ref{equations_solved} is,
\begin{equation}
Q=Q_{\rm Newton}=-\Pi\left(\frac{\theta-\theta_{\rm eq}}{\tau_{\rm
    rad}}\right),
  \label{newt_cool}
\end{equation}
where $\tau_{\rm rad}$ the characteristic radiative or relaxation
timescale. $\theta_{\rm eq}$ is the equilibrium potential temperature
and is derived from the equilibrium temperature ($T_{\rm eq}$) profile
using
\begin{equation}
\theta_{\rm eq}^{i}=\frac{T_{\rm eq}}{\Pi^{i}},
\end{equation}
where superscript $i$ denotes the current timestep. Practically, the
potential temperature is adjusted explicitly within the
semi--Lagrangian scheme using
\begin{equation}
\theta^{i+1}=\theta_{D}^{i}-\frac{\Delta t}{\tau_{\rm rad}}\left(\theta^{i}-\theta_{\rm eq}^{i}\right)_{D},
\end{equation}
where the superscript $i+1$ denotes the next timestep and $\Delta t$
is the length of the timestep. The subscript $D$ denotes a quantity at
the departure point of the fluid element \citep[see explanation in
  Section \ref{numerics} and][for a full discussion]{wood_2013}
\footnote{From the equations in this section one can recover, $Q_{\rm
    Newton}=\frac{T_{\rm eq}-T}{\tau_{\rm rad}}$ and
  $T^{i+1}=T^{i}-\frac{\Delta t}{\tau_{\rm rad}}(T^{i}-T_{\rm eq})$ as
  shown, for example in \citet{heng_2011}.}

We are currently completing the development of a two--stream,
correlated--k radiative transfer scheme. This will allow us to run
more realistic models and avoid the problems associated with
simplified radiative transfer schemes \citep[for instance the omission
  of thermal re--emission of heated gas, and the separation between
  the temperature adjustment and heat capacity of a given atmospheric
  fluid elements, see][for discussion]{showman_2009}.

\section{Test cases}
\label{test_cases}

We have performed simulations of a generic SHJ \citep[that prescribed
in][]{menou_2009} and HD 209458b \citep[as prescribed
in][]{heng_2011}. Table \ref{par_par} lists the general parameters
common for all of the SHJ or HD 209458b simulations. 

\begin{table*}
  \caption{Value of the general (i.e. set for a given test case) parameters for
    the test cases.}
\label{par_par}
\centering
\begin{tabular}{lcc}
  \hline\hline
  Quantity&SHJ&HD 209458b\\
  \hline
  Horizontal resolution&\multicolumn{2}{c}{$144_\lambda$, $90_\phi$}\\
  Standard vertical resolution&32&66\\
  Timestep (s)&\multicolumn{2}{c}{1200}\\
  Run length (Earth days)&\multicolumn{2}{c}{1200}\\
  Sampling rate, $\Delta t_{\rm s}$ (days)&\multicolumn{2}{c}{10}\\
  Initial inner boundary pressure, $p_{\rm s}$ (Pa)&$1\times 10^5$&$220\times 10^5$\\
  \multirow{2}{*}{Radiative timescale, $\tau_{\rm rad}$ (s)}&\multirow{2}{*}{$\frac{\pi}{\Omega_{\rm p}}\sim1.5\times 10^5$}&\citet{iro_2005},  where $p< 10\times10^5$ Pa ($\sim1\times 10^{3-8}$)\\
&&$\infty$, where $p\geq 10\times10^5$ Pa\\
  Initial temperature profile&Isothermal $1800$ K&$\frac{T_{\rm day}+T_{\rm night}}{2}$\\
  Equator--pole temperature difference, $\Delta T_{\rm EP}$ (K)&300&\multicolumn{1}{l}{\rdelim\{{5}{0mm}[Modified \citet{iro_2005} profiles]}\\
  Equatorial surface temperature, $T_{\rm surf}$ (K)&1600&\\
  Lapse rate, $\Gamma_{\rm trop}$ (Km$^{-1}$)&$2\times 10^{-4}$&\\
  Location of stratosphere ($z_{\rm stra}$, m \& $\sigma_{\rm stra}$)&$2\times10^6$, $\sim0.12$&\\
  Tropopause temperature increment , $\Delta T_{\rm strat}$ (K)&10&\\
  Rotation rate, $\Omega$ (s$^{-1}$)&$2.1\times 10^{-5}$&$2.06\times 10^{-5}$\\
  Radius, $R_{\rm p}$ (m)&$10^8$&$9.44\times10^7$\\
  Radius to outer boundary (m)&$3.29698\times 10^6$&$1.1\times 10^7$\\
  Surface gravity, $g_{\rm p}$ (ms$^{-2}$)&8&9.42\\
  Specific heat capacity (constant pressure), $c_{\rm p}$ (Jkg$^{-1}$K$^{-1}$)&13226.5&14308.4\\
  Ideal gas constant, $R$ (Jkg$^{-1}$K$^{-1}$)$^{(1)}$&3779&4593\\
  $K$, diffusion coefficient&0.495&0.158\\
  \hline
\end{tabular}
\tablebib{(1) The $R$ value is varied between simulations to attempt
  to represent differences in the molecular weight of the modeled portion
  of the atmosphere.}
\end{table*}

For each simulation we have followed \citet{held_1994} and
\citet{heng_2011} and run the simulations for 1200 days (here, and
throughout this work, `days' refers to Earth days). The first 200 days
are then discarded to allow for initial transients and `spin--up',
which is sufficient to span several relaxation times for the entire
atmosphere in the SHJ case and for the upper atmosphere down to a
pressure of $\sim 10^5$ Pa (or a few bar) for HD 209458b \citep[using
the radiative timescale of][]{iro_2005}. For the HD 209458b test case
1200 days is only sufficient to span $\sim 1$ radiative relaxation
time throughout the radiative zone. Additionally, as the HD 209458b
test case also includes a radiatively inactive region a significantly
longer time is required to reach a statistical steady state \citep[for
example][found after 5000 days the atmosphere had reached a steady
state down to $3\times 10^5$ Pa or $\sim$3 bar]{cooper_2005}. The
issue of whether the simulation has reached a statistically steady
state will be discussed in more detail in Section
\ref{hd209458b_discuss}. The solution from 200 to 1200 days is then
used to create zonally and temporally averaged temperature and zonal
wind plots, which we term `zonal mean' plots \citep[in a similar vein
to][]{heng_2011}.

As discussed in section \ref{vert}, to aid comparison with previous
works we present plots using $\sigma$ (SHJ) and $\log(p)$ (HD 209458b)
as our vertical coordinate, which we have created by interpolating the
values from the geometric grids onto the isobaric surface
required. The plots (throughout this work, for example Figure
\ref{shj_compare}) feature contour lines (solid for positive and
dashed for negative) that have been chosen, where applicable, to match
those in \citet{heng_2011}. These are complemented by colour scales,
where a greater number of divisions (than the line contours) are used
to aid qualitative interpretation of the data\footnote{The values of
  the labels for the colour scales have been rounded to integer
  values. Additionally, the total range used for the colour scale is
  larger than the range of the data.}. The colour scales chosen have
mostly been selected to match standard colour schemes in meteorology
(i.e. blue--red for temperature). For wind plots we have adopted a
blue--white--red colour scale where blue is retrograde or downdraft,
i.e. negative wind, red is prograde or updraft, i.e. positive wind and
white is positioned at zero\footnote{The splitting of the colour
  scales means that the colour scales need not be symmetric about
  zero.}.

\subsection{Shallow--Hot Jupiter}
\label{shj}

\subsubsection{Test case setup}
\label{shj_setup}
The SHJ test is that prescribed by \citet{menou_2009}, a thin layer of
a hypothetical tidally locked Jovian planet down to a depth of
$1\times 10^5$ Pa or 1 bar. The equilibrium temperature profile is,
\begin{equation}
T_{\rm eq}=T_{\rm vert}+\beta_{\rm trop}\Delta T_{\rm EP}\cos(\lambda - 180^{\circ})\cos(\phi),\\
\end{equation}
where $T_{\rm vert}$ is given by,
\begin{align}
  T_{\rm vert}=
\begin{cases}
    T_{\rm surf}-\Gamma_{\rm trop}(z_{\rm stra}+\frac{z-z_{\rm stra}}{2})\\
+\left( \left[ \frac{\Gamma_{\rm trop}(z-z_{\rm stra})}{2}\right] ^2+\Delta T_{\rm strat}^2\right) ^{\frac{1}{2}}\mbox{,}\, &z \leq z_{\rm stra}\mbox{,} \nonumber\\
    T_{\rm surf}-\Gamma_{\rm trop}z_{\rm stra}+\Delta T_{\rm strat}\mbox{,}\, &z > z_{\rm stra} \mbox{.} \end{cases}\\
\label{T_vert}
\end{align}
and $\beta_{\rm trop}$ is defined as
\begin{align}
  \beta_{\rm trop}&=\begin{cases}
    \sin\frac{\pi(\sigma-\sigma_{\rm stra})}{2(1-\sigma_{\rm stra})}\mbox{,}\, &z \leq z_{\rm stra} \,\mbox{ or }\, \sigma \geq\sigma_{\rm stra} \mbox{,} \nonumber\\
    0\mbox{,}\, &z > z_{\rm stra} \, \mbox{ or }\, \sigma < \sigma_{\rm stra} \mbox{.} \end{cases}\\
\end{align}
The values for the parameters featured in these equations are
presented in Table \ref{par_par}. The radiative relaxation timescale
throughout the entire atmosphere is set to $\tau_{\rm
  rad}=\pi / \Omega_{\rm p}\sim 1.731$ days.

We have run this test case using the ``Full'', ``Deep'' and
``Shallow'' equation sets (see Table \ref{model_names} for
explanation), with the rest of the setup the same for each
simulation. The number of vertical levels is 32 and the level top is
placed at $3.29698\times 10^6$ m, no sponge layer was necessary and
the diffusion has been applied to $u$, $v$ and $\theta$\footnote{We
  have performed a simulation incorporating a sponge and found no
  significant differences in the results from those presented in
  Figure \ref{shj_compare}.}.

We started the atmosphere initially at rest and in vertical
hydrostatic equilibrium using an isothermal temperature profile set at
$1800$ K as used by \citet{heng_2011}.

\subsubsection{Results}
\label{shj_results}

\begin{figure*}
\centering
  \includegraphics[width=8.5cm,angle=0,origin=c]{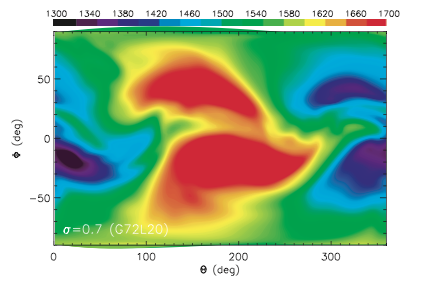}
  \hspace*{-0.7cm}\includegraphics[width=7.0cm,angle=90]{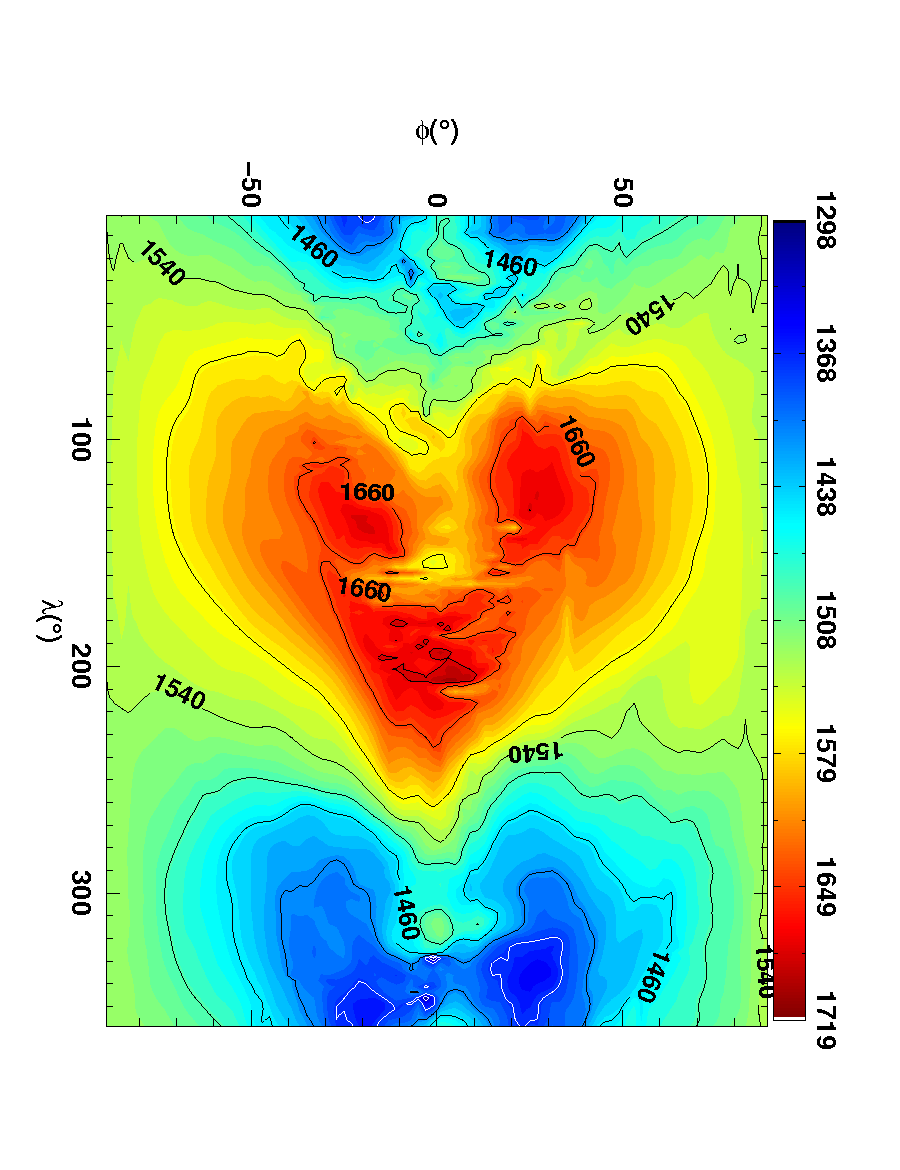}
  \includegraphics[width=8.5cm,angle=0,origin=c]{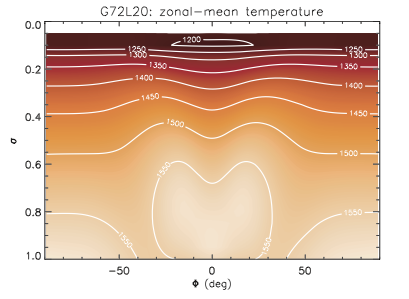}
  \hspace*{-0.7cm}\includegraphics[width=7.0cm,angle=90]{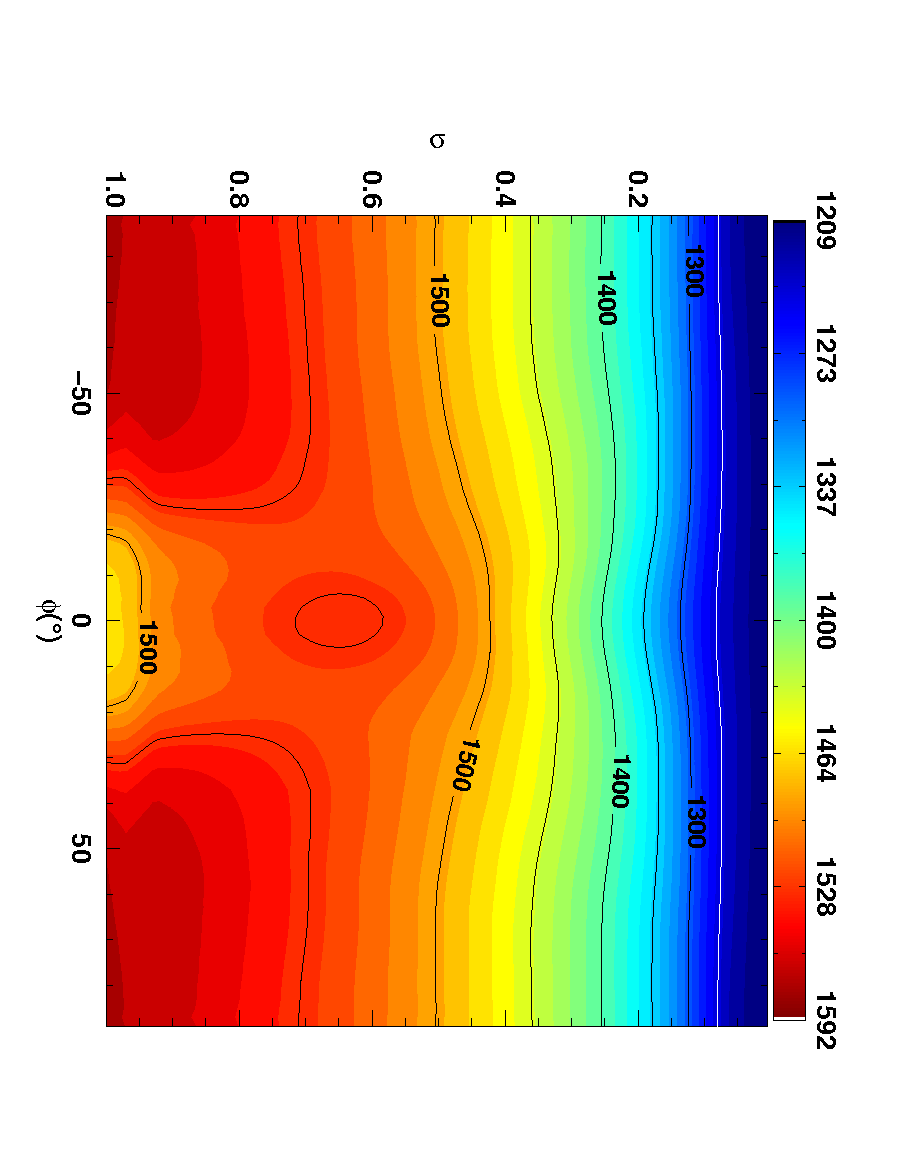}
  \includegraphics[width=8.5cm,angle=0,origin=c]{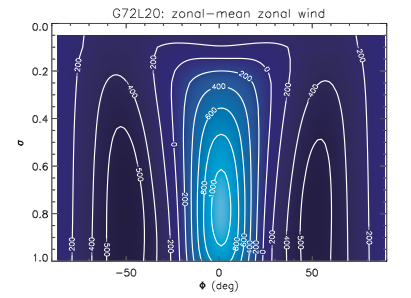}
  \hspace*{-0.7cm}\includegraphics[width=7.0cm,angle=90]{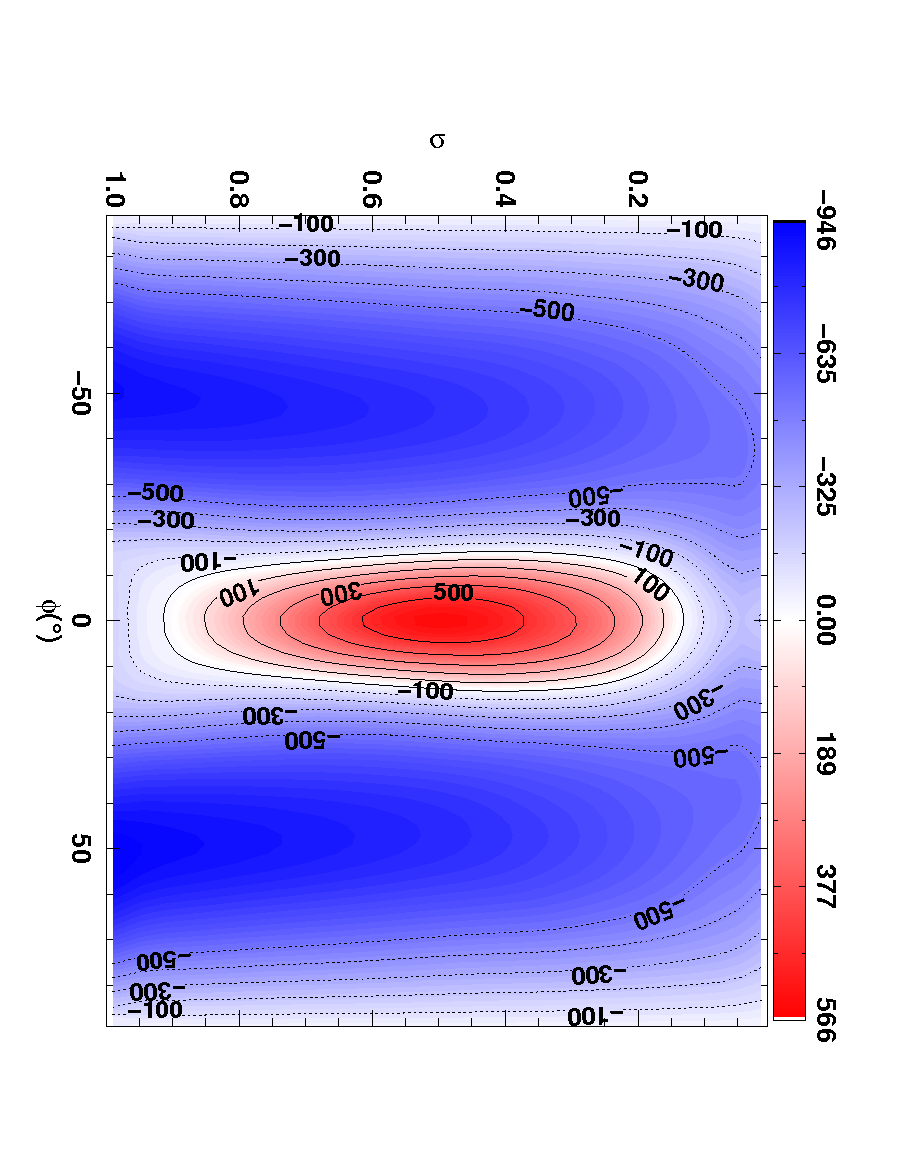}
  \caption{Figure showing the solutions to the SHJ test
    case. \textit{Left panels} are figures reproduced from
    \citet{heng_2011} using the finite--difference model (reproduced
    by permission of Oxford University Press), and the \textit{right
      panels} are results from this work for the ``Shallow'' case (see
    Table \ref{model_names} for explanation). The \textit{top row}
    shows the temperature field at $\sigma=0.675$ and 346 days. The
    \textit{middle} and \textit{bottom rows} show the zonal mean plots
    for temperature and wind respectively (i.e. zonally and
    temporally, from 200-1200 days, averaged).}
  \label{shj_compare}
\end{figure*}

The resulting flow and temperature of the ``Shallow'' SHJ test case at
the $\sigma=0.675$ surface after 346 days, as well as the zonal mean
plots are shown alongside the figures from \citet{heng_2011} (using
their finite--grid model) in Figure \ref{shj_compare}. We present the
instantaneous temperature field at $\sigma=0.675$ instead of the
quoted value of 0.7 in \citet{heng_2011} as this quoted value does not
represent the actual value of the surface, but the half--level just
above it (i.e. at lower sigma and greater height). Therefore, the real
$\sigma$ value is half the vertical resolution below the quoted
$\sigma$ value \citep[see][for a full discussion of this in regards to
Earth--like tests]{mayne_2013}.

\begin{figure}
\centering
  \hspace*{-0.7cm}\includegraphics[width=7.0cm,angle=90]{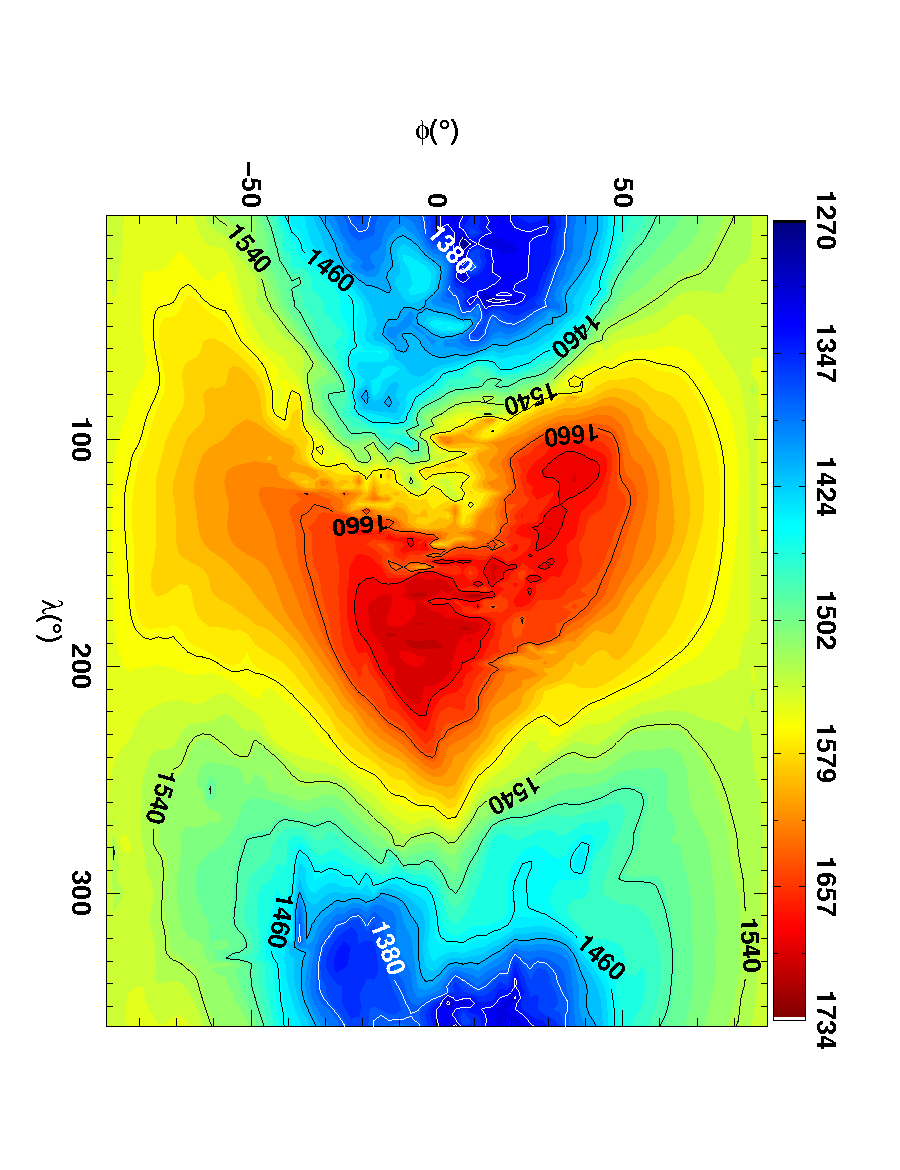}
  \hspace*{-0.7cm}\includegraphics[width=7.0cm,angle=90]{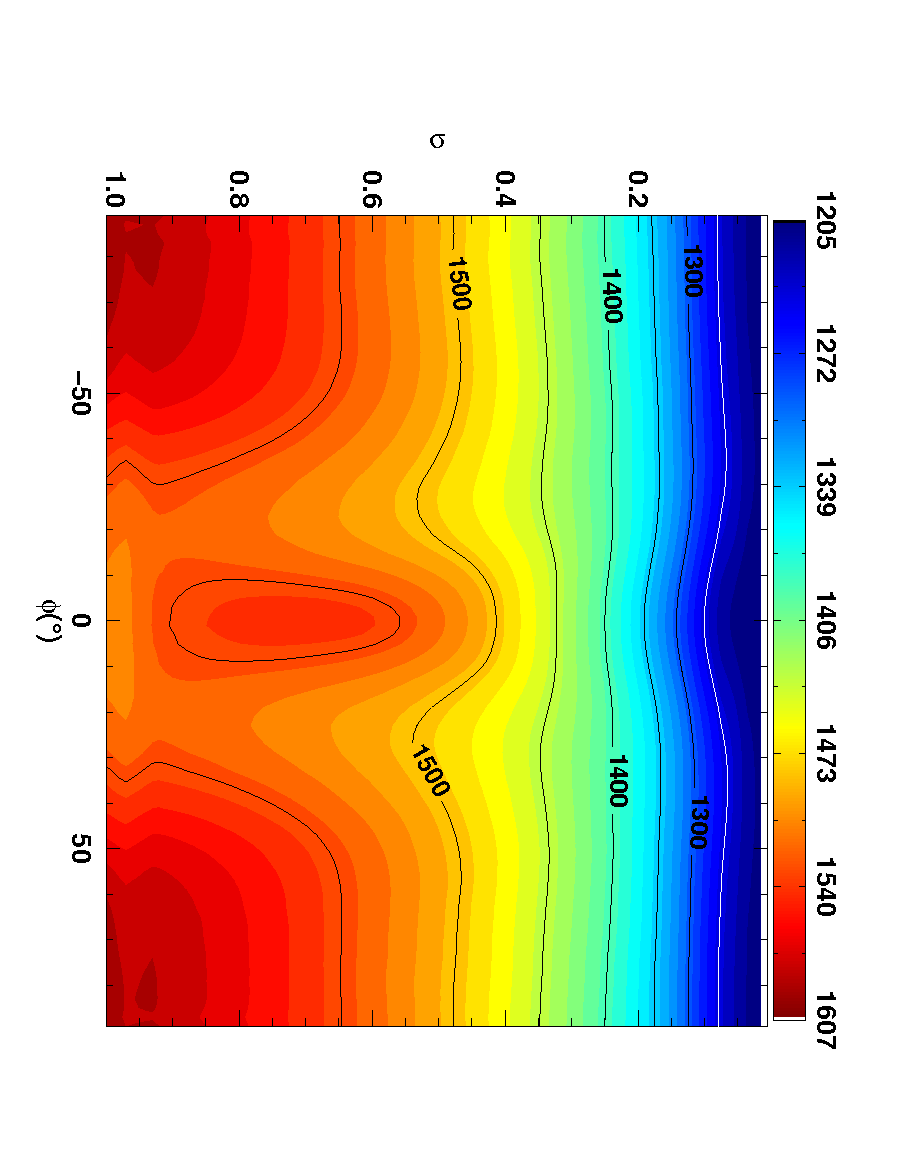}
  \hspace*{-0.7cm}\includegraphics[width=7.0cm,angle=90]{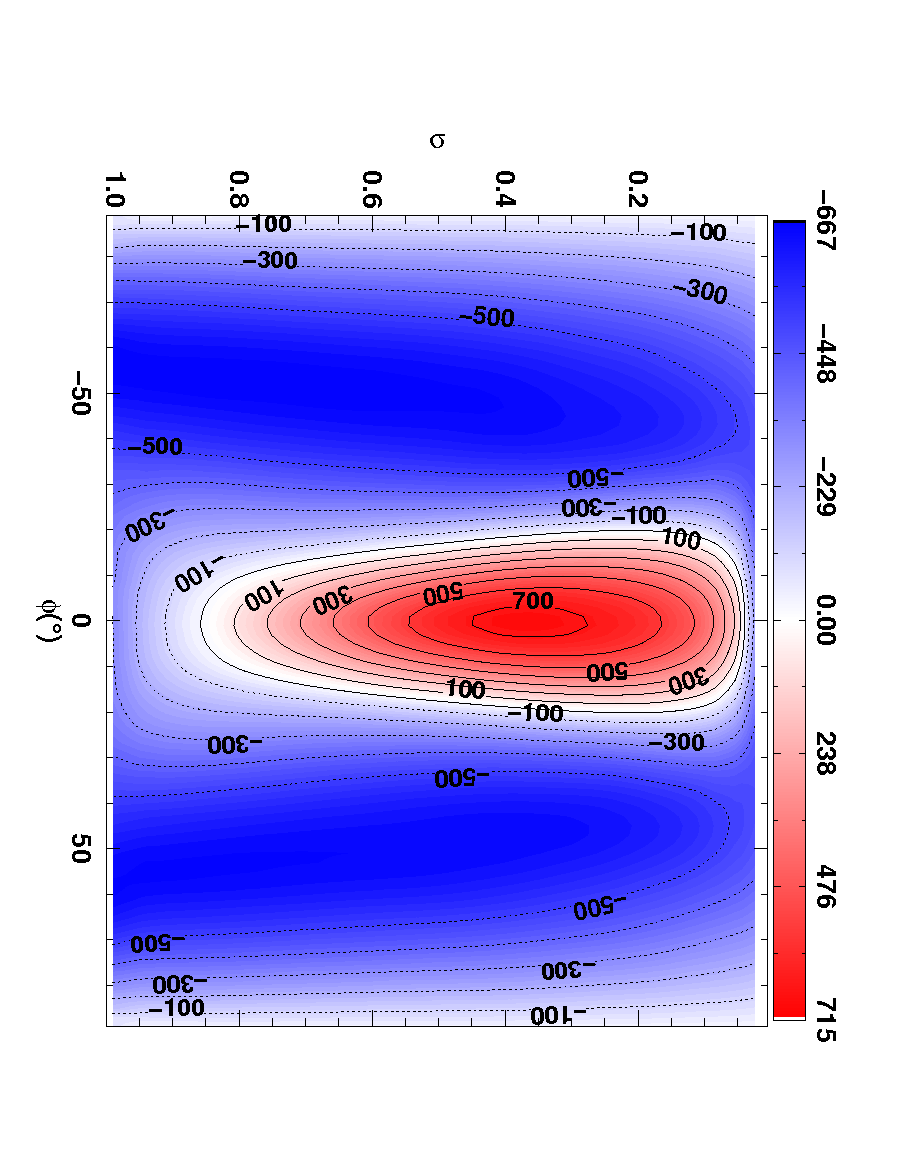}
  \caption{Figure matching those described in Figure \ref{shj_compare}
    but for the ``Full'' case (see Table \ref{model_names} for
    explanation).}
\label{shj_compare_deep}
\end{figure}

Figure \ref{shj_compare} shows that, qualitatively, we match the broad
characteristics of the flow. Figure \ref{shj_compare_deep} then shows
the same plots but for the ``Full'' case (the ``Deep'' case is omitted
as it is virtually identical to the ``Full''). Figure
\ref{shj_compare_deep} shows an atmospheric structure broadly
consistent with both the ``Shallow'' case and that of
\citet{heng_2011}. As the atmosphere of the SHJ is only $1\times 10^5$
Pa or 1 bar in extent its height is $\sim4\times10^6$m, and as the
planetary radius is $R_{\rm p}=10^8$ (see Table \ref{par_par}), it is
unsurprising that no difference is found when relaxing the
`shallow--atmosphere' approximation (see Table
\ref{model_names}). Indeed the resulting flow is very similar in all
cases. Some slight differences are present which will be discussed
briefly in Section \ref{shj_discuss}, but for now we move on to a more
physically interesting test case.

\subsection{HD 209458b}
\label{hd209458b}

\subsubsection{Test case setup}
\label{hd209458b_setup}
The test case for HD 209458b is a slightly adjusted version of that
prescribed in \citet{heng_2011} \citep[similar to that described
in][]{rauscher_2010}, where the temperature and relaxation profiles
are taken from the radiative equilibrium models of
\citet{iro_2005}. The domain encompasses a radiatively inactive region
from $2.2\times 10^7$ to $1\times 10^6$ Pa (or 220 to 10 bar)
\citep[where $\tau_{\rm rad}=\infty$, termed `inactive'
in][]{heng_2011} with a radiative zone above this.

\begin{figure}
\centering
  \hspace*{-0.7cm}\includegraphics[width=7.0cm,angle=90.0]{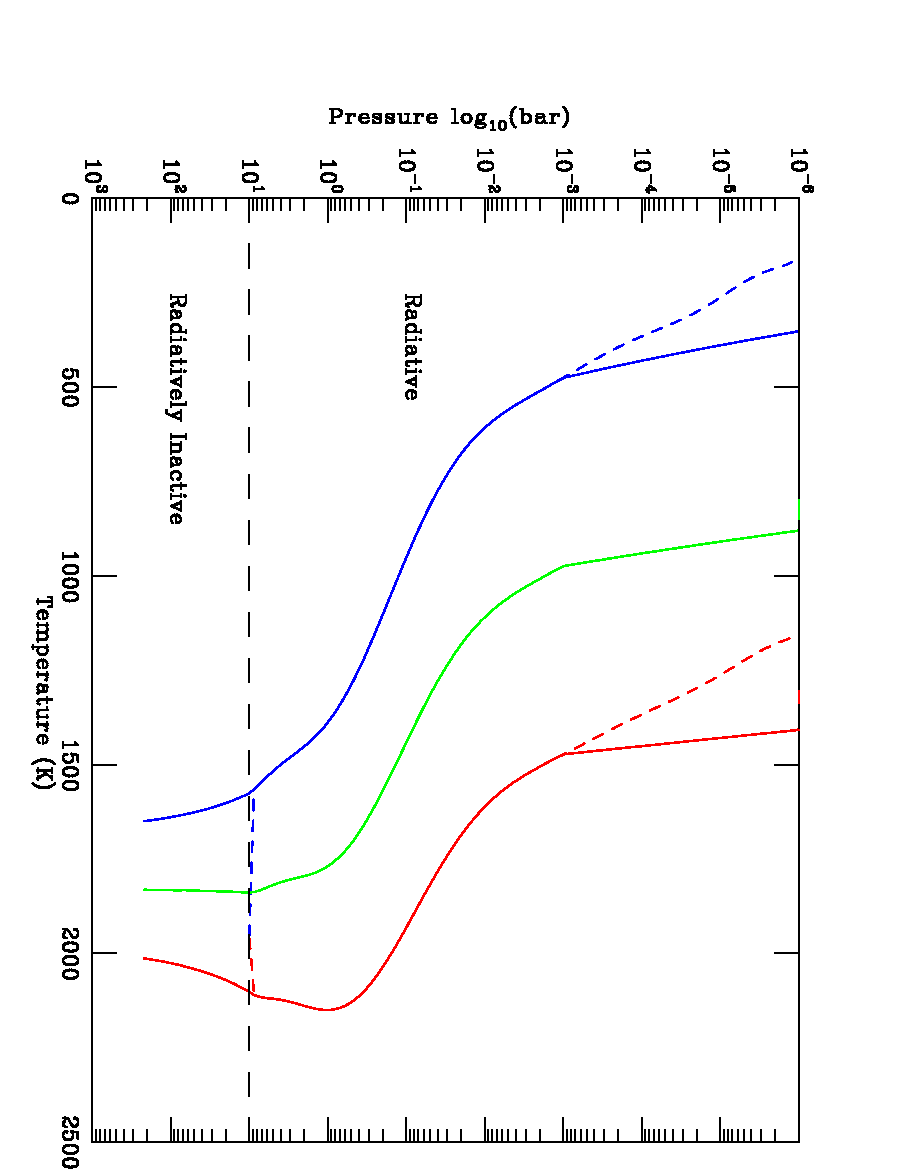}
  \caption{Temperature--pressure profiles used for HD 209458b. The
    solid lines are from this work, and the dashed lines are the
    polynomial fits of \citet{heng_2011} to the models of
    \citet{iro_2005}. The blue lines are the night side profiles, the
    red lines the day side profiles (i.e. $T_{\rm night}$ and $T_{\rm
      day}$, respectively) and the green line is the initial
    profile. The horizontal dashed line demarks the radiatively
    inactive and radiative regions.}
\label{hd209_tprof}
\end{figure}

As discussed in Section \ref{vert} due to the horizontal gradients in
pressure in the upper atmosphere, as we are using a height based
approach and matching a test case performed in pressure coordinates we
are including, necessarily, an extra section of computational domain
i.e. the low pressure night side region. We found for our
non--hydrostatic code the model was extremely unstable on the night
side in this very cool low pressure region, leading to exponential
growth of vertical velocities under small perturbations. Additionally,
we found that the discontinuities in temperature across the $1\times
10^6$ Pa (or 10 bar) boundary found in the profile described in
\citet{heng_2011} also led to instability \citep[as discussed
in][]{rauscher_2010}. Therefore, we have slightly adjusted the
profiles of \citet{heng_2011}. The most significant change, a modest
heating of around 150 K, is performed in the region above $10^{-3}$
bar. This region is not included in the model of \citet{heng_2011}, as
their upper boundary is placed at this pressure. The altered
temperature profiles are shown in Equations \ref{hd209_force_1} and
\ref{hd209_force_2} and plotted in Figure \ref{hd209_tprof} (with the
radiative and radiatively inactive regions also indicated).
\begin{equation}
T_{\rm night}=\begin{cases}
T_{\rm night}^{\prime}\rvert^{p_{\rm high}}+100 {\rm K}\left(1.0-e^{-(\log(p)-\log(p_{\rm high}))}\right) \mbox{,}\,& p\geq p_{\rm high}\\
{\rm MAX}\left( T_{\rm night}^{\prime}\rvert^{p_{\rm low}}\times e^{0.10(\log(p)-\log(p_{\rm low}))}, 250\right)\mbox{,}\,& p< p_{\rm low}\\
T_{\rm night}^{\prime}\rvert^p\, &\mbox{otherwise}
\end{cases}
\label{hd209_force_1}
\end{equation}
\begin{equation}
T_{\rm day}=\begin{cases}
T_{\rm day}^{\prime}\rvert^{p_{\rm high}}-120.0 {\rm K}\left(1.0-e^{-(\log(p)-\log(p_{\rm high}))}\right) \mbox{,}\,& p\geq p_{\rm high}\\
{\rm MAX}\left( T_{\rm day}^{\prime}\rvert^{p_{\rm low}}\times e^{0.015(\log(p)-\log(p_{\rm low}))}, 1000\right)\mbox{,}\,& p< p_{\rm low}\\
T_{\rm day}^{\prime}\rvert^p\, &\mbox{otherwise}
\end{cases}
\label{hd209_force_2}
\end{equation}
$T_{\rm day}$ and $T_{\rm night}$ are the day and night side
temperature profiles and $p$ is the pressure. $T_{\rm
  night}^\prime$ and $T_{\rm day}^\prime$ are the polynomial fits of
\citet{heng_2011} to the day and night side profiles of
\citet{iro_2005}, and $p_{\rm low}$ and $p_{\rm high}$ are $100$ and
$1\times 10^6$ Pa respectively (or $1\times 10^{-3}$ and $10$ bar).

The resulting profiles in Equations \ref{hd209_force_1} and
\ref{hd209_force_2} are then combined to create a temperature map of
the planet's atmosphere using,
\begin{multline}
T_{\rm eq}=\\
\begin{cases}
\left[ T_{\rm night}^4+ (T_{\rm day}^4 - T_{\rm night}^4)\cos(\lambda-180^\circ)\cos\phi\right]^\frac{1}{4}\mbox{,}\,& 90^\circ\leq \lambda \leq 270^\circ\\
T_{\rm night}\mbox{,}\, &\mbox{otherwise.}\\
\end{cases}
\label{hd209_temp}
\end{multline}
We have run this test case using the ``Full'', ``Deep'' and
``Shallow'' equations sets with the top boundary placed at $1.1\times
10^7$ m and use 66 vertical levels (distributed uniformly in
height). For this test case we require a sponge layer and minimise
this for each simulation, where $\eta_{\rm s}=0.75$ in all cases. $C$
is 0.20 for both the ``Deep'' and ``Shallow'' case but 0.15 for the
``Full'' case. The effect of both the sponge layer and the use of
uniform vertical levels (as opposed to those sampling, for instance,
uniform $\log(p)$) have been explored and are briefly discussed in
\ref{hd209458b_discuss}.

Each of the simulations has been initialised in hydrostatic equilibrium
using a temperature profile midway between the day and night profiles
(i.e. $(T_{\rm day}+T_{\rm night}) / 2$) and zero initial winds. As we
are trying reproduce the results of a test case, we postpone a
detailed exploration of the effect of varying initial conditions for
later work (Mayne et al, in preparation).

\subsubsection{Results}
\label{hd209458b_results}

In general our resulting large--scale, long--term flows and those of
\citet{heng_2011} for HD 209458b are qualitatively very
similar. 

\begin{figure*}
\centering
  \includegraphics[width=8.5cm,angle=0.0,origin=c]{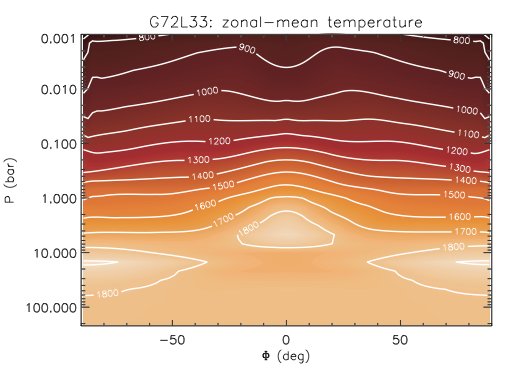}
  \includegraphics[width=8.5cm,angle=0.0,origin=c]{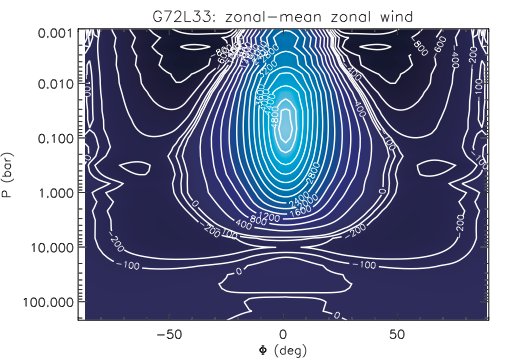}
  \includegraphics[width=8.5cm,angle=0,origin=c]{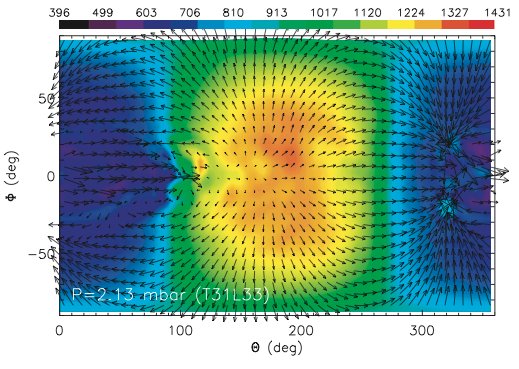}
  \includegraphics[width=8.5cm,angle=0,origin=c]{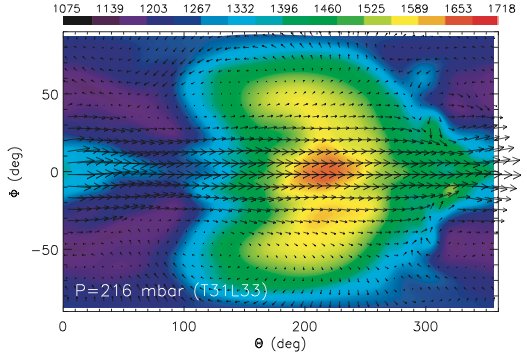}
  \includegraphics[width=8.5cm,angle=0.0,origin=c]{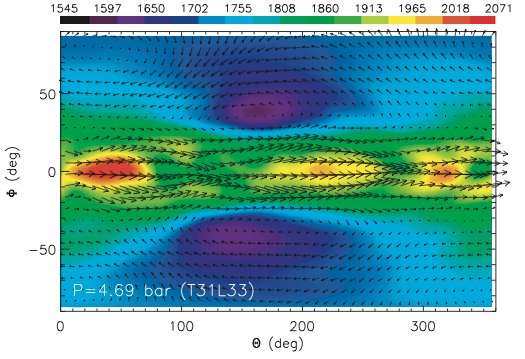}
  \includegraphics[width=8.5cm,angle=0.0,origin=c]{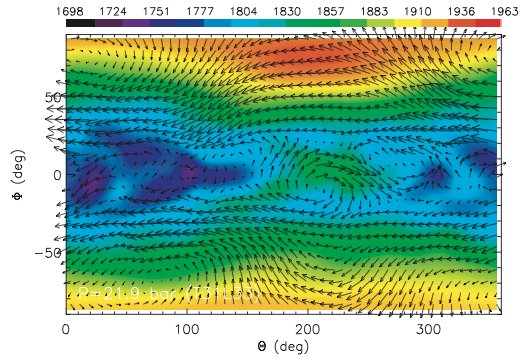}
  \caption{Figure showing results for the HD 209458b test case
    reproduced from \citet{heng_2011} (reproduced by permission of
    Oxford University Press). The \textit{top row} shows the
    zonal mean plots (i.e. zonally and temporally, from 200-1200
    days, averaged, using bar as the unit of pressure) of temperature
    (\textit{left}) and zonal wind (\textit{right}). The
    \textit{middle and bottom rows} show the temperature (colour) and
    horizontal velocities (vectors) at pressures 213 (\textit{middle
      left}), 21$\,$600 (\textit{middle right}), 4.69$\times 10^5$
    (\textit{bottom left}) and 21.9$\times 10^5$ Pa (\textit{bottom
      right}) after 1200 days.}
\label{hd209_bench_heng}
\end{figure*}

\begin{figure*}
\centering
  \hspace*{-0.7cm}\includegraphics[width=7.0cm,angle=90]{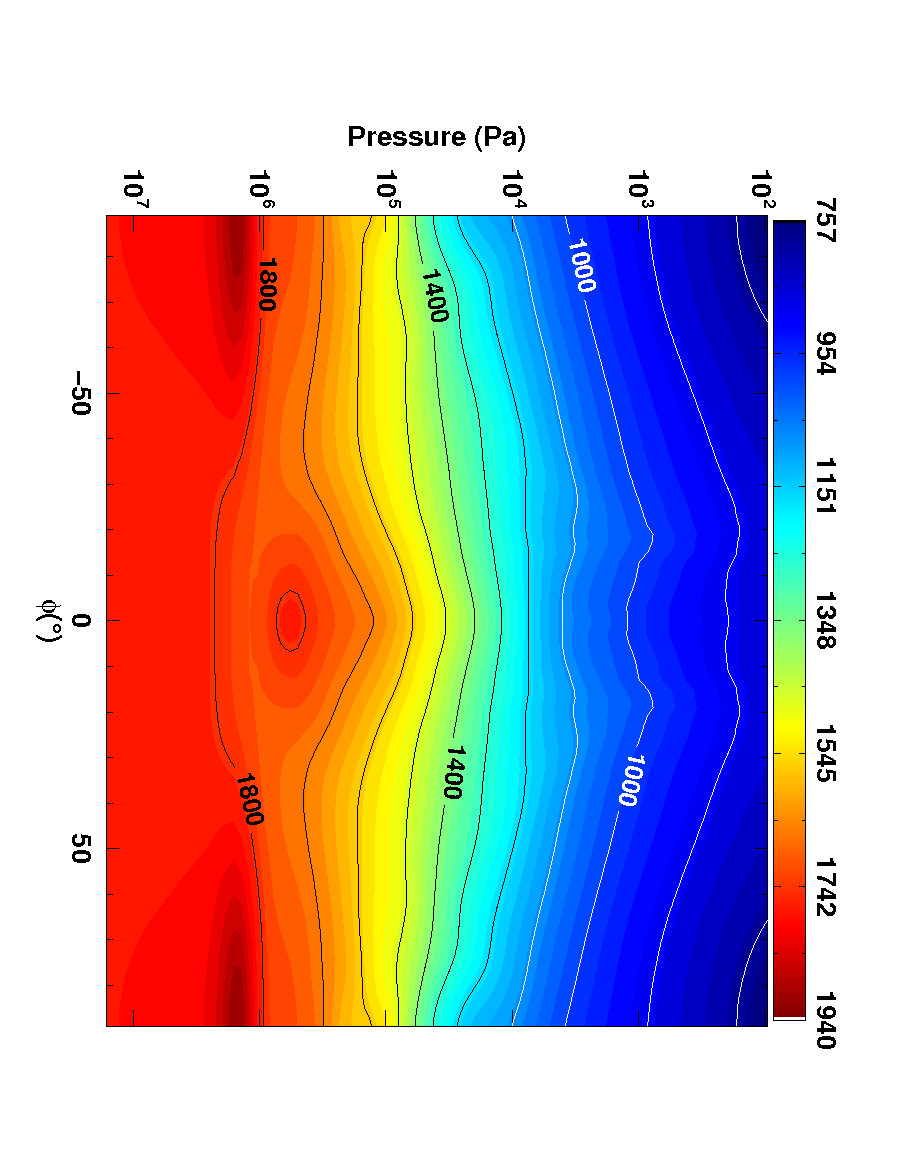}
  \hspace*{-0.7cm}\includegraphics[width=7.0cm,angle=90]{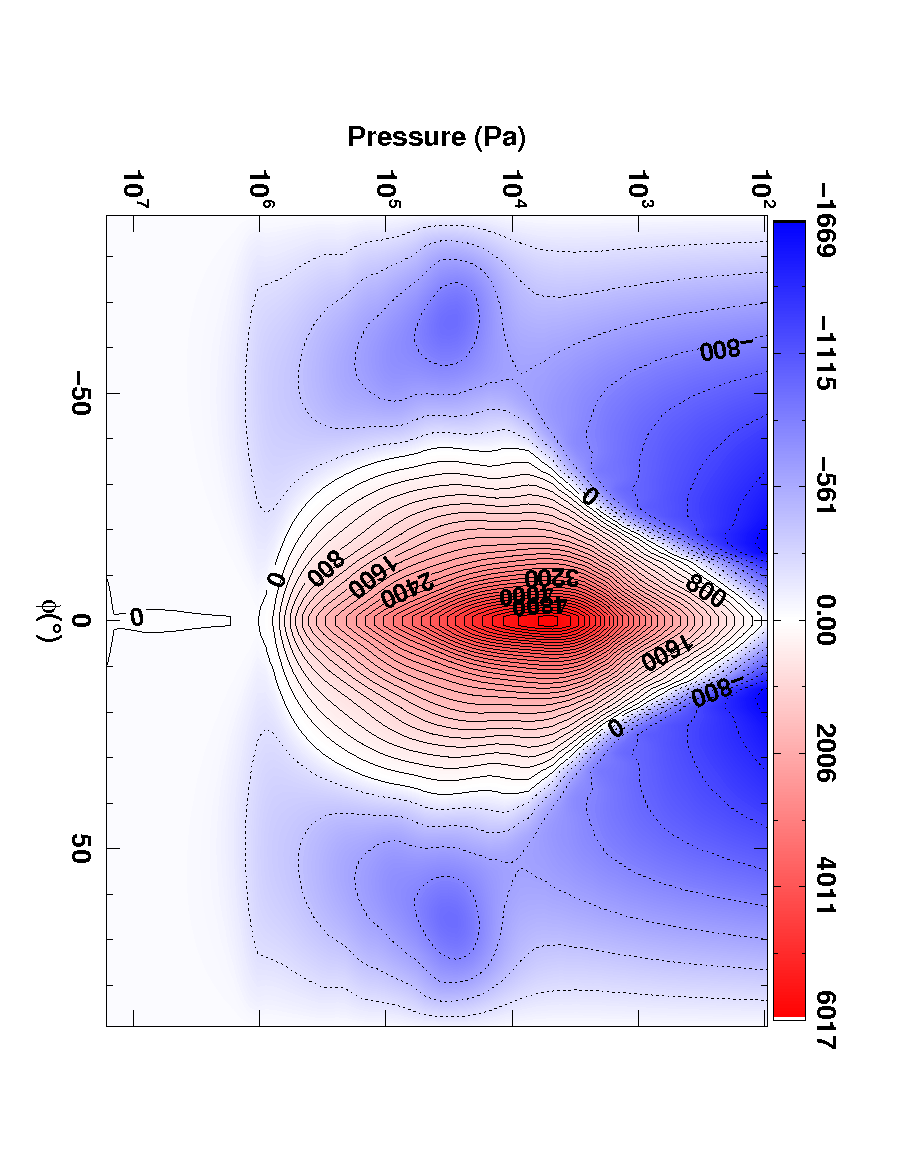}
  \hspace*{-0.7cm}\includegraphics[width=7.0cm,angle=90]{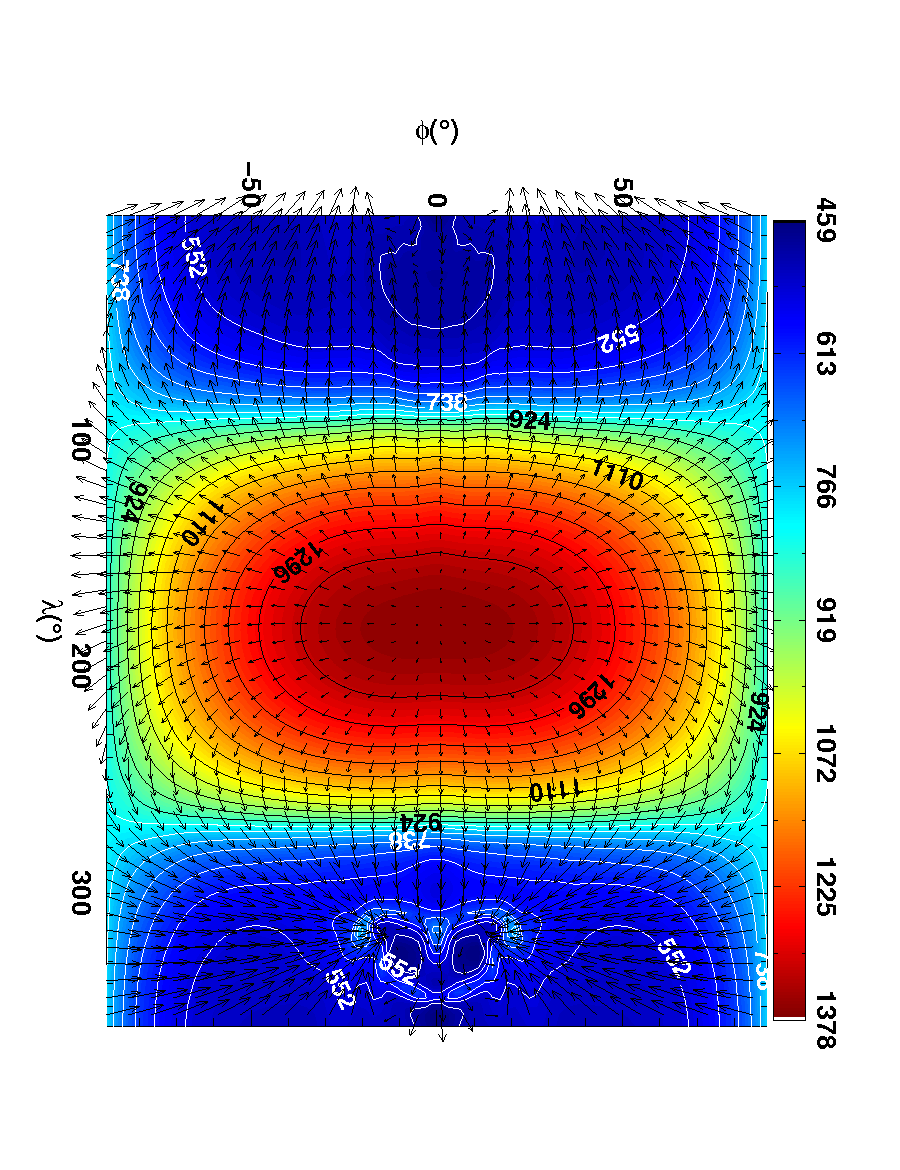}
  \hspace*{-0.7cm}\includegraphics[width=7.0cm,angle=90]{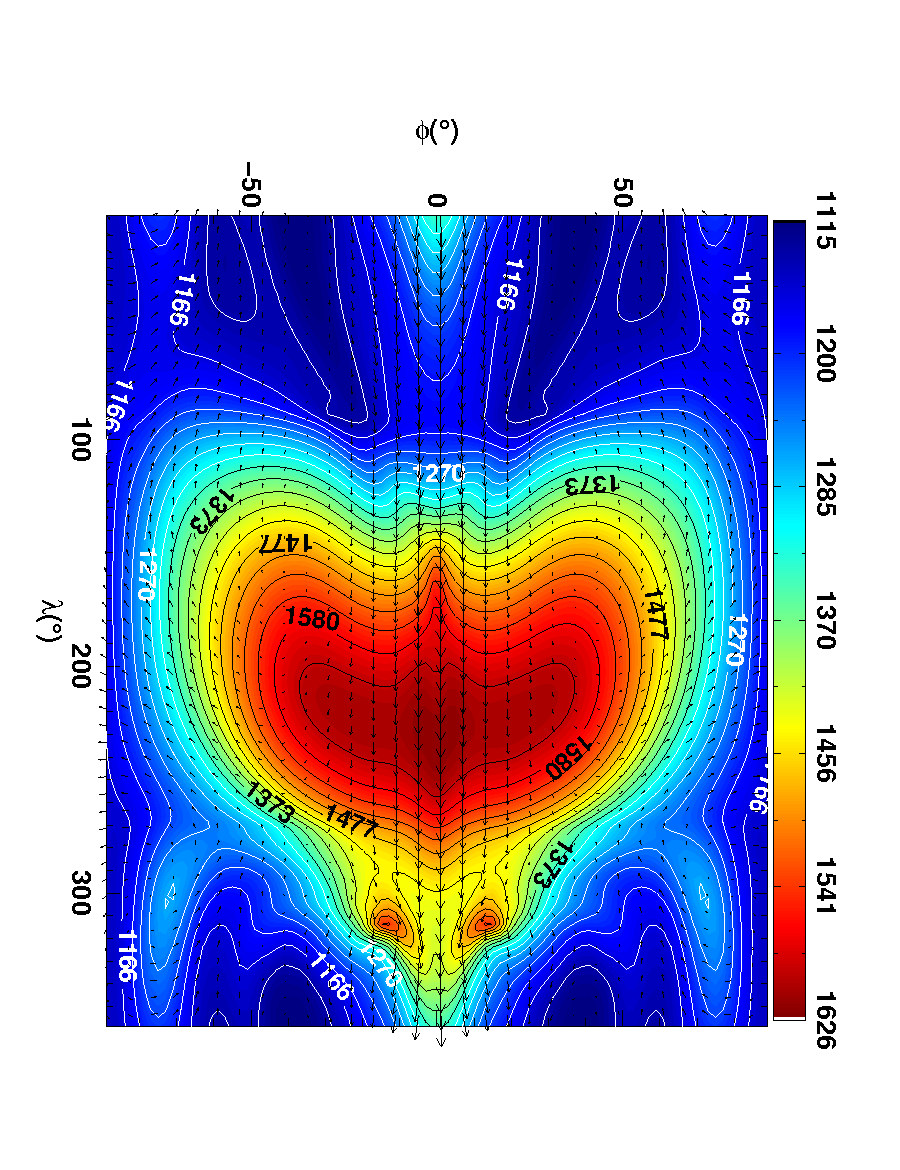}
  \hspace*{-0.7cm}\includegraphics[width=7.0cm,angle=90]{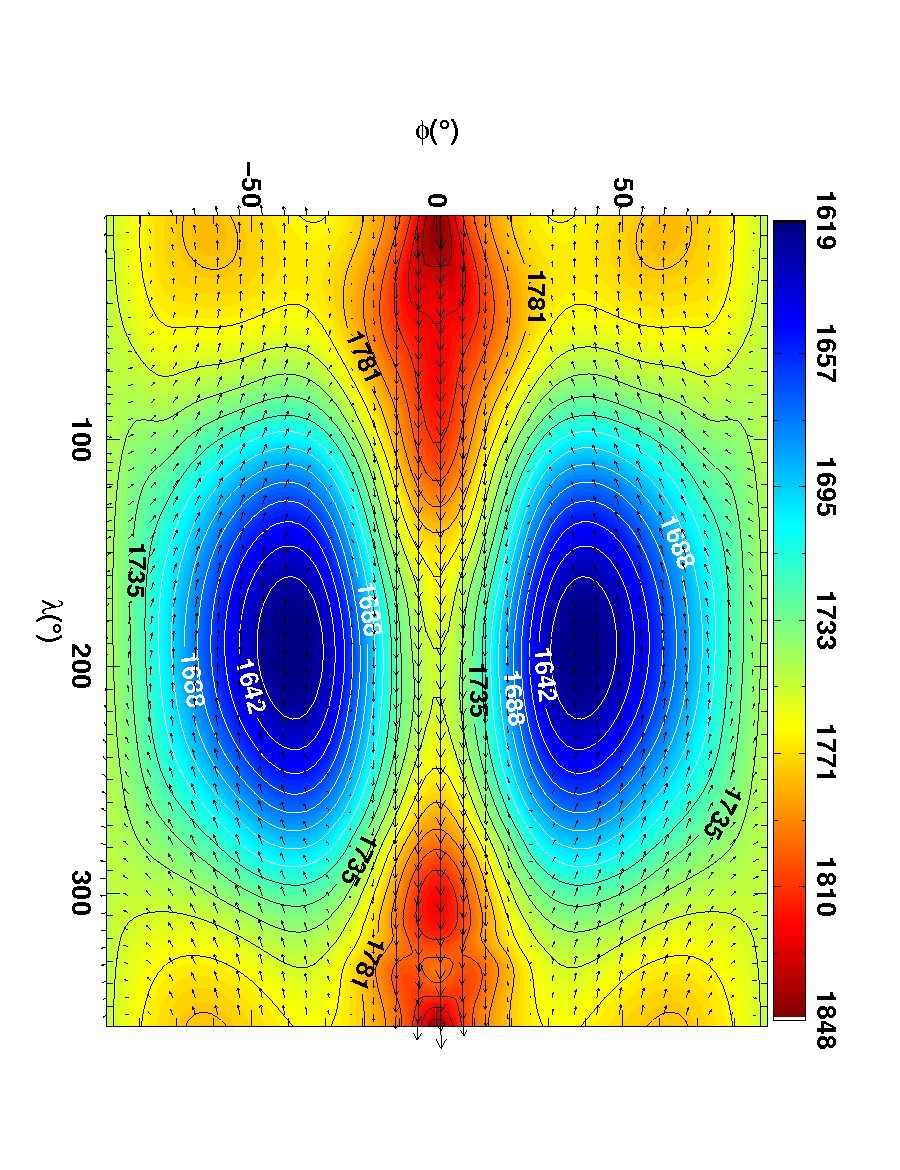}
  \hspace*{-0.7cm}\includegraphics[width=7.0cm,angle=90]{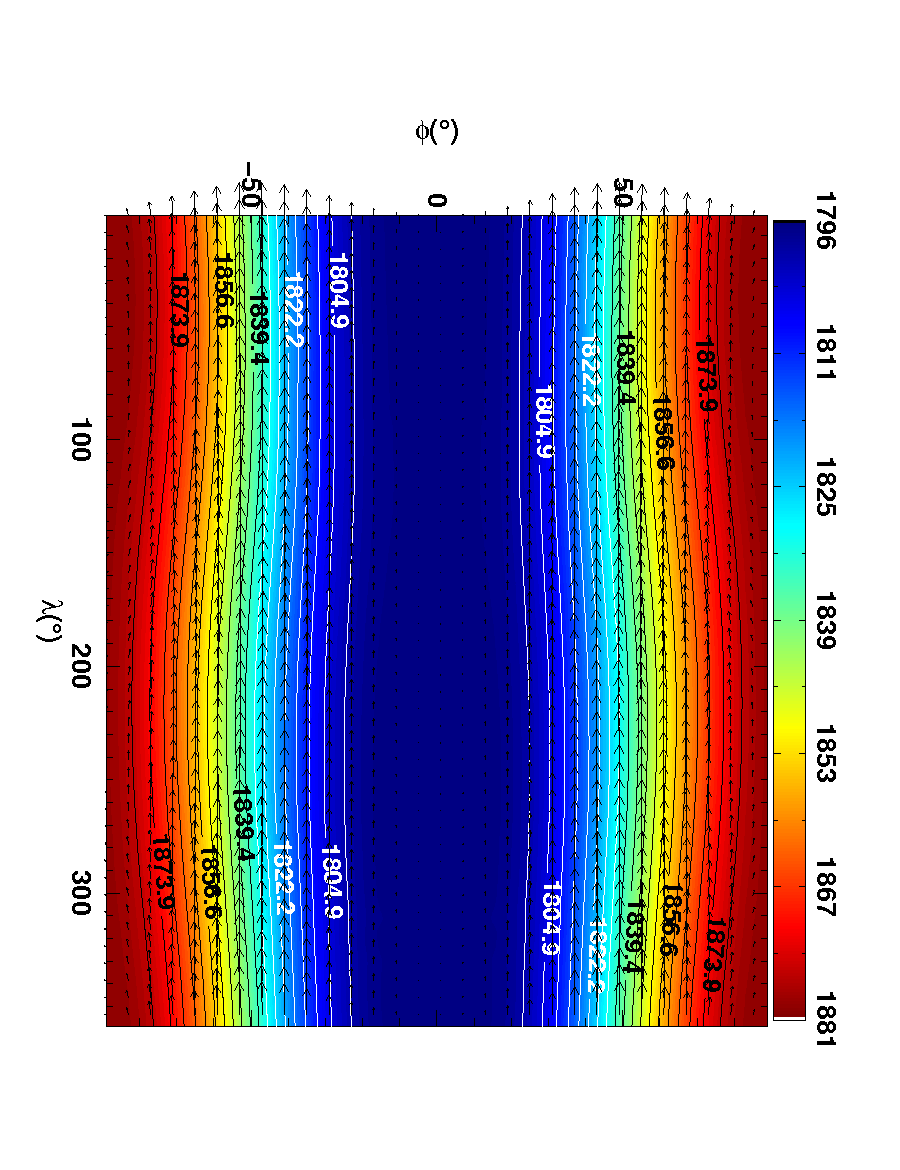}
  \caption{Figure matching those described in Figure
    \ref{hd209_bench_heng} but for our ``Shallow'' case. The
    zonal mean plots present pressure in Pa (SI unit, where $1$
    bar=$1\times10^5$ Pa).}
\label{hd209_bench_shallow}
\end{figure*}

In order to aid comparison Figure \ref{hd209_bench_heng} reproduces
the results of \citet{heng_2011}. Figure \ref{hd209_bench_heng} shows
snapshots of temperature and horizontal velocity for the same pressure
levels (i.e. 213, 21$\,$6000, 4.69$\times 10^5$ and 21.9$\times 10^5$
Pa) as in \citet{heng_2011} at 1200 days as found using their spectral
code\footnote{We do not compare to the finite--difference model as the
  full set of snapshots for this case are not presented in
  \cite{heng_2011}.}. Figure \ref{hd209_bench_heng} also shows the
zonal mean plots for the finite difference model of
\citet{heng_2011}. The same plots for our ``Shallow'' case are
presented in Figure \ref{hd209_bench_shallow}. We note that
\citet{heng_2011} uses the pressure unit of bar, whereas we use SI
units, Pa (where $1$ bar is $1\times10^5$ Pa).

\begin{figure*}
\centering
  \hspace*{-0.7cm}\includegraphics[width=7.0cm,angle=90]{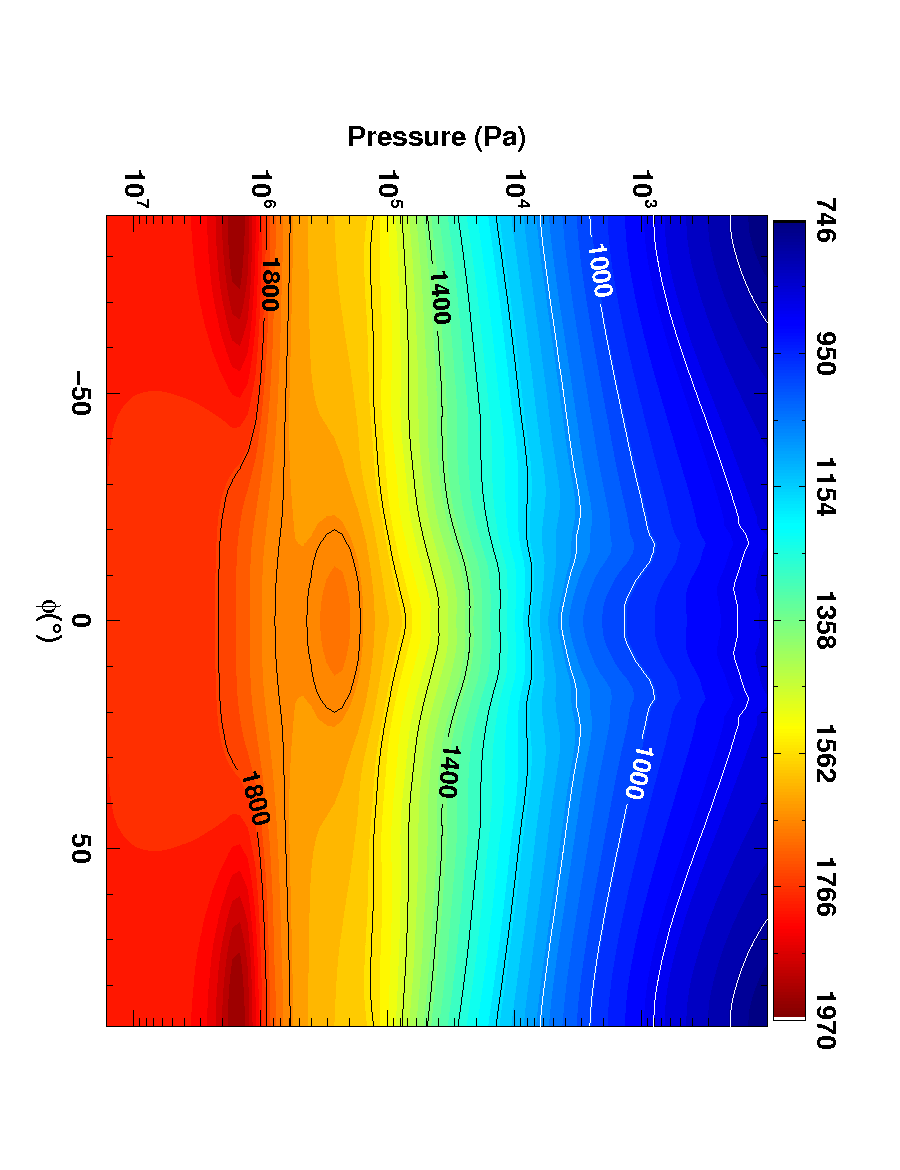}
  \hspace*{-0.7cm}\includegraphics[width=7.0cm,angle=90]{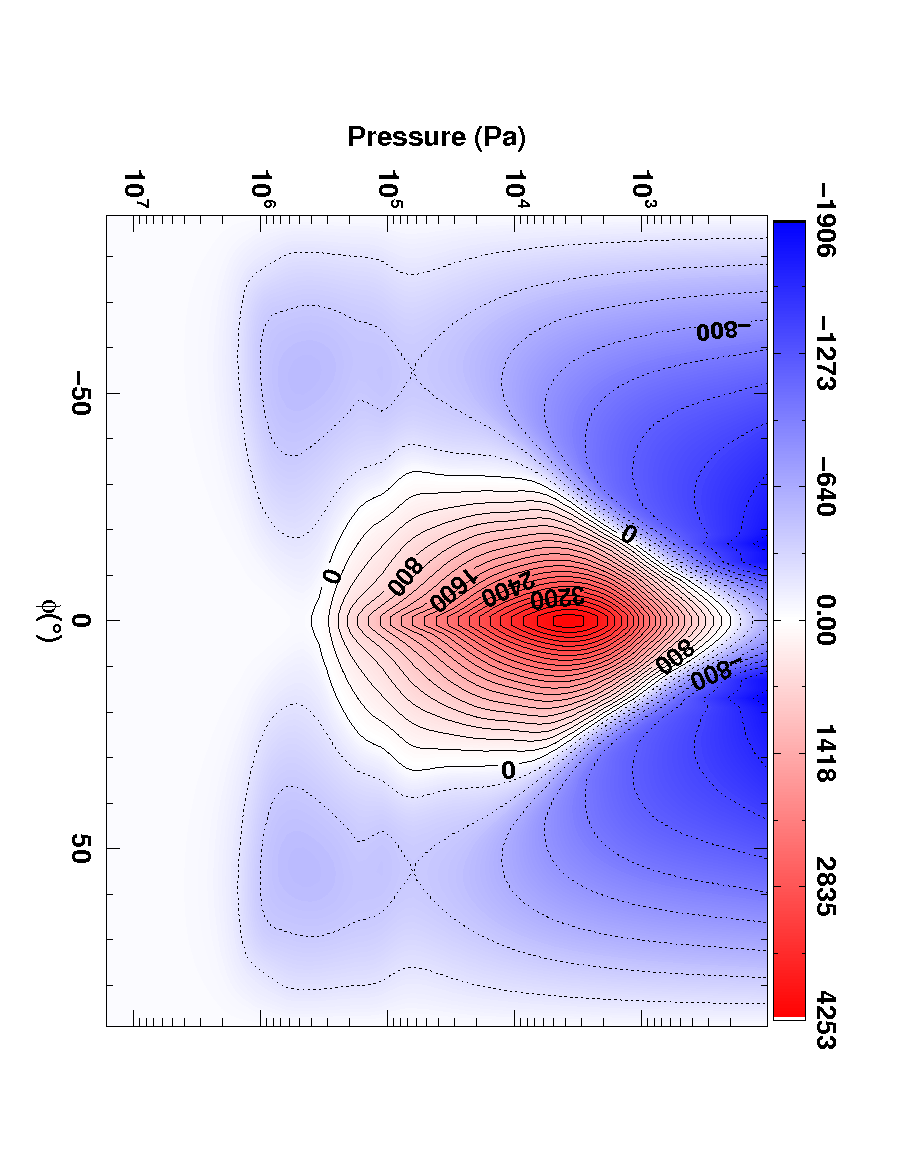}
  \hspace*{-0.7cm}\includegraphics[width=7.0cm,angle=90]{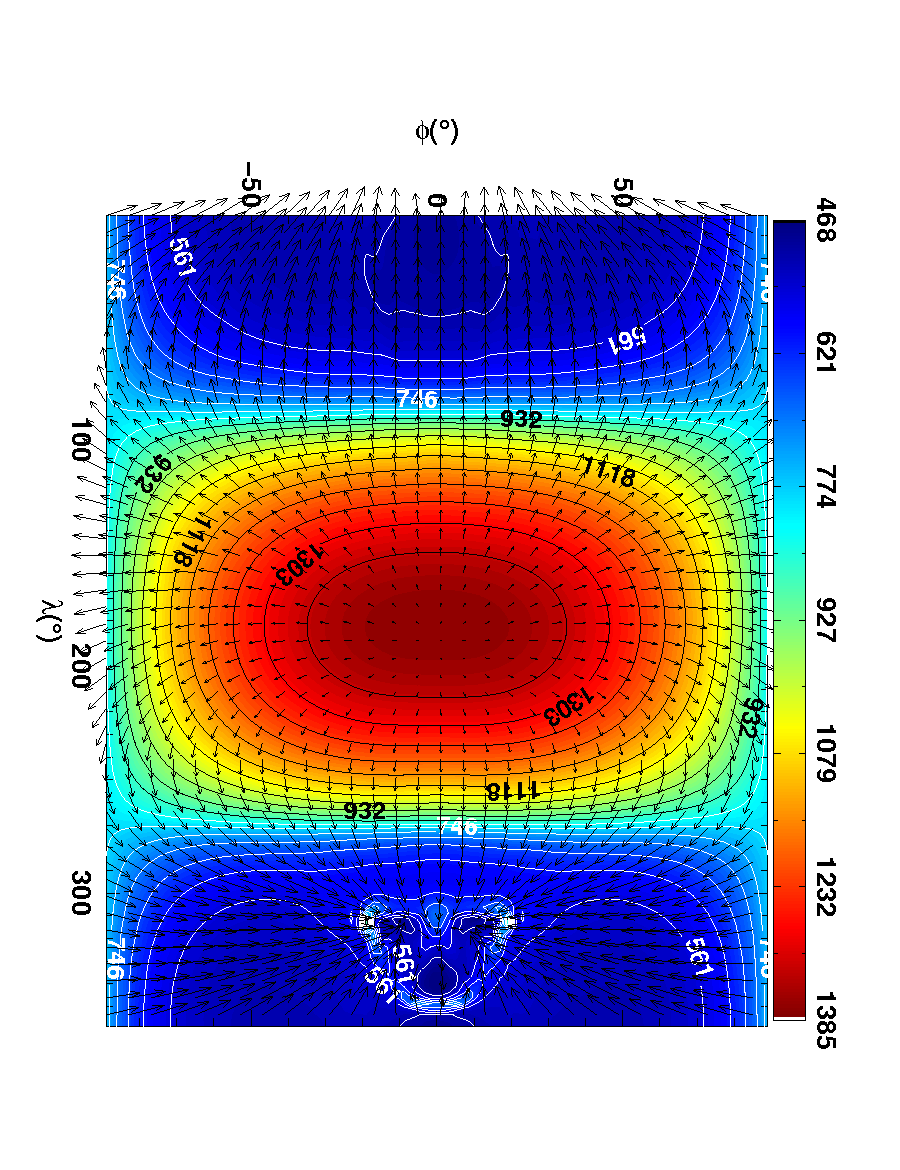}
  \hspace*{-0.7cm}\includegraphics[width=7.0cm,angle=90]{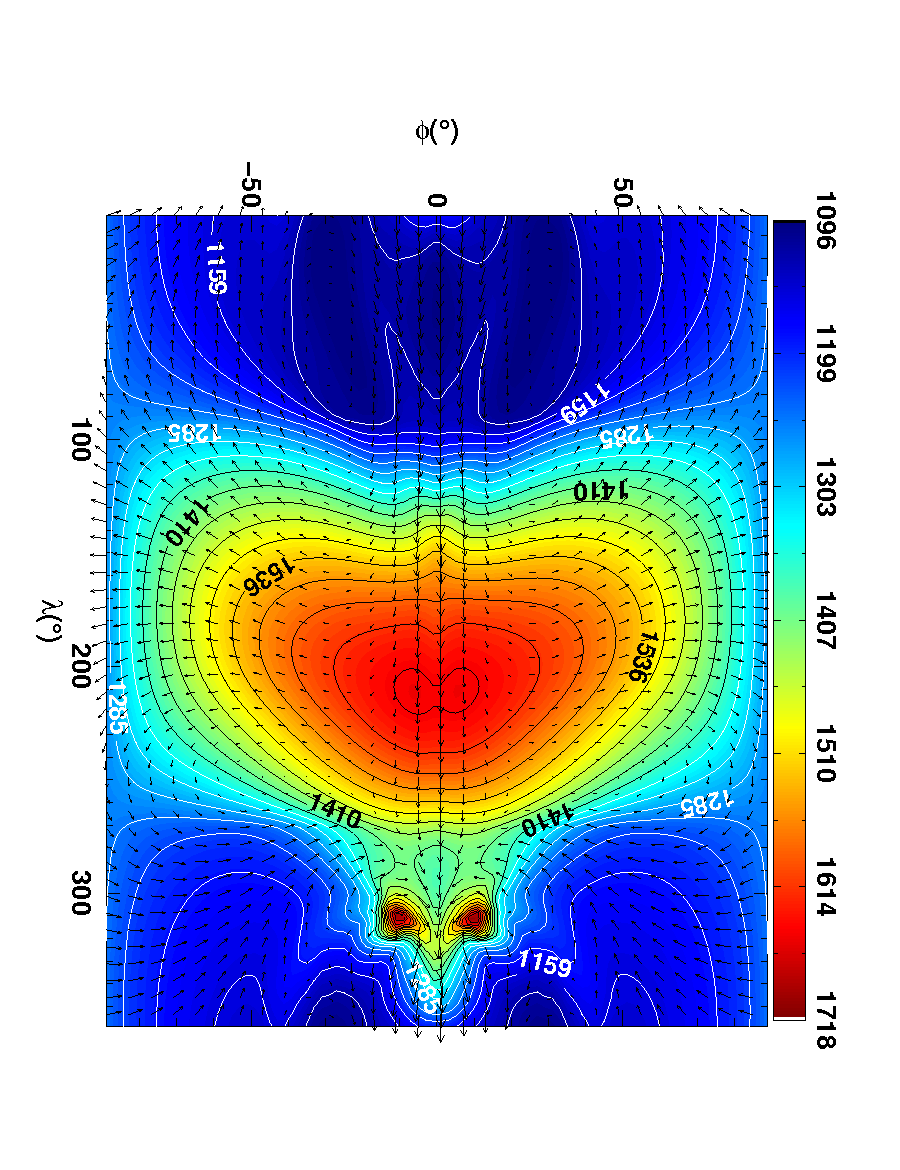}
  \hspace*{-0.7cm}\includegraphics[width=7.0cm,angle=90]{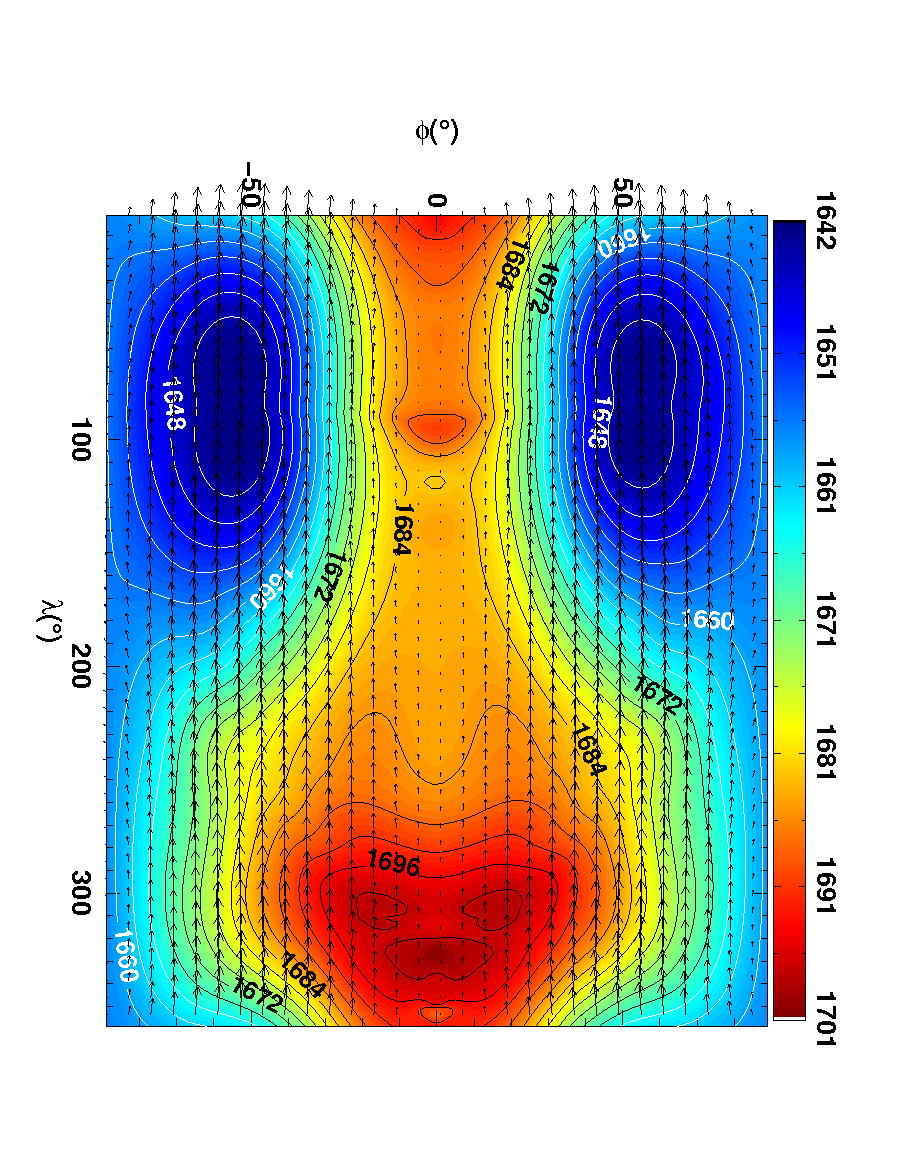}
  \hspace*{-0.7cm}\includegraphics[width=7.0cm,angle=90]{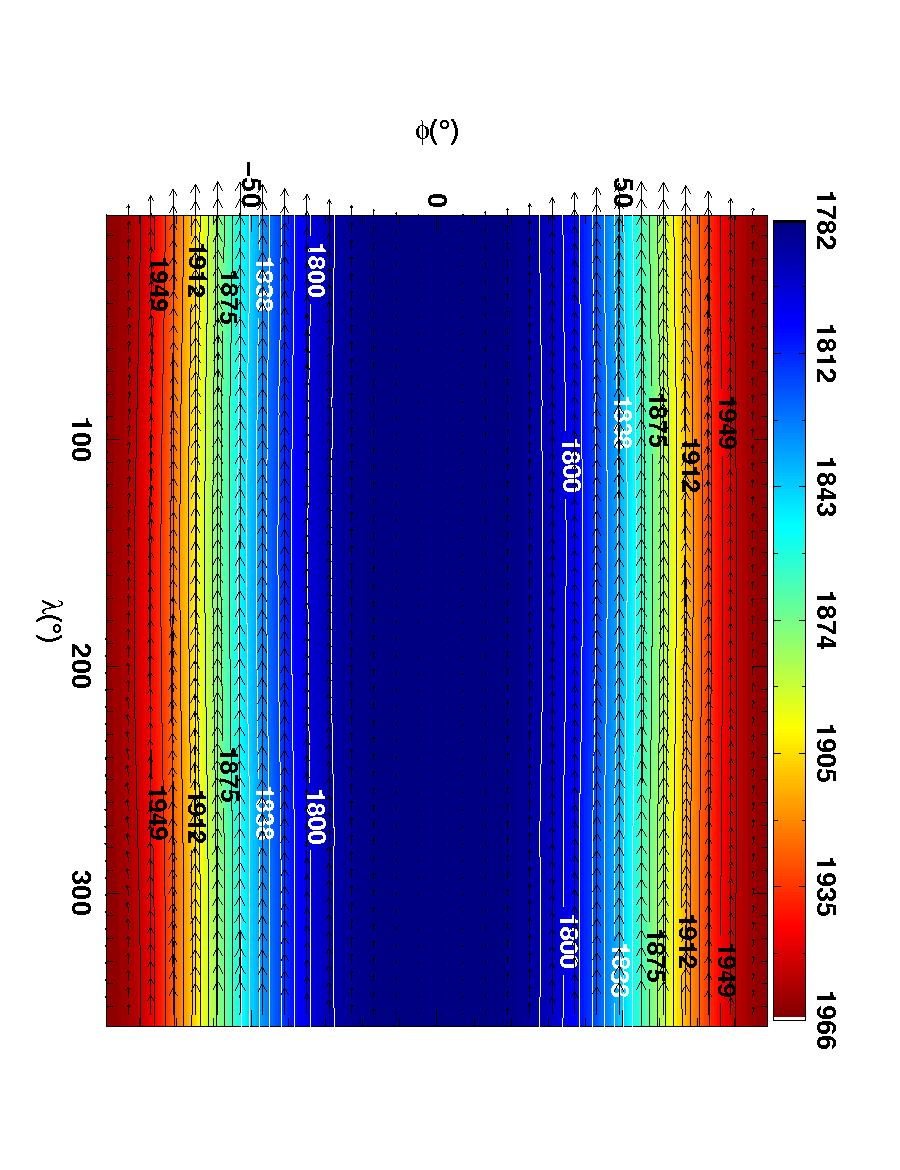}
  \caption{Figure matching those described in Figure
    \ref{hd209_bench_heng} but for our ``Deep'' case. The zonal mean
    plots present pressure in Pa (SI unit, where $1$ bar=$1\times10^5$
    Pa).}
\label{hd209_bench_deep}
\end{figure*}

\begin{figure*}
\centering
  \hspace*{-0.7cm}\includegraphics[width=7.0cm,angle=90]{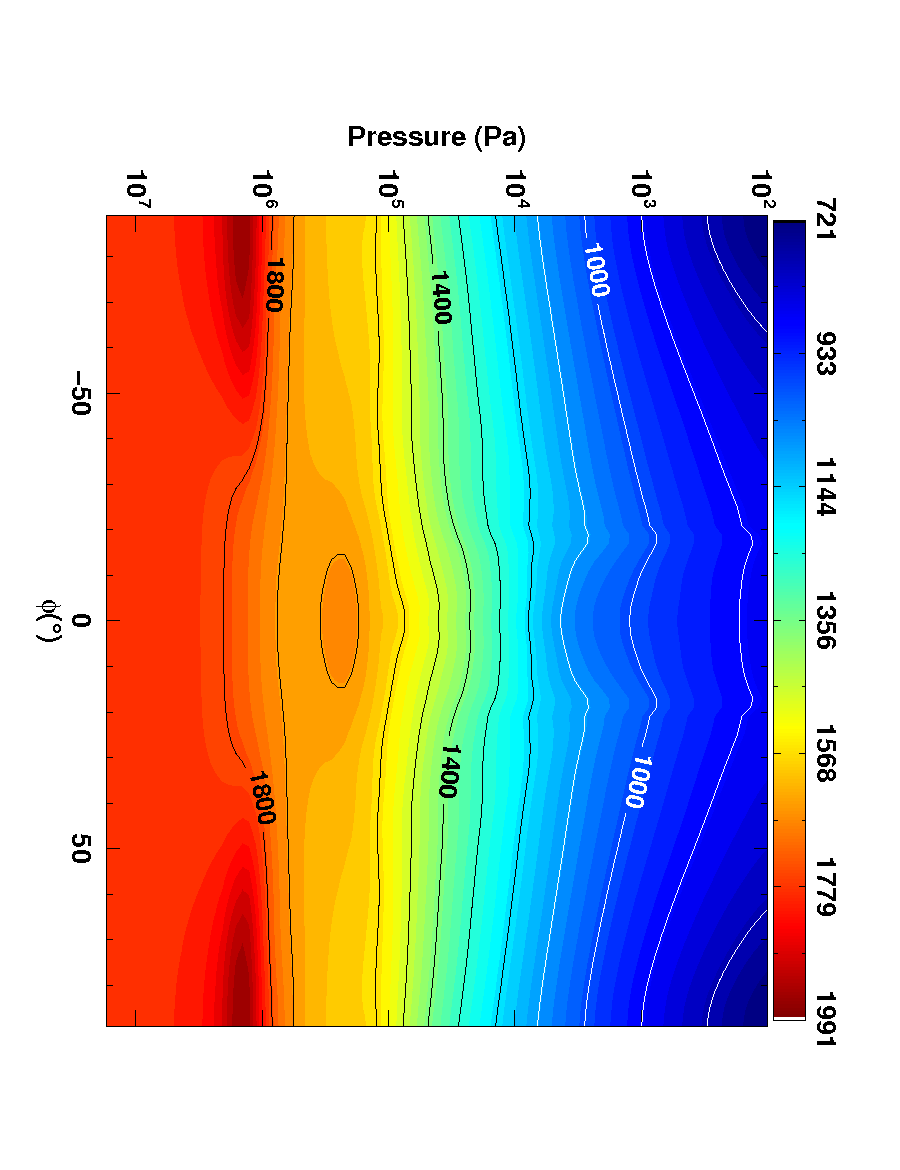}
  \hspace*{-0.7cm}\includegraphics[width=7.0cm,angle=90]{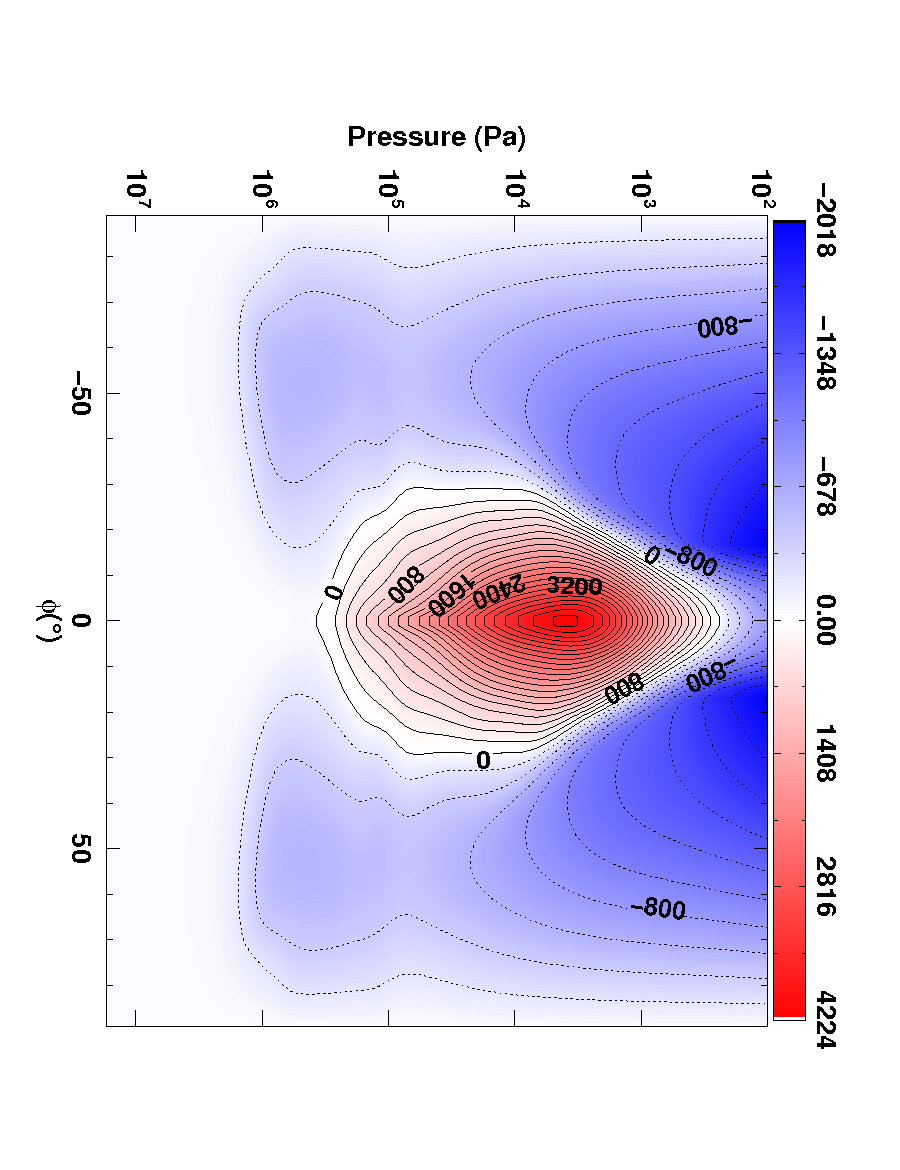}
  \hspace*{-0.7cm}\includegraphics[width=7.0cm,angle=90]{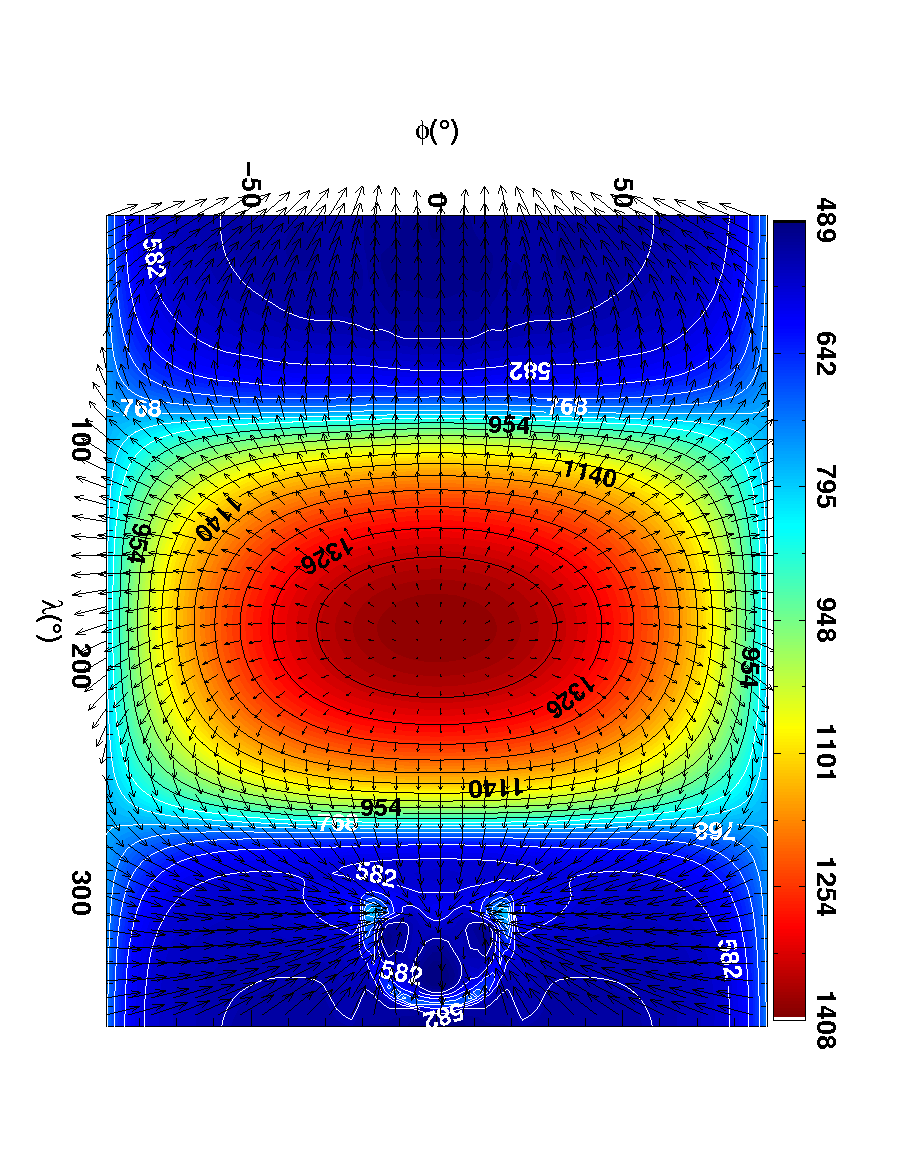}
  \hspace*{-0.7cm}\includegraphics[width=7.0cm,angle=90]{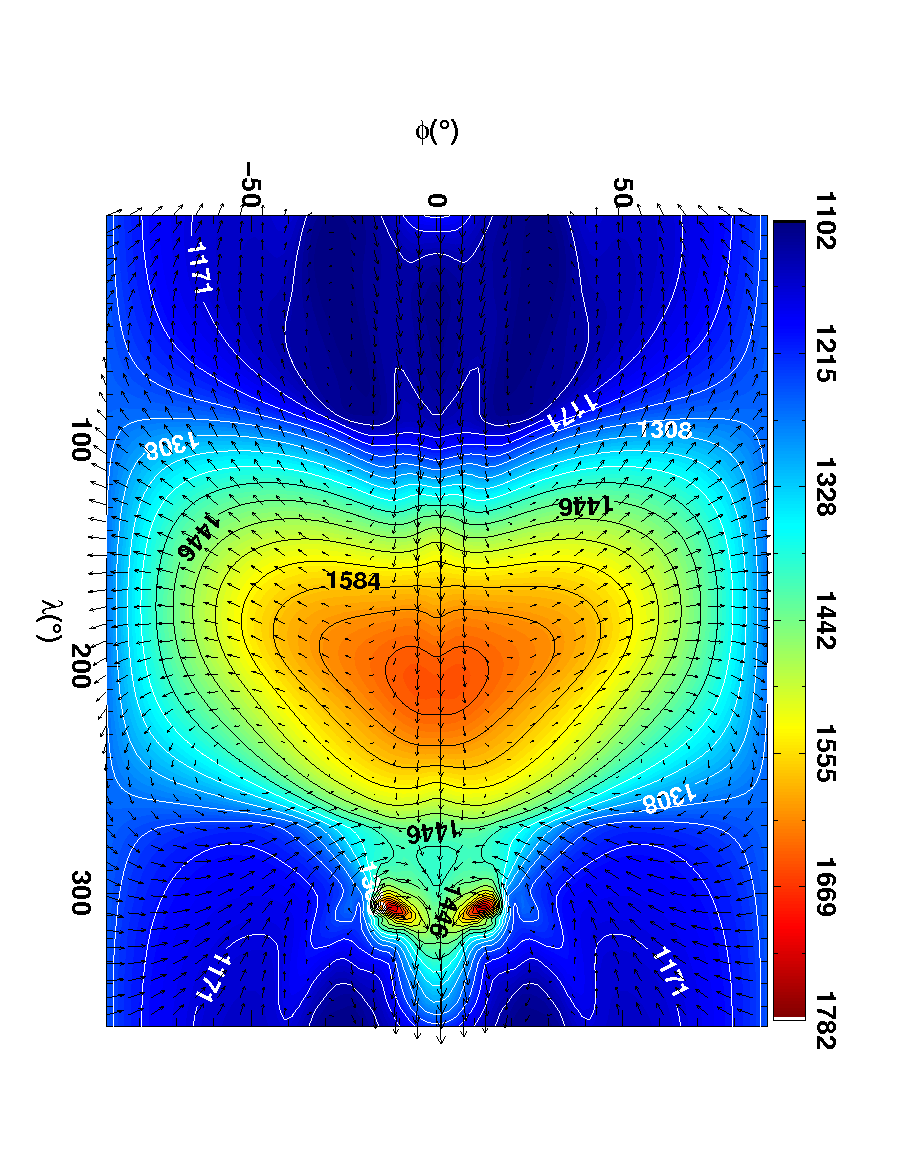}
  \hspace*{-0.7cm}\includegraphics[width=7.0cm,angle=90]{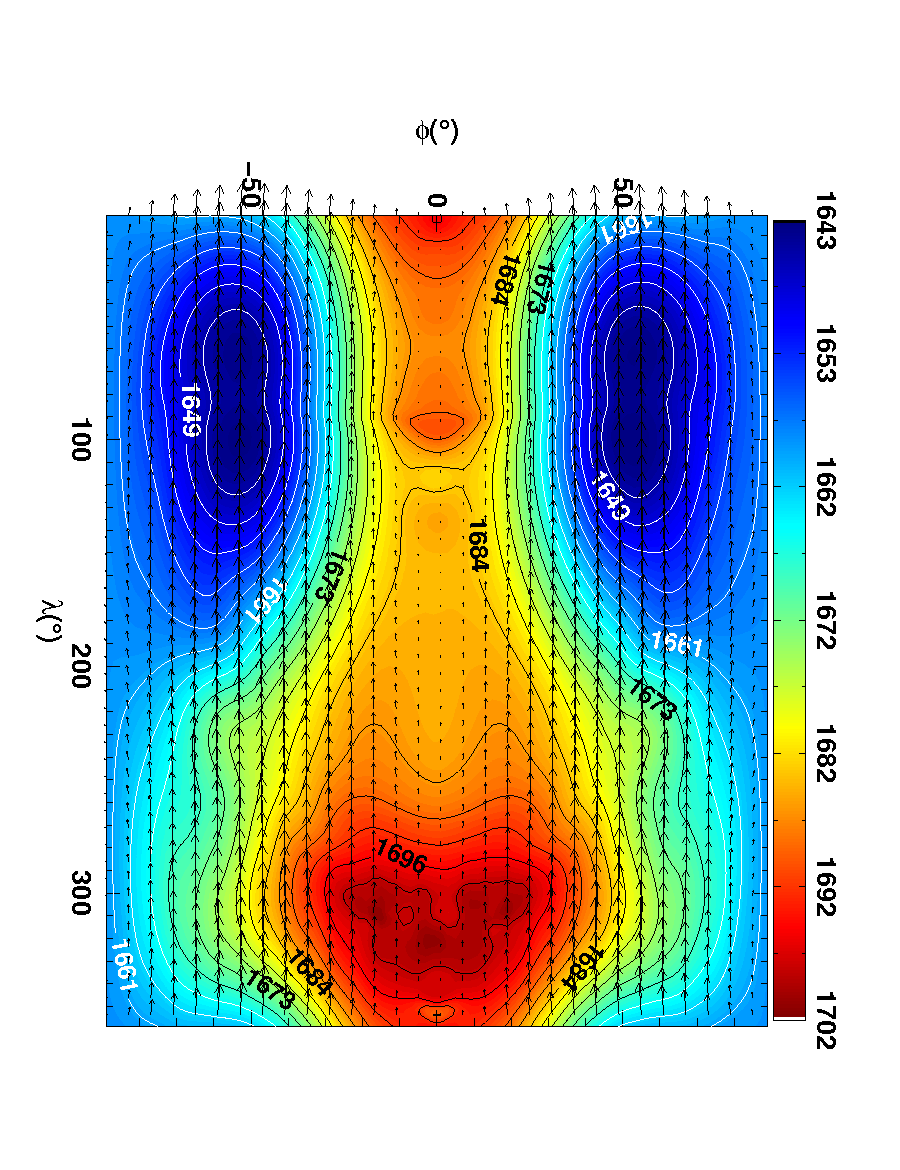}
  \hspace*{-0.7cm}\includegraphics[width=7.0cm,angle=90]{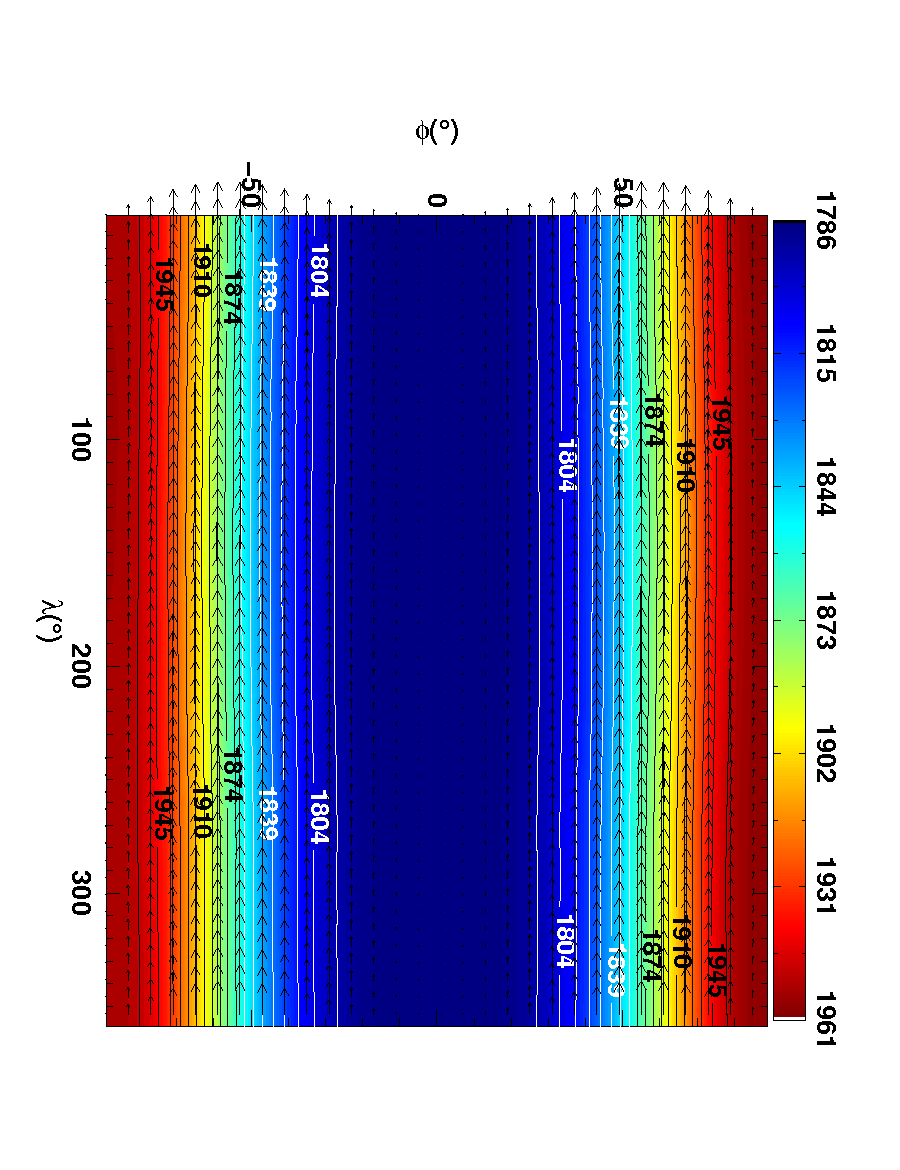}
  \caption{Figure matching those described in Figure
    \ref{hd209_bench_heng} but for our ``Full'' case. The zonal mean
    plots present pressure in Pa (SI unit, where $1$ bar=$1\times10^5$
    Pa).}
\label{hd209_bench_full}
\end{figure*}

Figures \ref{hd209_bench_deep} and \ref{hd209_bench_full} show the
same plots as Figures \ref{hd209_bench_heng} and
\ref{hd209_bench_shallow} but for the ``Deep'' and ``Full'' cases,
respectively. Comparing the results of \citet{heng_2011} reproduced in
Figure \ref{hd209_bench_heng} with our own results shown in Figures
\ref{hd209_bench_shallow}, \ref{hd209_bench_deep} and
\ref{hd209_bench_full}, shows good, qualitative, agreement. In all
cases we produce a wide, in latitude, prograde equatorial jet
extending throughout the upper atmosphere from about $5\times 10^5$ Pa
($5$ bar) to $100$ Pa (or 1 mbar), flanked by retrograde winds. The
temperature distribution also matches across the radiative zone. The
jet does sharpen slightly, in latitude, and move to higher altitudes
and lower pressures, as well as reducing in magnitude, when moving to
the more sophisticated equation sets (i.e. ``Shallow'' to
``Full''). 

The instantaneous slices through the atmosphere at 213 and 21$\,$600
Pa are also consistent across the figures presented. The 213 and
21$\,$600 Pa isobaric surfaces exhibit diverging flow at the lower
pressures and the development of a circumplanetary jet, with an
associated shift in the temperature distribution at the higher
pressure of the two surfaces. The temperature distributions also show
little variation ($\Delta T\lesssim 150$K) across all simulations,
which is unsurprising given the short radiative timescale at these
pressures. At the higher pressure of 21.9$\times 10^5$ Pa the flow,
morphologically, is still very similar, however the flow of
\citet{heng_2011} appears less coherent. Additionally, slightly larger
differences in temperature (than those found at the lower pressures)
across the simulations appear, for the deepest isobaric surface. The
pole, at depths, in the radiatively inactive region appears to become
warmer as we move to the more complete (i.e. ``Deep'' and ``Full'')
dynamical equations. 

The isobaric slice which shows the most difference between simulations
is at 4.69$\times 10^5$ Pa. Here the flow morphology of the
instantaneous field at 1200 days is quite different across the
simulations, as is the associated temperature structure. Both the
``Deep'' and ``Full'' cases show a counter rotating, or westward
moving flow at all latitudes. There is also a shift in the temperature
distribution, with the regions of lowest temperature shifted to lower
longitudes (i.e. westward). Despite the differences in the
instantaneous slices at 4.69$\times 10^5$ Pa, the overall flow
morphology is qualitatively very similar through each of
simulations. Moreover, the time averaged flow and temperature
structure, for all simulations, shows very little difference, despite
the differences in numerical scheme, initial conditions and the
equations solved.

\begin{figure*}
\centering
  \hspace*{-0.7cm}\includegraphics[width=7.0cm,angle=90]{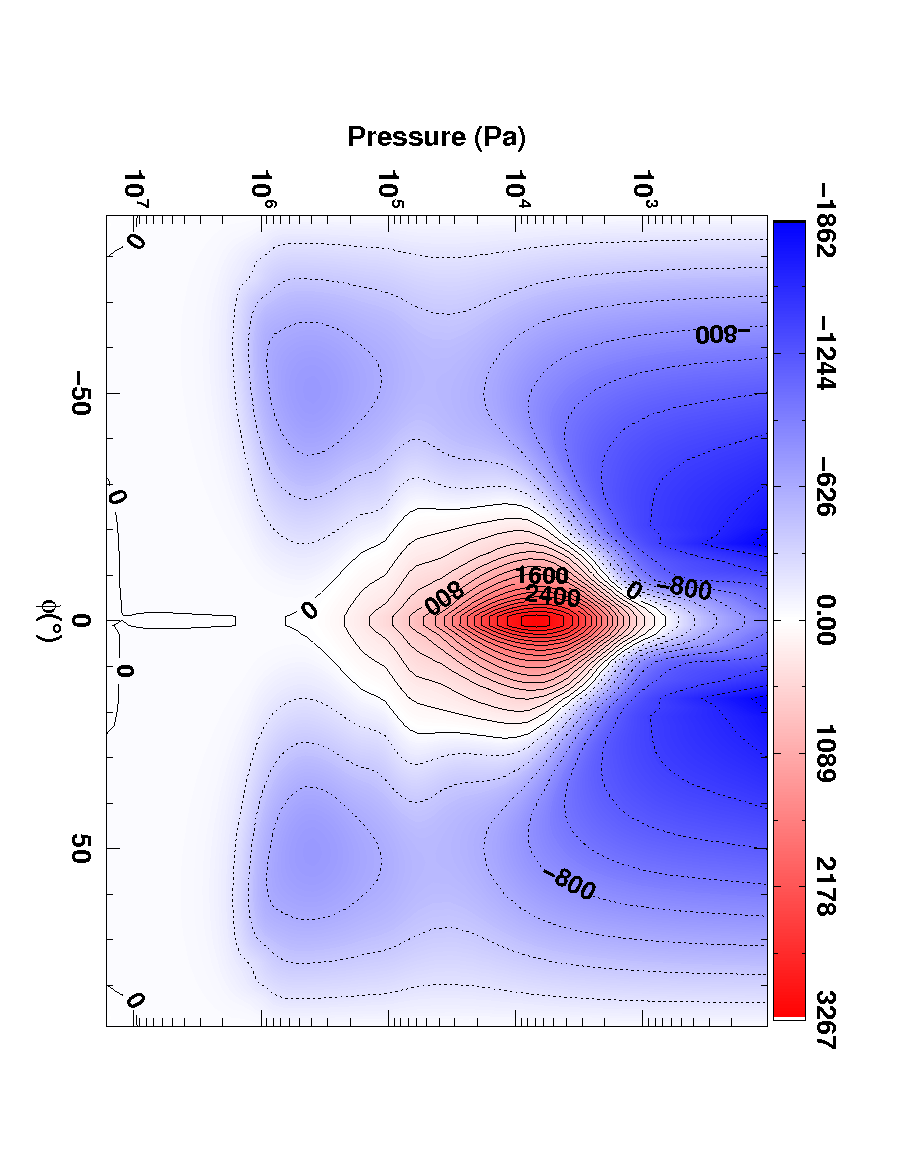}
 \hspace*{-0.7cm}\includegraphics[width=7.0cm,angle=90]{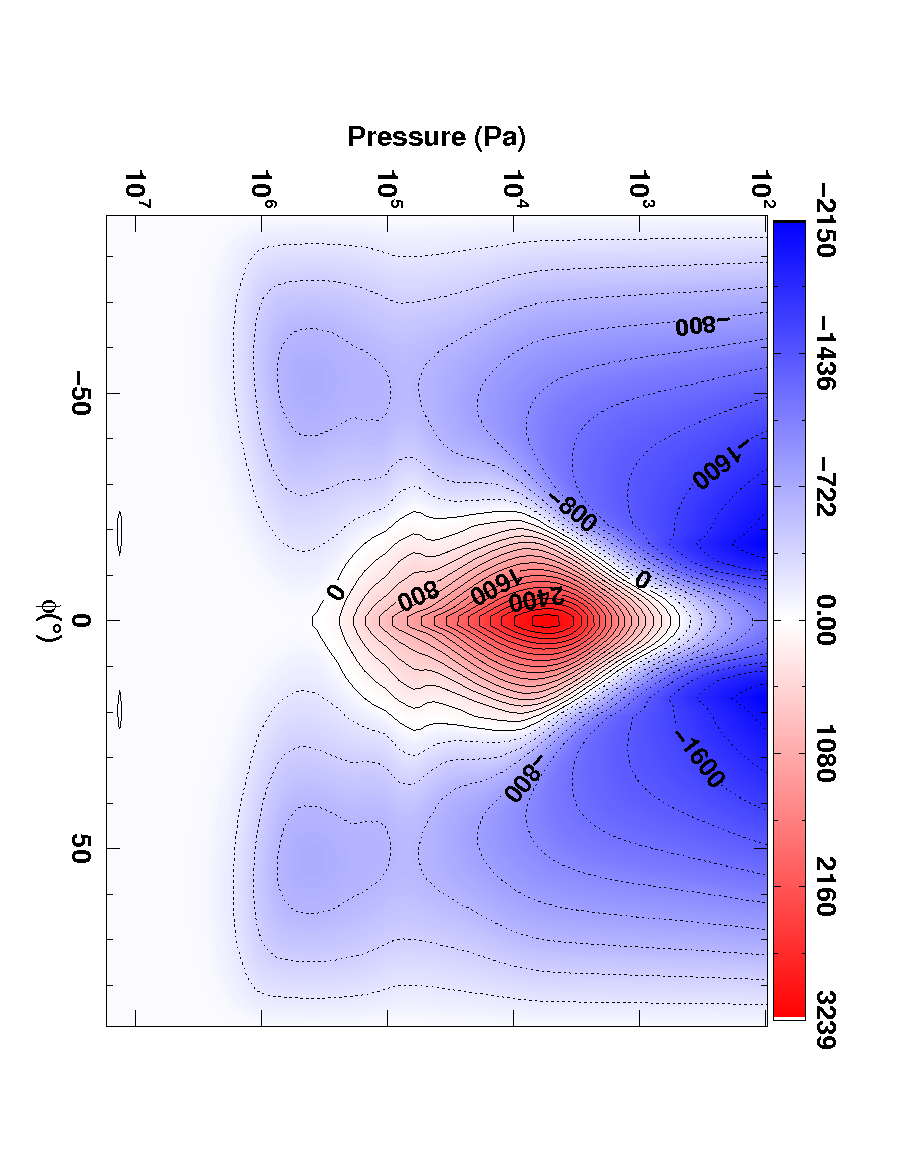}
 \hspace*{-0.7cm}\includegraphics[width=7.0cm,angle=90]{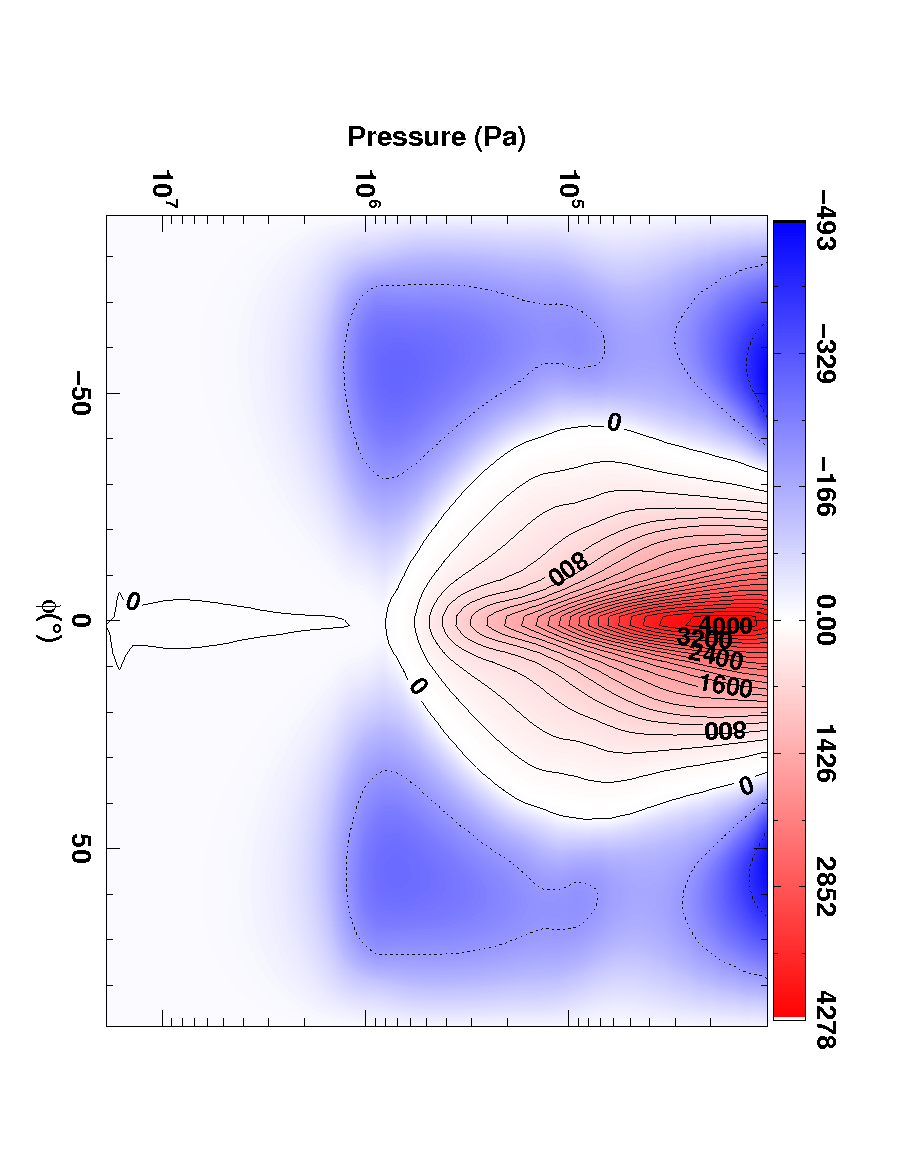}
 \hspace*{-0.7cm}\includegraphics[width=7.0cm,angle=90]{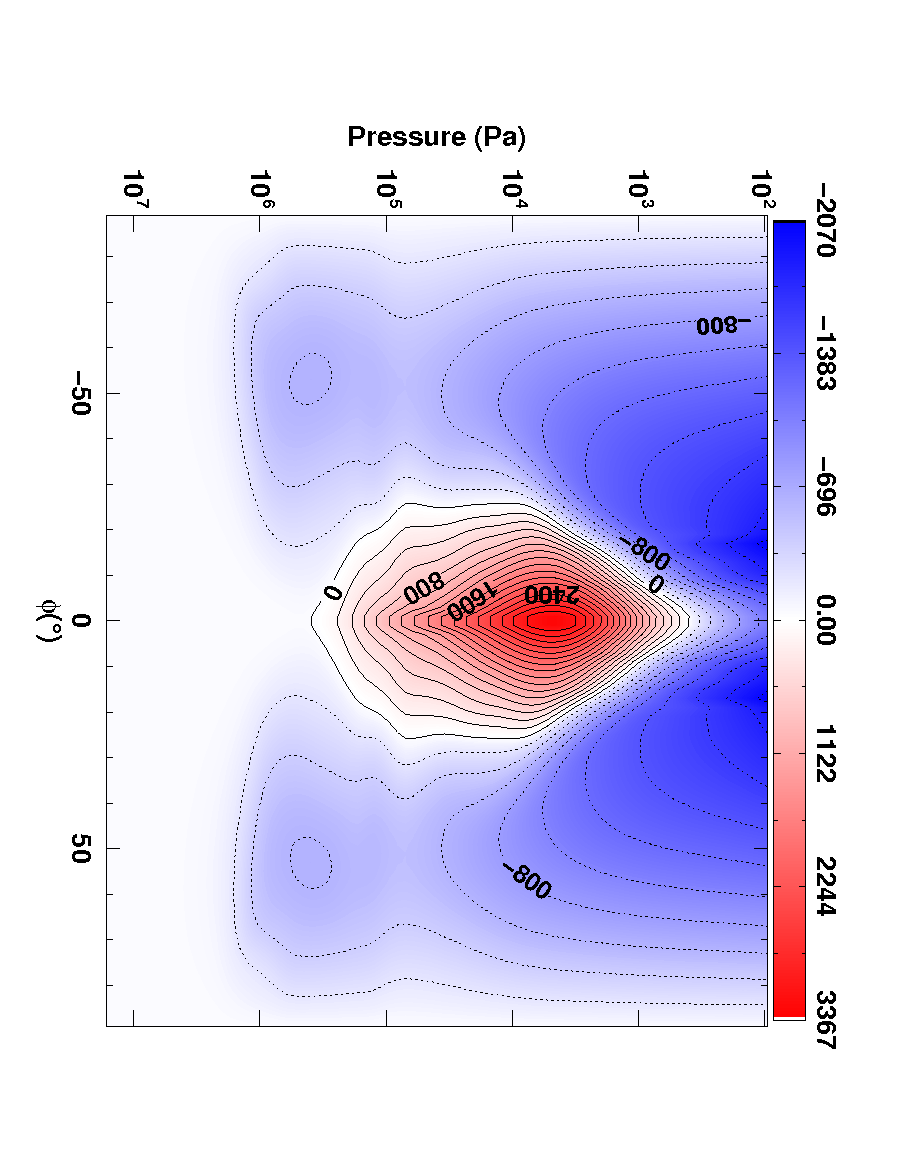}
  \caption{Figure showing the insensitivity of the zonally and
    temporally averaged zonal wind (ms$^{-1}$) to the different
    modelling choices. The simulations in the \textit{left panels} use
    the ``Shallow'' and the \textit{right panels} the ``Full''
    equation set. The \textit{top left panel} shows a simulations
    where diffusion is applied to $\theta$ in addition to $u$ and
    $v$. The \textit{top right panel} shows a simulations with
    non--uniform vertical level placement to optimise sampling of the
    \textit{local} scaleheight. The \textit{bottom left panel} shows
    the results when the atmospheric height is decreased from
    $H=1.1\times 10^7$ m to $H=6.7\times 10^6$ m, and the
    \textit{bottom right panel} when it is increased to $1.25\times
    10^7$ m (although only the overlapping pressure domain of these
    simulations with that of the models in \citet{heng_2011}, shown in
    Figure \ref{hd209_bench_heng}, is displayed to aid comparison).}
\label{hd209_bench_variants}
\end{figure*}

In Figure \ref{hd209_bench_variants} we present the results from a
subset of the simulations we have run demonstrating the relative
invariance of the derived flow structure for this test case, over 1200
days. Here we term the standard simulations as those presented in
Figures \ref{hd209_bench_shallow}, \ref{hd209_bench_deep} and
\ref{hd209_bench_full}. We have then run a set of simulations where we
have changed individual parameters or settings to explore their
effect, using each of the ``Shallow'' and ``Full'' equation sets. Here
we present only a subset in order to sample the whole `parameter
space' with as few figures as possible.

As discussed in Section \ref{diffusion} we apply diffusion to the $u$
and $v$ fields only for this test case, in order to simply reproduce a
more consistent result with that of \citet{heng_2011}. The \textit{top
  left panel} for Figure \ref{hd209_bench_variants} shows the results
for a simulations with exactly the same setup as the ``Shallow'' case
shown in the \textit{top right} panel of Figure
\ref{hd209_bench_shallow}, but with diffusion additionally applied to
the $\theta$ field. The jet structure still persists but has shifted
to higher pressures, sharpened in latitude and diminished, slightly
diverging from the results of \citet{heng_2011} (as discussed in
Section \ref{diffusion}). This change is likely due to the effect of
diffusion of the potential temperature on the baroclinically unstable
regions flanking the equatorial jet. The details and changes in the
underlying mechanism which `pumps' the jet will be explored in a
future publication (Mayne et al, in preparation). Despite the
differences, the flanking retrograde jets are still present and the
relative prograde to retrograde motions are similar to the previous
simulations.

As mentioned in Section \ref{vert} we also adopt uniformly distributed
(in geometric height) vertical levels for our standard HD 209458b
simulations, as opposed to those uniform in $\log(p)$ \citep[as
  adopted by][]{heng_2011}. The \textit{top right panel} of Figure
\ref{hd209_bench_variants} shows the resulting flow for a simulations
matching the ``Full'' case presented in the \textit{top right panel}
of Figure \ref{hd209_bench_full} but with only the vertical level
distribution altered. The non-uniform level distribution used has been
chosen to sample the \textit{local minimum} atmospheric
scaleheight. At each height, starting at the inner boundary, the
smallest scaleheight (usually on the cooler night side) was found and
three levels were placed within this scaleheight. The process was
repeated till the height domain of the atmosphere ($1.1\times 10^7$ m)
was reached (and resulted in a total of 78 vertical levels, compared
to 66 for the standard simulations). Again, a similar flow morphology
is found with a prograde jet flanked by retrograde jets, with only a
modest sharpening of the jet apparent when compared to the standard
``Full'' case.

Finally, we have also, as detailed in Section \ref{boundary} included
a sponge layer in our upper boundary condition. Therefore, to explore
the effect of this damping we have run two further simulations. The
\textit{bottom right panel} of Figure \ref{hd209_bench_variants} shows
a simulations where only the upper boundary has been altered from the
standard ``Full'' case presented in Figure \ref{hd209_bench_full}, and
is placed at $1.25\times 10^7$ m above the inner boundary (using 80
vertical levels to retain a similar vertical resolution). As we
increase the size of the domain, our upper boundary moves to lower
pressures, however, in Figure \ref{hd209_bench_variants} we only
present the vertical section of the domain matching that encompassed
by the standard ``Full'' case to aid comparison\footnote{The flow
  above this, at lower pressures, is just an obvious extension of the
  retrograde flow shown at the top of the figure.}. For this
simulation the damping layer only becomes non--negligible for pressure
lower than $<100$ Pa, i.e. above the plotted region. As before, the
flow does not diverge significantly from what one would expect of the
simulations both of \citet{heng_2011} and others in this
work. 

Secondly, in the \textit{bottom left panel} we present a simulation
matching the standard ``Shallow'' case, presented in \textit{top right
  panel} of Figure \ref{hd209_bench_shallow}, except the upper
boundary has been placed at $6.7 \times10^6$ m above the inner
boundary (using 40 vertical levels, again to retain a similar vertical
resolution), and does not include a damping layer in the upper
boundary condition. Once more, the flow is approximately what one
might expect from inspection of our standard case and those of
\citet{heng_2011}. The results presented in the \textit{bottom panels}
of Figure \ref{hd209_bench_variants} indicate that the damping layer
is not significantly altering our results besides its inclusion being
physically preferable (by preventing reflection of gravity waves back
into the domain at the upper boundary).

Figure \ref{hd209_bench_variants} represents only a subset of the
simulations we have run to explore the sensitivity to the parameters
and numerical choices. However, all of the simulations not presented
here show a similar qualitative flow structure, i.e. a prograde
equatorial jet flanked by retrograde winds, over the 1200 day test
period.

The key conclusion one can draw from these results is that the general
atmospheric structure is relatively invariant over 1200 days under a
range of model choices. Therefore, the resulting zonal mean
diagnostics plots for the HD 209458b test case \citep[as presented
in][]{heng_2011} are qualitatively very similar for all models. When
comparing our ``Shallow'' case with the primitive model of
\citet{heng_2011} the agreement suggests that, for this test, the
relaxation of the hydrostatic approximation and change in vertical
coordinate (from $\sigma$ to height) is unimportant. Furthermore,
although deviation is present in the snapshots and detail of the jet
structures, further relaxation of the `shallow--fluid' and
`traditional' approximations does not significantly alter the results
(our ``Deep'' case). Finally, the additional relaxation of the
constant gravity assumption (as represented by our ``Full'' case) also
does not cause the long--term, large--scale atmospheric structure to
change dramatically (i.e. the zonal mean plots). We have also shown
that the results are relatively invariant to our numerical choices
associated with diffusion, vertical resolution (and level placement)
and the upper boundary sponge or damping layer. However, again slight
differences in the detail of the flow structures are apparent.

\section{Discussion}
\label{discussion}

Despite the general concordance of our results with literature
results, and across our different model types, some differences are
apparent which we briefly discuss in this section. The zonal and
temporal averaging involved in the zonal mean plots is intended to
provide a robust way to compare the long--term and large--scale
structure of the model atmospheres. Therefore, by design these plots
are relatively insensitive to the more detailed differences in the
atmospheres.

\subsection{Shallow--Hot Jupiter}
\label{shj_discuss}

As discussed in Section \ref{test_cases} we have placed our vertical
levels for the SHJ test case at positions emulating the $\sigma$
levels described in \citet{heng_2011}. This process involved running a
SHJ test case, with uniformly distributed vertical levels, to
completion and zonally and temporally averaging the pressure structure
to provide a mapping from height to pressure. During this process, the
largest $\sigma$ value, i.e. the level closest to the inner boundary,
leads to a very small (in vertical size) grid cell, which, even with a
semi--implicit scheme, led to a numerical instability of the vertical
velocity. Therefore, the lowest level was adjusted to create a larger
(in vertical extent) bottom cell more numerically stable for a
non--hydrostatic code.

Although our results for the SHJ are qualitatively similar to those of
\citet{heng_2011} some differences are apparent. Most notably,
perhaps, is the fact that our jets (prograde or retrograde) do not
intersect either the boundary. No sponge layer is incorporated in the
upper boundary for this test, but the result is unchanged when it
is. This slight discrepancy between our results and those of
\citet{menou_2009} and \citet{heng_2011} is most likely caused by
differences in domain or boundary conditions, as both boundaries
intersect the flow features we are trying to capture. In this respect,
i.e. likely dependency of the results on the domain and boundary
conditions, the SHJ test is a poor benchmark.

\subsection{HD 209458b}
\label{hd209458b_discuss}

As explained in Sections \ref{approx} and \ref{test_cases}, the
prescribed test duration of 1200 days is only sufficient to span
approximately one relaxation time for the deeper regions of the
radiative zone. This, in addition to the fact it includes a
radiatively inactive region which can only reach a relaxed or steady
state through dynamical processes, suggests that 1200 days is
insufficient for this case to reach a statistical steady state. Models
based on the primitive equations have already shown that the deeper
atmosphere will not reach a steady state in 1200 days. Both
\citet{cooper_2005} and \citet{rauscher_2010} present evidence
indicating that the atmosphere down to only $\sim 3\times 10^5$ Pa (or
$\sim$3 bar) had relaxed in their models after 5000 and $\sim$600
days, respectively. \citet{rauscher_2010} additionally, explicitly
show that the kinetic energy is still evolving in the deeper regions
of their modeled atmosphere after 1200 days. Additionally, models from
the literature which include a more complete dynamical description,
have been run for much shorter times than 1200 days. For example
\citet{dobbs_dixon_2010} run their simulations, which include the full
dynamical equations, for only $\sim$ 100 days. 

As suggested by \citet{showman_2002} a downward flux of kinetic energy
was found in models of HD 209458b by \citet{cooper_2005}, therefore
energy is transported into the deeper radiatively inactive
region. Energy is also injected by the compressional heating. As
discussed in Section \ref{approx} if one compares the equation sets
used in our different models, as presented in Table \ref{eqn_sets},
the terms affected as we move from a ``Shallow'' to a ``Deep'' and on
to a ``Full'' equation set involve `exchange' between the components
of momentum, and importantly the vertical and horizontal
components. Moreover, relaxing the constant gravity assumption, in
particular, acts to weaken the stratification of the
atmosphere. Therefore, it is plausible that as one moves to a more
complete dynamical description, one allows the transfer of energy and
momentum between the upper radiative atmosphere with short relaxation
time (see discussion in Section \ref{equations_solved}), with both the
deeper longer timescale radiative and the even deeper radiatively
inactive regions.

\begin{figure}
\centering
  \hspace*{-0.7cm}\includegraphics[width=7.0cm,angle=90]{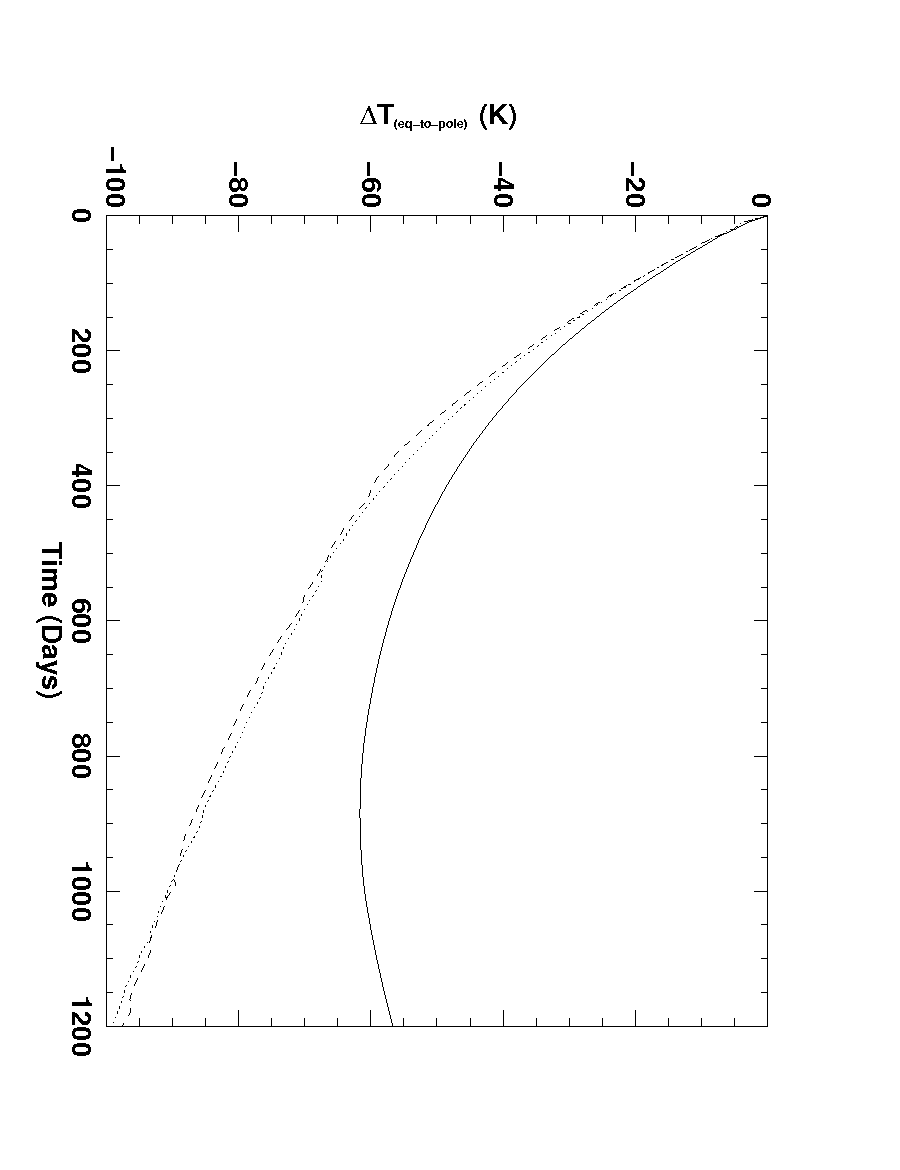}
  \hspace*{-0.7cm}\includegraphics[width=7.0cm,angle=90]{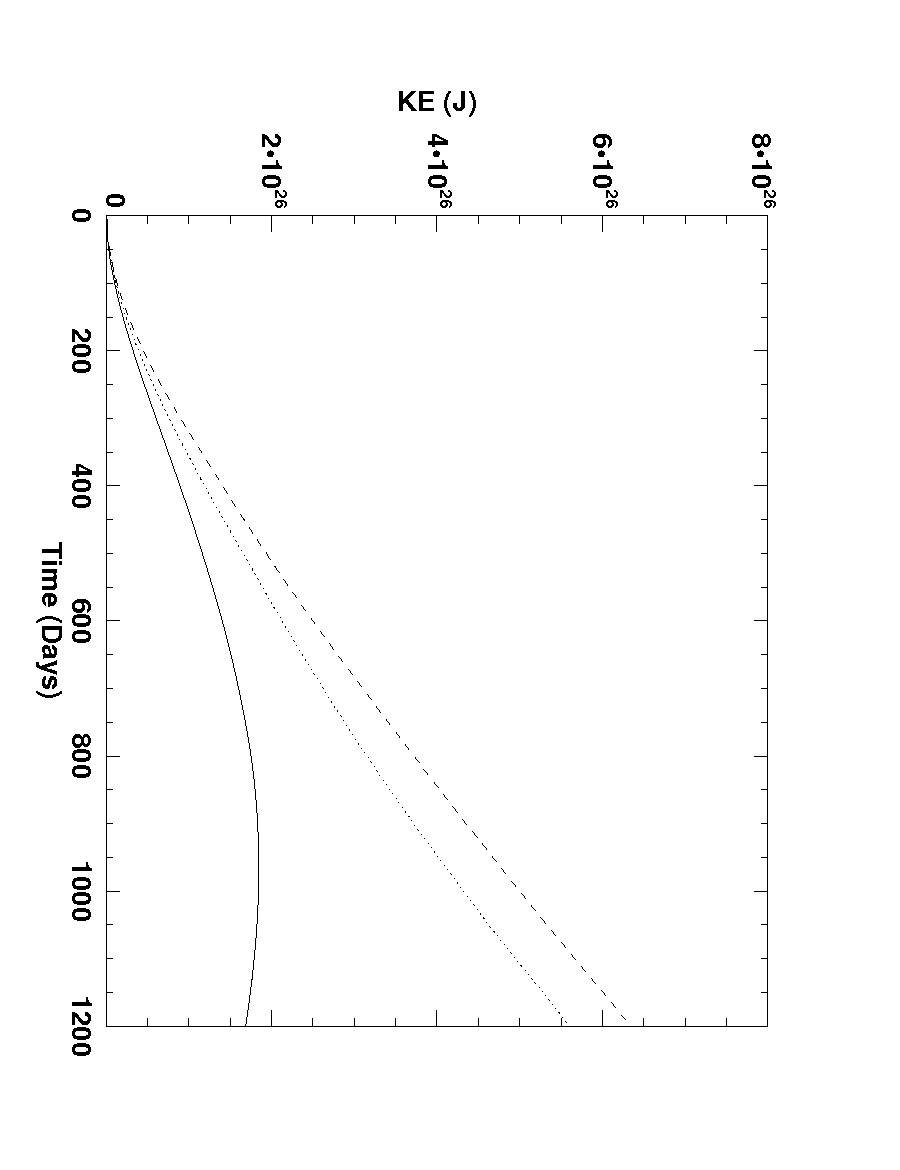}
  \caption{Figure showing the zonally and vertically averaged
    equator--to--pole temperature difference (\textit{top panel}), and
    total kinetic energy (\textit{bottom panel}) for the radiatively
    inactive region (i.e. $p\geq 1\times10^6$ Pa or 10 bar), for the
    HD 209458b test case. The ``Shallow'', ``Deep'' and ``Full'' cases
    are shown as the solid, dotted and dashed lines, respectively.}
 \label{move_deep}
\end{figure}

A retrograde flow in the deep region of the atmosphere must arise
through an equatorward meridional flow (by conservation of angular
momentum) or by vertical transport of angular momentum by waves or
eddies, and must be accompanied by a warming of the polar regions
relative to the equator (by thermal wind balance), which itself can
only arise through a meridional circulation with descent near the
poles and ascent near the equator. Figure \ref{move_deep} shows the
vertically and zonally averaged equator--to--pole temperature
difference (in the sense $T_{\rm equator}-T_{\rm pole}$), and total
kinetic energy, for the radiatively inactive region (i.e. $p\geq
1\times10^6$ Pa or 10 bar), for the HD 209458b test case and each
equation set. Figure \ref{move_deep} shows evidence that the
latitudinal temperature gradient in the deep atmosphere, and the
kinetic energy, are approaching a steady state in the ``Shallow''
case. However, for both the ``Deep'' and ``Full'' cases the average
latitudinal temperature gradient and total kinetic energy, are both
still increasing by the end of the simulation. Additionally, Figure
\ref{move_deep} shows that the average temperature difference between
the equator and pole is larger in the ``Deep'' and ``Full'' cases than
in the ``Shallow'' case, as is the total kinetic energy, in the
radiatively inactive region. Figure \ref{move_deep}, therefore, gives
a strong indication both that the more simplified equation sets poorly
represent the dynamical evolution of the deep atmosphere, and that the
more sophisticated cases require a longer time to reach a
statistically steady state.

The radiative timescale at the bottom of the radiative zone is
$\tau_{\rm rad}\sim 500$ days, for HD 209458b. Therefore, given that
below this the radiative timescale is infinite one would expect the
timescale for relaxation of the radiatively inactive region to be
$>>500$ days. The total elapsed time for the test cases performed in
this work is 1200 days, suggesting it is unlikely the deep atmosphere
will reach a relaxed state (an estimate supported by the data
presented in Figure \ref{move_deep}). Given that the angular momentum,
and kinetic energy budget of the atmosphere can potentially be
dominated by the deepest regions of the atmosphere, and the relaxation
time of the deeper layers is long (compared to model elapsed times),
it suggests that partially relaxed solutions to the entire atmospheric
flow may not be persistent equilibrium states. It has been shown, for
models solving the HPEs, that the results of such simulations are
invariant to initial conditions \citep[for discussion
see][]{liu_2012}. However, as discussed in Section \ref{approx} the
NHD equations include terms which act to exchange momentum between the
vertical and horizontal motion. This exchange couples the shallow and
deep atmosphere over long timescales meaning invariance to initial
conditions cannot be proven until a statistical steady state
\emph{throughout} the model domain is reached. The alteration of the
flow as the deeper layers become activated may lead to the
establishment of a different equilibrium state \citep[multiple
equilibria are discussed in][]{liu_2012}, or it may move through a
transient phase.

Problematically, for models such as the UM, and more specifically the
ENDGame dynamical core, which solve the NHD equations, the interaction
with the deeper layers is extremely slow and therefore exploration of
this element may require huge computer resources \citep[i.e. long
integration times as mentioned in][]{showman_2008}. As a note of
warning \citet{viallet_2008} demonstrate that for simulations of dwarf
novae, where one side is strongly irradiated by the primary star,
divergent flow is found, but no statistical steady state has been
reached.

\begin{figure*}
\centering
  \hspace*{-0.7cm}\includegraphics[width=7.0cm,angle=90]{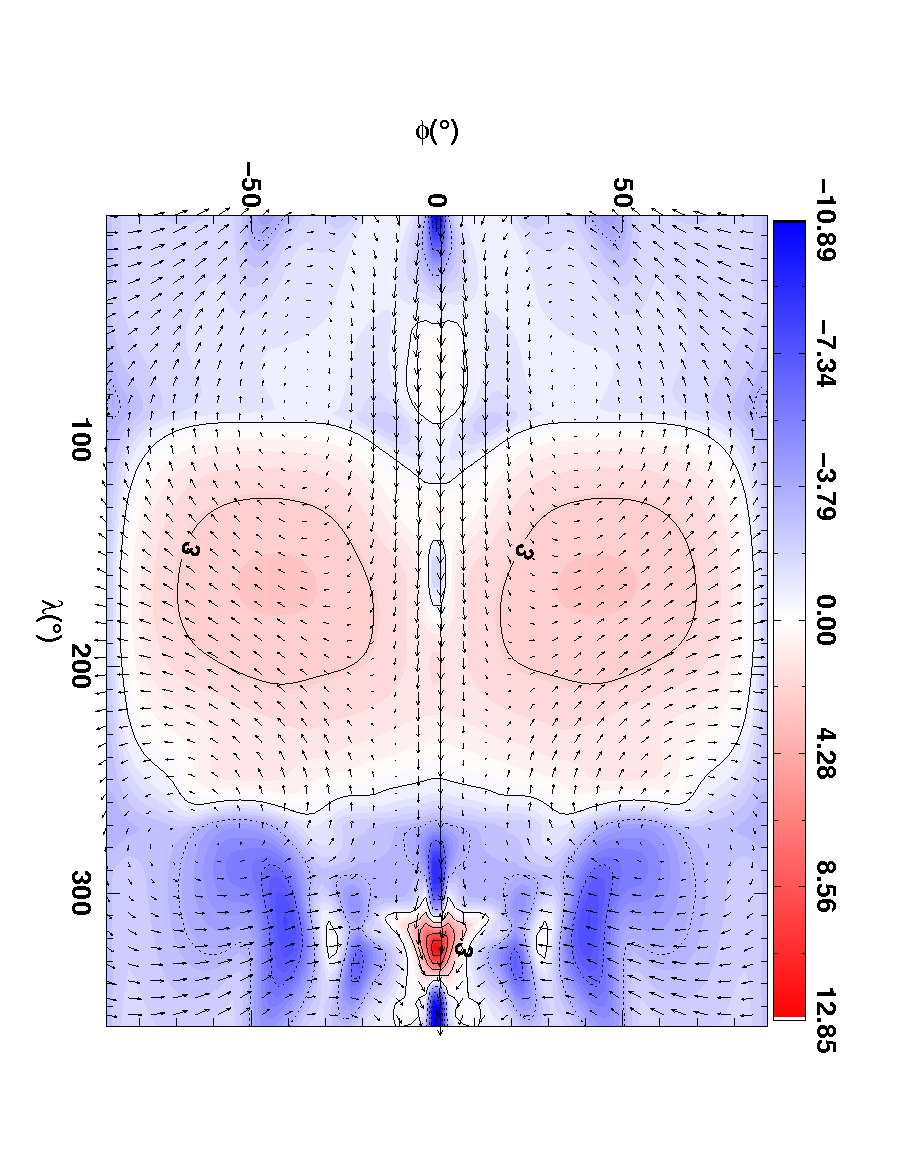}
  \hspace*{-0.7cm}\includegraphics[width=7.0cm,angle=90]{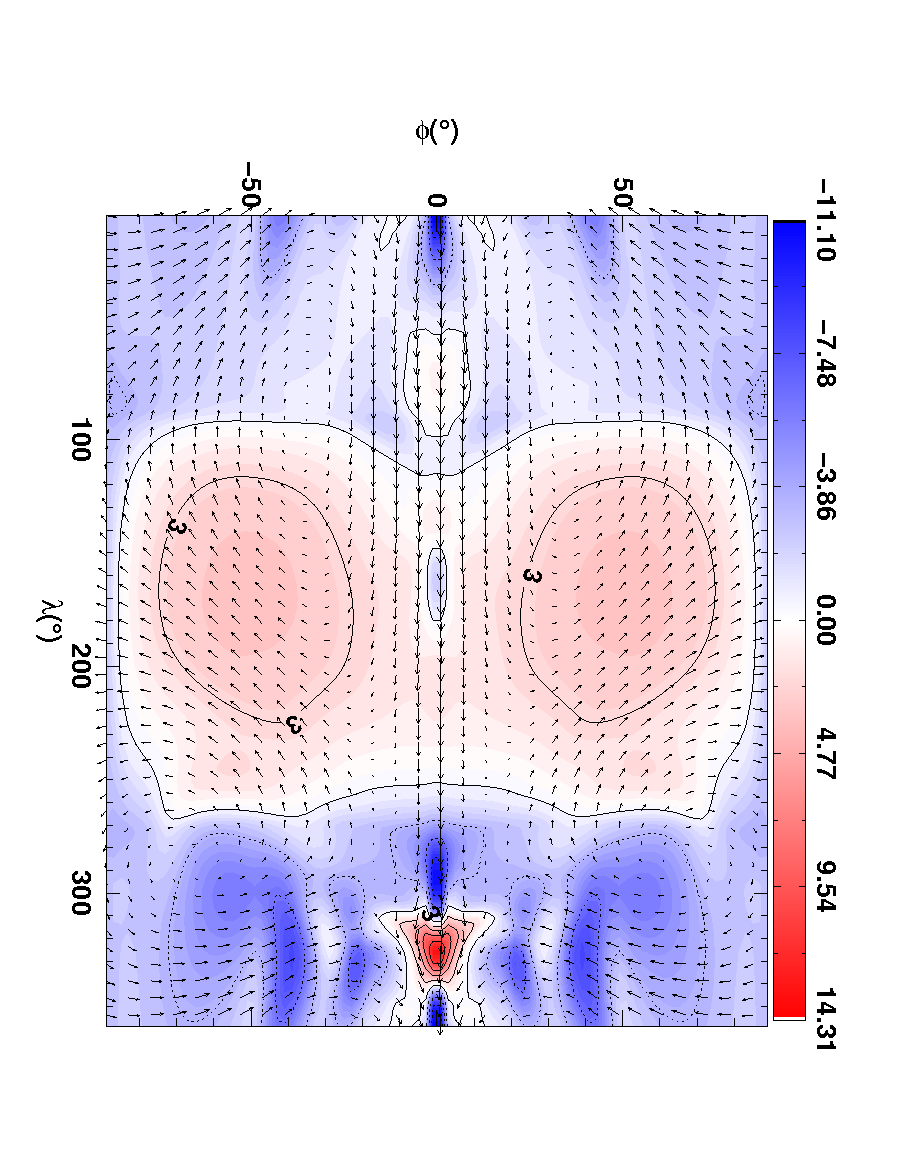}
  \hspace*{-0.7cm}\includegraphics[width=7.0cm,angle=90]{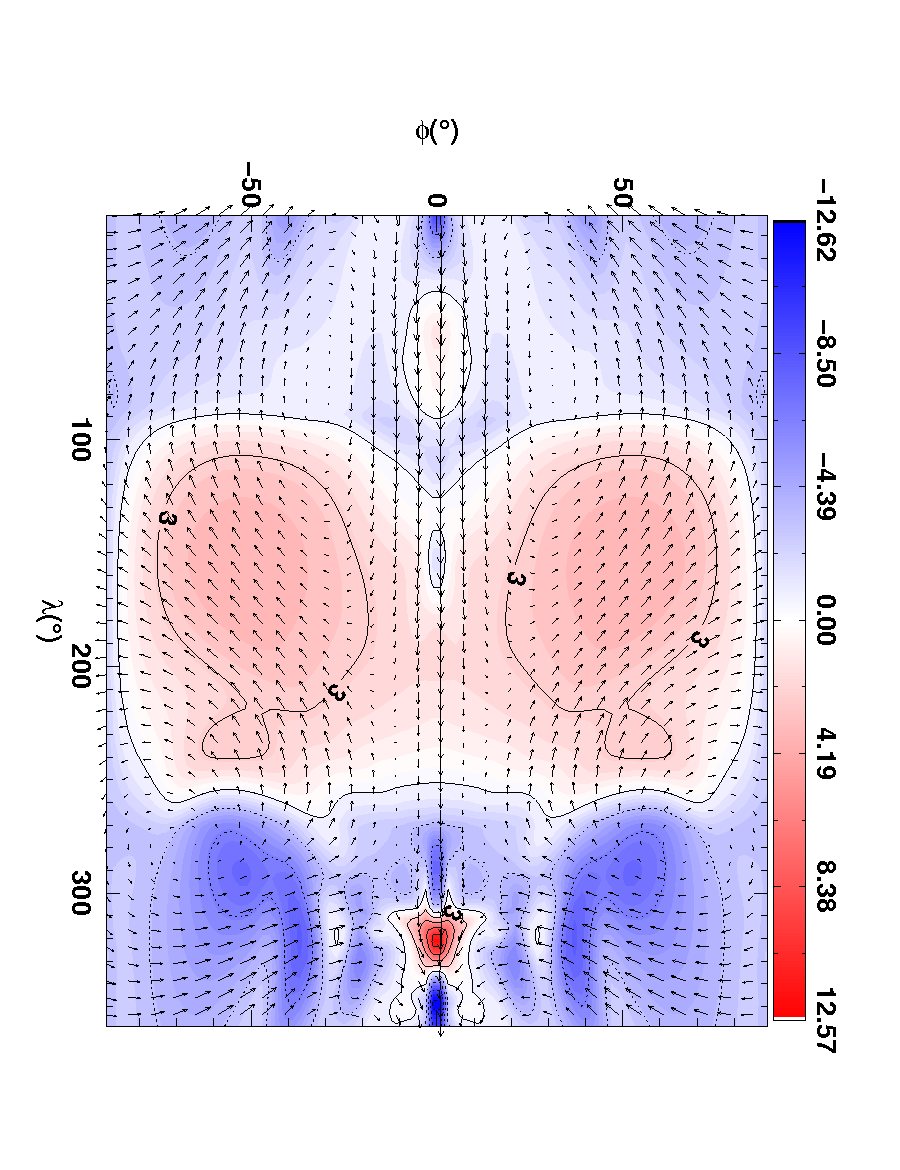}
  \hspace*{-0.7cm}\includegraphics[width=7.0cm,angle=90]{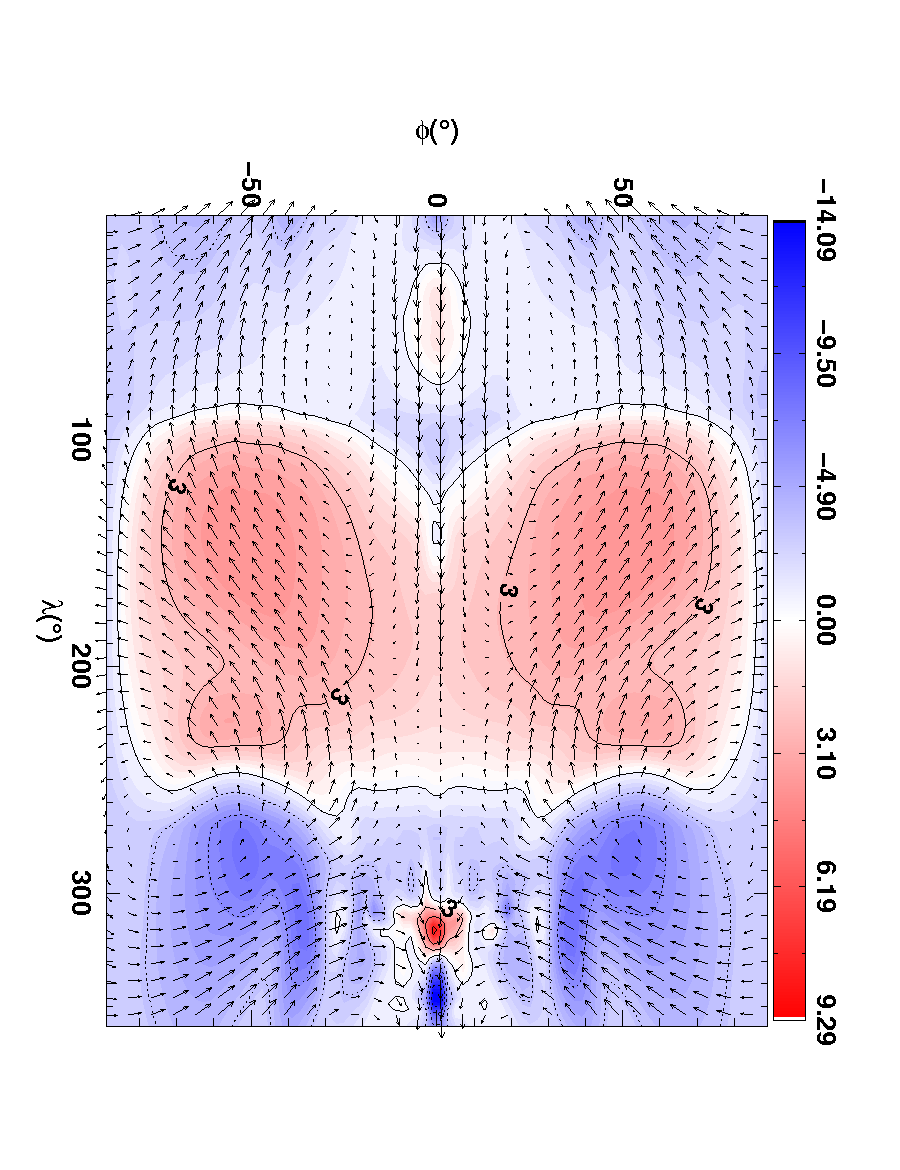}
  \caption{Figure showing the horizontal velocity (vector arrows) and
    vertical velocity (colour scale) for the ``Full'' case (see Table
    \ref{model_names} for explanation) at $1\times 10^5$ Pa (1 bar)
    and after 100 (\textit{top left}), 400 (\textit{top right}), 800
    (\textit{bottom left}) and 1200 (\textit{bottom right})
    days. Although the colour scales differ, the contour lines are the
    same for \textit{all panels}.}
 \label{jet_decay}
\end{figure*}

For our simulations, the zonal mean plots all show a prograde
equatorial jet, demonstrating insensitivity of the mechanism which
produces this feature to the dynamical equations used, over 1200
days. However, given that the radiatively inactive region, for the
``Deep'' and ``Full'' cases is clearly still evolving, one might
expect the lower pressure regions of the atmosphere to also
demonstrate evolution. The zonal mean plots show that the prograde
equatorial jet is thinned (in latitude), contracted in height and
diminished in magnitude, in the ``Deep'' and ``Full'' cases when
compared to the ``Shallow'' case. Looking in detail at the time
evolution of the flow one finds a largely invariant structure
throughout most of the atmosphere in the ``Shallow'' case, where
exchange between the vertical and horizontal momentum is
inhibited. However, both the ``Deep'' and ``Full'' cases exhibit a
varying large--scale flow structures. Figure \ref{jet_decay} shows
slices through the ``Full'' case at $1\times 10^5$ Pa (or 1 bar)
at 100, 400, 800 and 1200 days (\textit{top left}, \textit{top right},
\textit{bottom left} and \textit{bottom right panels},
respectively). 

The slices in Figure \ref{jet_decay} show horizontal velocity
(vectors) and the vertical velocity (colour scale). In this case (as
is evident to a lesser degree in the ``Deep'' case) the large--scale
flow is clearly still evolving. As the simulations runs the eastward
jet, which is `spun--up' in the first tens of days, gradually degrades
and westward flow encroaches across the equator. This effect is seen,
to differing degrees, throughout the atmosphere and leads to the
thinning and diminishing of the jet when performing a zonal
average\footnote{We perform zonal averaging after our data have been
  transformed into pressure space to match the models of
  \citet{heng_2011}, and thereby avoid problems of comparison of
  quantities zonally averaged along different iso--surfaces \citep[as
  described in the appendix of][]{hardiman_2010}.}. It is intriguing,
that the departure of our results from the results of
\citet{heng_2011} is most apparent when the constant gravity
approximation is relaxed. This change acts to weaken the
stratification and thereby increase the efficiency of vertical
transport via, for instance, gravity waves.

\begin{figure*}
\centering
  \hspace*{-0.7cm}\includegraphics[width=7.0cm,angle=90]{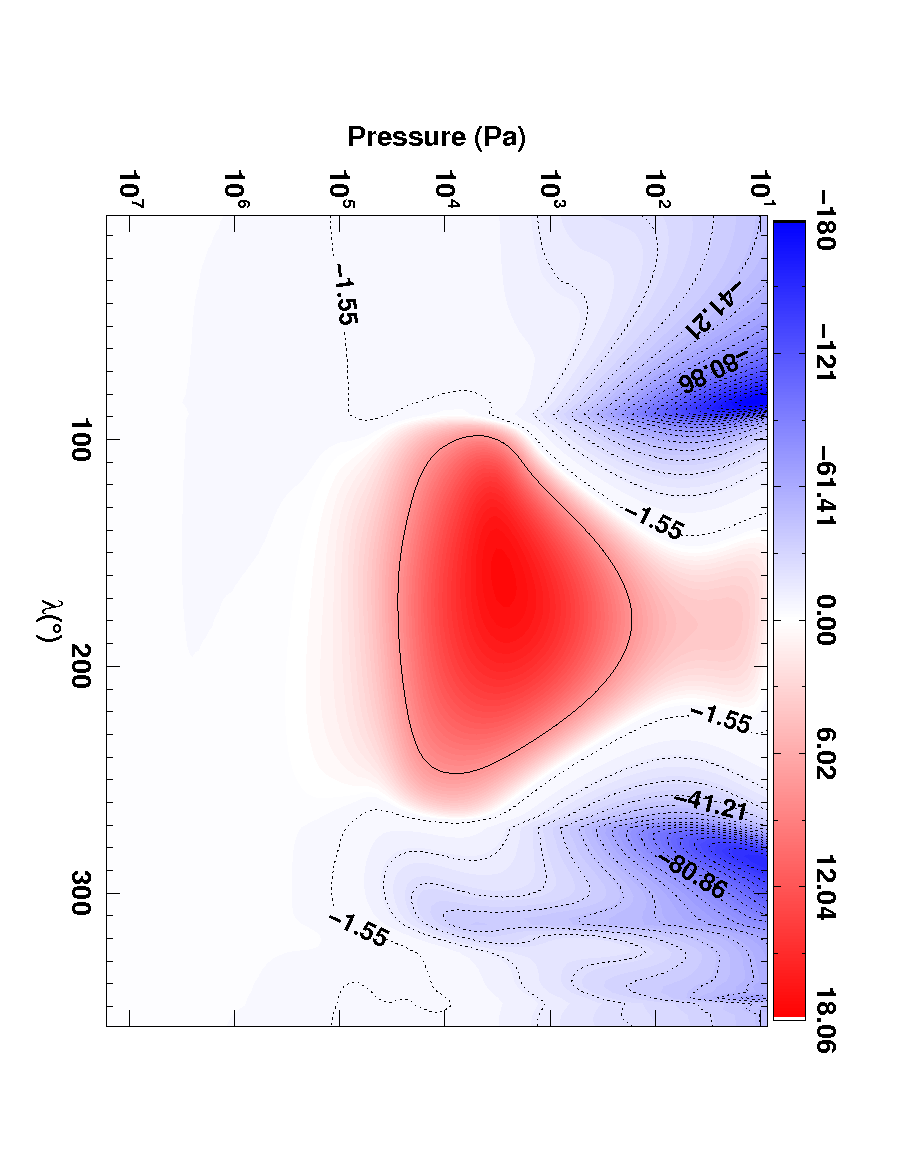}
  \hspace*{-0.7cm}\includegraphics[width=7.0cm,angle=90]{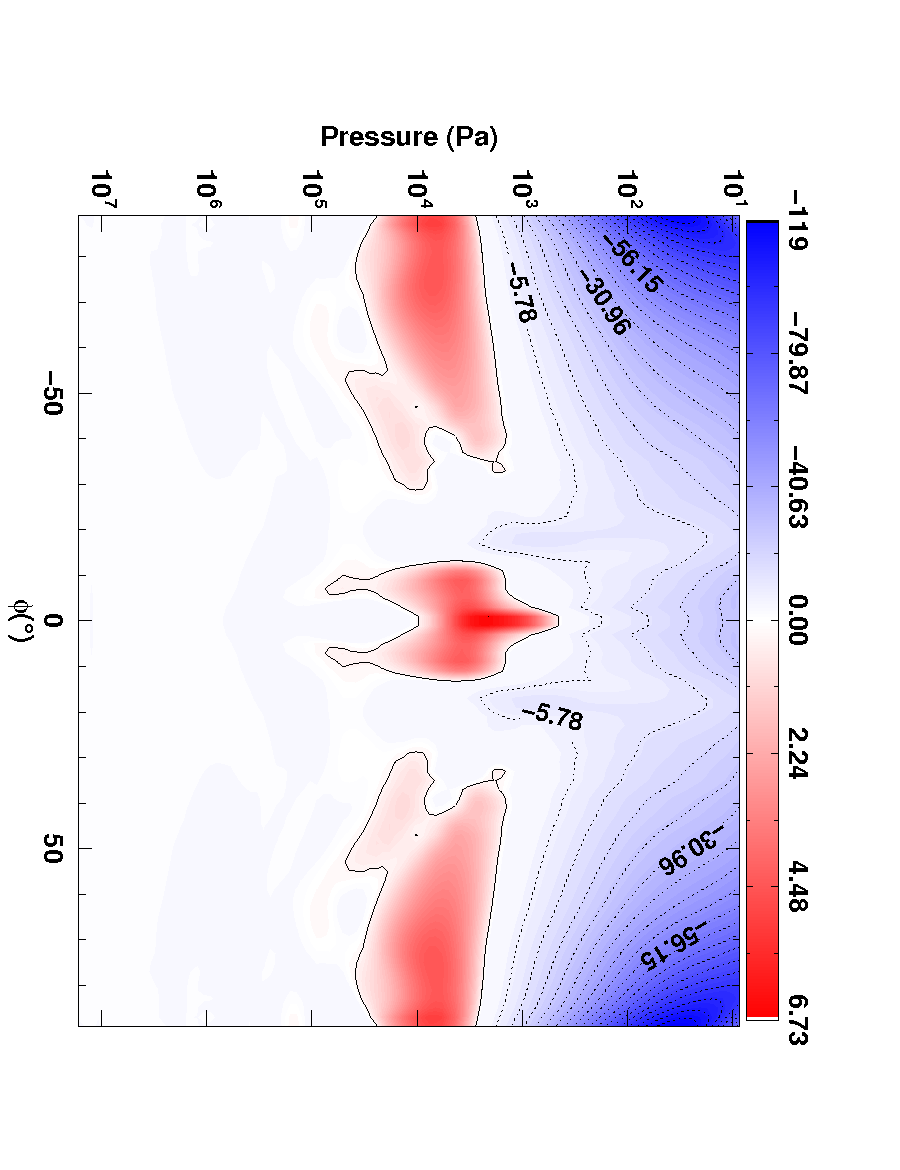}
  \hspace*{-0.7cm}\includegraphics[width=7.0cm,angle=90]{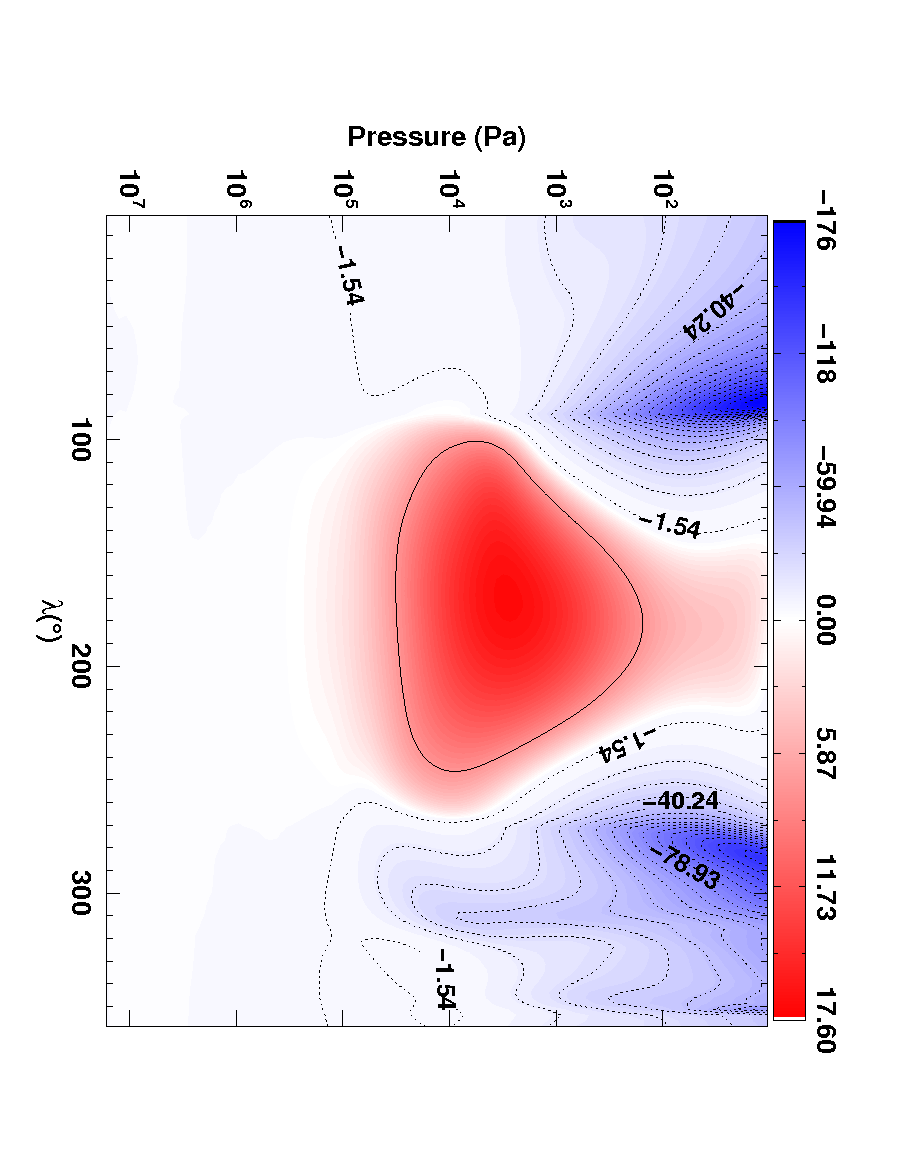}
  \hspace*{-0.7cm}\includegraphics[width=7.0cm,angle=90]{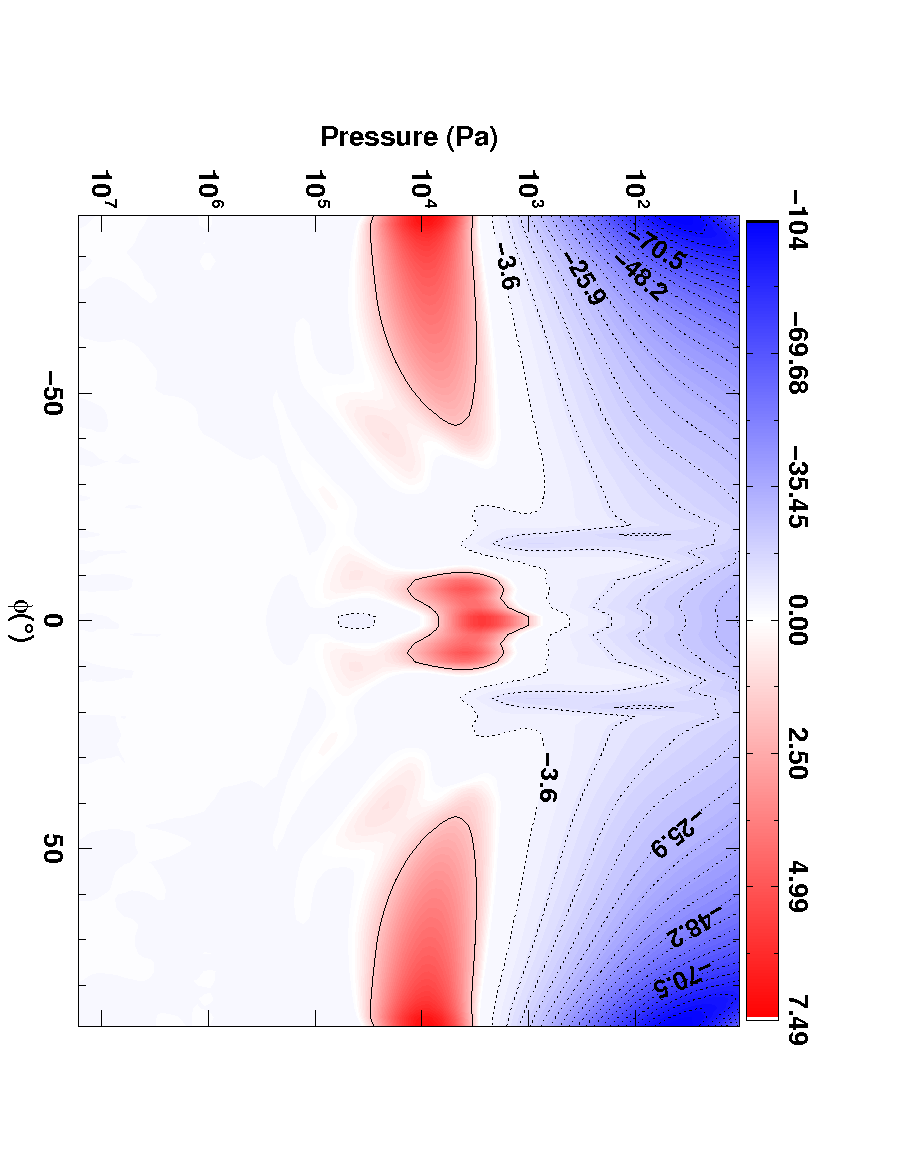}
  \hspace*{-0.7cm}\includegraphics[width=7.0cm,angle=90]{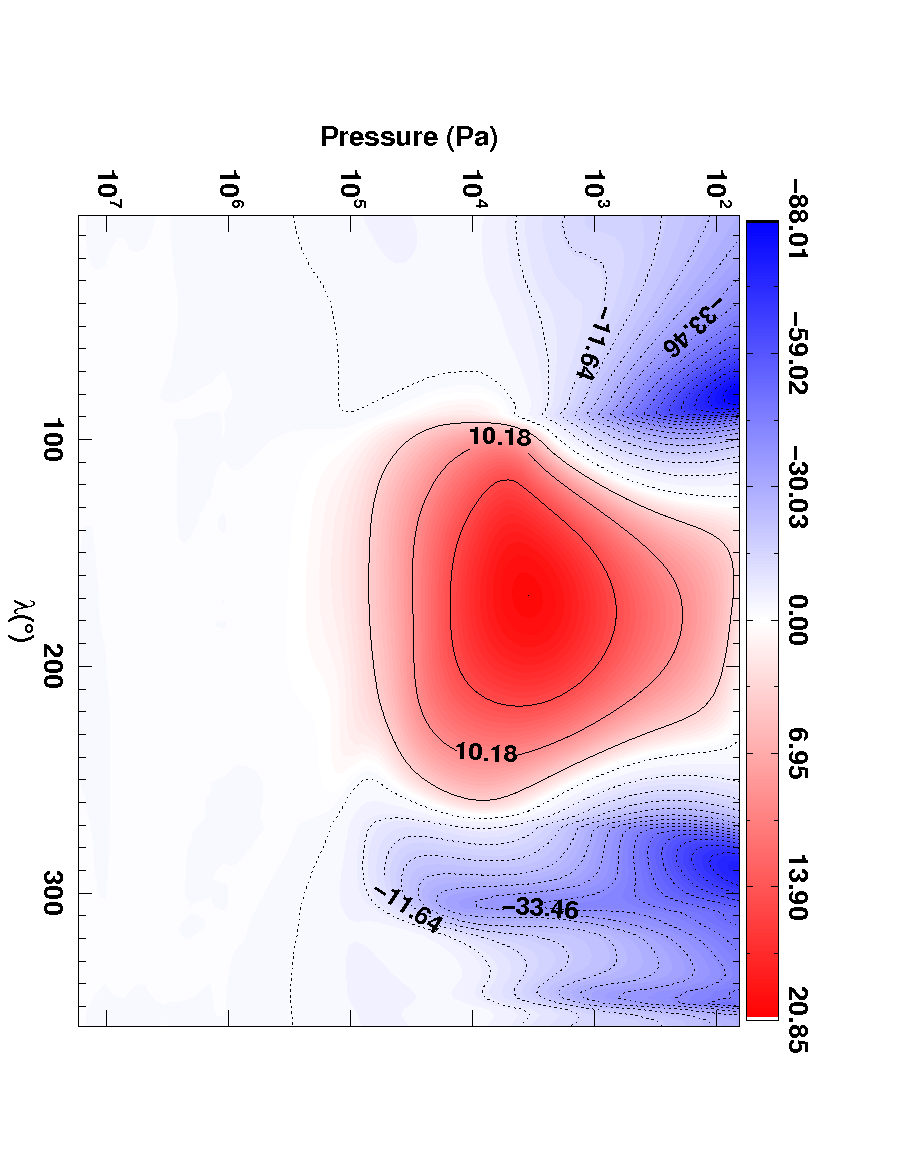}
  \hspace*{-0.7cm}\includegraphics[width=7.0cm,angle=90]{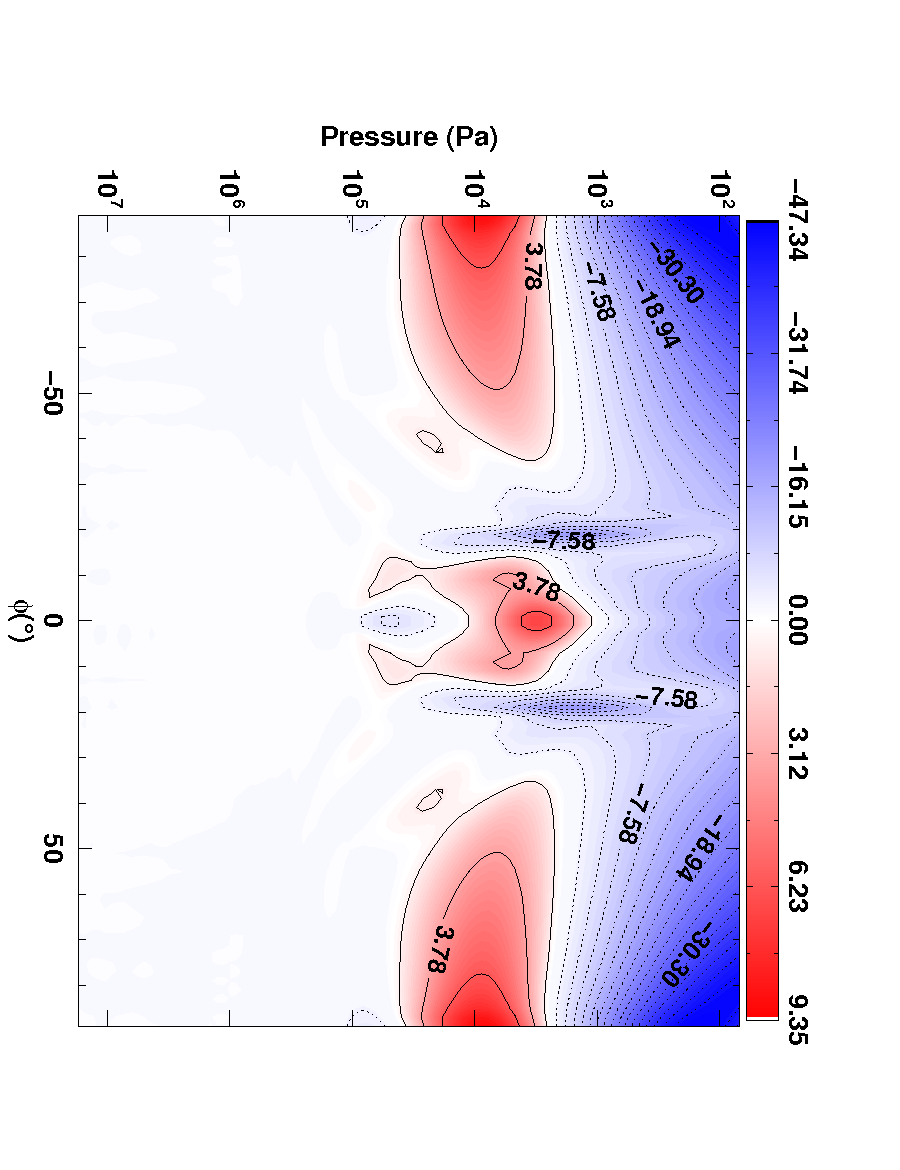}
  \caption{Figure showing vertical velocity, as a function of
    pressure, for the ``Shallow'', ``Deep'' and ``Full'' cases (see
    Table \ref{model_names} for explanation) as the \textit{top},
    \textit{middle} and \textit{bottom panels} respectively. The
    \textit{left} and \textit{right panels} show vertical velocity as
    a function of longitude where a meridional average (performed in a
    point--wise fashion, i.e. $\int\,vd\,\phi$ as opposed to
    $\int\,\cos\phi vd\,\phi$, to emphasise differences in the
    vertical flow towards the polar regions) has been performed, and
    of latitude where a zonal average has been performed,
    respectively.}
  \label{wvel_avg}
\end{figure*}

Figure \ref{wvel_avg} shows the time averaged (from 200 to 1200 days)
vertical velocities for the ``Shallow'', ``Deep'' and ``Full'' cases,
as a function of pressure and either longitude (\textit{left panels})
or latitude (\textit{right panels}). In each case the field has been
averaged in the horizontal dimension not plotted, i.e. if plotted as a
function of latitude it has been zonally averaged. The meridional
average is performed in a point--wise fashion, i.e. $\int\,vd\,\phi$
as opposed to $\int\,\cos\phi vd\,\phi$, to emphasise differences in
vertical flow towards the polar regions. Figure \ref{wvel_avg} shows
some significant differences in the vertical velocity profiles between
the simulations. Firstly, the \textit{left panels} show the
meridionally averaged updraft is stronger, broader (in longitude) and
larger in vertical extent in the ``Full'' case. However, the ``Deep''
case appears similar to the ``Shallow''. Secondly, the \textit{right
  panels} show, for the zonally averaged vertical circulation, as we
move from the ``Shallow'' to ``Deep'' to ``Full'' cases, the updraft
at the equator, and over the poles, strengthens, and the fast flowing
downdrafts flanking the jet (in latitude), become stronger. Similar to
jets on Earth, the regions flanking the jet are baroclinically
unstable and will, therefore, generate eddies and perturbations, such
as atmospheric Rossby waves. The interaction of these perturbations
with the mean flow provides a mechanism to move energy (and angular
momentum) up--gradient, i.e. into the jet, and therefore sustain the
jets against dissipation.
 
\citet{showman_2011b} show that the jet pumping mechanism for hot
Jupiters is unlikely to be similar to that relevant to Earth's
mid--latitude jets, i.e. the poleward motion of atmospheric Rossby
waves. In fact the likely culprit, given the planetary scale of Rossby
waves for hot Jupiters, is the interaction between standing
atmospheric waves and the mean flow. Such standing waves are planetary
in scale, and therefore are certainly poorly represented by any model
which adopts the `traditional' approximation \citep[as discussed
in][]{white_1995}. Additionally, \citet{showman_2011b} show that the
vertical transport of eddy momentum is a vital ingredient in the
balance of superrotation at the equator. Therefore, it is clear that
altering the efficiency of vertical transport will affect this
mechanism, leading to a change in the balance of the pumping of the
jet. Work is in progress to fully investigate this issue, which
requires simulations run for a significantly longer integration time
(Mayne et al, in preparation).

\section{Conclusion}
\label{conclusions}

We have presented the first application of the UK Met Office global
circulation model, the Unified Model, to hot Jupiters. In this work we
have tested the ENDGame dynamical core (the component of a GCM which
solves the equations of motion of the atmosphere) using a shallow--hot
Jupiter \citep[SHJ, as prescribed in][]{menou_2009} and a HD 209458b
test case \citep{heng_2011}. This work represents the first results of
such test cases using a meteorological GCM solving the
non--hydrostatic, deep--atmosphere equations. We have also completed
the test case using the same code under varying levels of
simplification to the governing dynamical equations. This work is
complementary to the testing we have performed modelling Earth--like
systems \citep{mayne_2013}.

In this work we suggest that, when relaxing the canonical
simplifications made to the dynamical equations, the deeper regions of
the radiative atmosphere, and the radiatively inactive regions, do not
reach a steady state and are still evolving throughout the 1200 day
test case. We have found that moving to a more complete description of
the dynamics activates exchange between the vertical and horizontal
momentum, and the deeper and shallower atmosphere. This leads to a
degradation of the eastward prograde equatorial jet, and could
represent either the beginnings of a new equilibrium state or multiple
states, which may be dependent on the initial conditions of the
radiatively inactive region of the atmosphere. In a future work we
will investigate longer integration times, and explore the effect of
simplifications to the dynamical equations on examples of jet pumping
mechanisms in these objects. These results suggest that the test cases
performed are not necessarily good benchmarks for a model solving the
non--hydrostatic, deep--atmosphere equations.

We also aim to investigate the importance of the deeper atmosphere,
and therefore, move the inner boundary for a HD 209458b simulation
much deeper to $\sim 10^8$ Pa (or kbar) levels. This will require
adaptation of the equation of state and increased flexibility in the
prescription of $c_p$. These test cases have been performed using a
Newtonian cooling scheme. As discussed in \citet{showman_2009} such a
scheme does not include blackbody thermal emission of the gas itself,
which can be significant when a region of heated material is advected
into a region of net cooling. Using such a scheme the gas is just
arbitrarily heated or cooled without taking into account its
re-radiation into the surrounding area. In fact, as only the
temperature is adjusted without knowledge of the specific heat
capacity or quantity of material in a given cell (nor its optical
properties) the energy deposited (or removed) from the system is
unrepresentative. These problems can lead to regions where the heating
or cooling is artificially high. To correct this we are adapting a
non--grey radiative transfer scheme, under the two--stream
approximation, which will be coupled to the UM dynamical core,
ENDgame, under hot Jupiter conditions. The subsequent comparison to
observations will be performed with a more physically meaningful
model, once the coupling of our adapted schemes is complete.

The UM GCM is a powerful tool with which to study the effect, on the
predicted states of exoplanet atmospheres, of both the interaction
between the observable and deep atmosphere, and canonically made
approximations to the governing dynamical equations. The ability to
alter the level of simplification of the underlying dynamical
equations (as provided by the ENDGame dynamical core) will prove vital
as we explore exotic climate regimes where assumptions based on
Earth's atmosphere cannot \textit{a priori} be assumed valid.

\begin{acknowledgements}
  Firstly, we would like to extend our gratitude to the referee whose
  thoughtful and clearly expounded comments greatly improved this
  manuscript. We would also like to thank Tom Melvin for his expert
  advice. This work is supported by the European Research Council
  under the European Communitys Seventh Framework Programme
  (FP7/2007-2013 Grant Agreement No. 247060) and by the Consolidated
  STFC grant ST/J001627/1.This work is also partly supported by the
  Royal Society award WM090065. The calculations for this paper were
  performed on the DiRAC Facility jointly funded by STFC, the Large
  Facilities Capital Fund of BIS, and the University of Exeter.
\end{acknowledgements}

%-------------------------------------------------------------------
\bibliographystyle{aa}
\bibliography{references}

\end{document}